\documentclass[prresearch, 10pt, twocolumn, superscriptaddress]{revtex4-2} 

\usepackage[T1]{fontenc}
\usepackage[utf8]{inputenc}
\usepackage{multirow}
\usepackage{dcolumn}
\usepackage{amsmath}
\usepackage{amsfonts}
\usepackage{amssymb}
\usepackage{tabularx}
\usepackage{bbm, dsfont}
\usepackage[breaklinks,colorlinks,bookmarks=false,citecolor=blue,linkcolor=blue,urlcolor=blue]{hyperref}

\usepackage{pst-plot}

\usepackage{orcidlink}

\makeatletter
\newcommand*{\rom}[1]{\expandafter\@slowromancap\romannumeral#1@}
\makeatother

\newcommand{\nn}{\nonumber}

\newcommand{\vb}[0]{V$_\text{B}^-$}
\newcommand{\cv}[1]{\left(#1\right)}
\newcommand{\cvb}[1]{\left[#1\right]}
\newcommand{\cvc}[1]{\left\{#1\right\}}
\newcommand{\cvv}[1]{\left\vert #1\right\vert}
\newcommand{\ket}[1]{\vert#1\rangle}
\newcommand{\bra}[1]{\langle#1\vert}
\newcommand{\ketbra}[2]{\ket{#1}\bra{#2}}

\newcommand{\iden}{\mathbbm{1}}

\newcommand{\tra}{\text{Tr}}
\newcommand{\sinc}{\text{sinc}}

\newcommand{\add}[1]{#1}%{\textcolor{blue}{#1}}
\newcommand{\highlight}[1]{#1}%{\textcolor{blue}{#1}}

\begin{document}

\title{Arbitrary Manipulation of Nuclear Spins in hexagonal Boron Nitride}
\author{Fattah Sakuldee\,\orcidlink{0000-0001-8756-7904}}
\email{fattah.sakuldee@sjtu.edu.cn}
\affiliation{Wilczek Quantum Center, School of Physics and Astronomy, Shanghai Jiao Tong University, 200240 Shanghai, China}
\author{Mehdi Abdi\,\orcidlink{0000-0002-9681-5751}}
\email{mehabdi@gmail.com}
\affiliation{Department of Physics, Isfahan University of Technology, Isfahan 84156-83111, Iran}
\affiliation{Wilczek Quantum Center, School of Physics and Astronomy, Shanghai Jiao Tong University, 200240 Shanghai, China}

\begin{abstract}
Due to its localized nature and controllability, the negatively charged boron vacancy centers (\vb) in hexagonal boron nitride (hBN) are a promising spin platform for accessing its neighboring nuclei with potential for performing quantum computational tasks. However, the methods of utilizing and manipulating the nuclear spins are still lacking. In this work, we propose a protocol for the preparation of single- and multi-qubit gates on the nuclear spins, utilizing the electron spin as an auxiliary qubit. By applying a background magnetic field and a multi-tone continuous drive, we show that the electron spin coupling to the nuclei can be efficiently engineered. This allows for suppressing the undesired electron-nuclear interactions through the Hahn echo pulse sequence. The target gates are then implemented by employing proper RF drives. Our numerical results for realistic parameters show gate fidelities as high as $99\%$ for single-qubit and $95\%$ for multi-qubit gates. With the gate execution durations being less than $300$ ns, our protocol evades electron spin decoherence effects. Therefore, our scheme sets the stage for the practical application of \vb\ in hBN for quantum computation purposes.
\end{abstract}

\date{\today}
        
\maketitle

\section{Introduction}
%%%%%
%\emph{\textbf{Introduction}.}---
In the past decade, color centers have been observed experimentally in the layered hexagonal boron nitride (hBN) structures~\cite{Tran2015, Gottscholl2021, Stern2022, Mu2022} and the subsequent theoretical investigations~\cite{Abdi2018, Sajid2020} suggested the highly controllable electronic spin states; the accessibility for initialization and control via optical and microwave pulses of the negatively charged boron vacancy centers \vb\ (which we refer to as VB-center) has also been reported in experiments~\cite{Gottscholl2020}. 
It has then become an interesting quest to utilize these advantages for quantum applications. Potential physical implications are raging from a host for quantum emitters~\cite{Cakan2025, Ahmadi2024, Cholsuk2022, Wolfowicz2021, Chejanovsky2016, Proscia2018, Exarhos2019} to quantum manipulation of other degrees of freedom and quantum sensing~\cite{Tabesh2021, Liu2019, Abdi2021, Kianinia2020, Shaik2021, Gao2021, Mendelson2021, Guo2022, Vaidya2023, Das2024}.
The basic initialization, control, and readout of the electron spin of a VB-center in hBN has been well established, e.g., in Refs.~\cite{Gottscholl2021, Gottscholl2020, Wong2015}; however, the extension to incorporate other degrees of freedom, including the nuclear spins in hBN, into the manipulation scheme in the system is limited. 

\add{
A promising candidate for quantum computation is the cluster of a VB-center and its three neighboring nitrogen nuclear spins, \highlight{which here we refer to as} VB3N. The strong coupling of the nuclei to the electron spin, along with the natural lattice symmetry and orderly distribution of the clusters through the lattice, promises a potentially scalable platform. The nuclear spins can be efficiently initialized~\cite{Gao2022, Tabesh2023}, and synchronous operations can be performed \highlight{by using dynamical decoupling techniques}~\cite{Sakuldee2025}. However, due to the high symmetry and strong isotropic interactions, precise control over individual nuclear spins in the cluster remains challenging, especially when the magnetic field is aligned with the defect symmetry axis~\cite{Sakuldee2025}. 
}

\add{
In this \highlight{article}, we propose a protocol for a wide class of quantum operations including single- and multi-qubit gates, on the nuclear spins in a VB3N cluster, using the interaction asymmetry induced by a proper background magnetic field orientation.
Such operations are particularly feasible thanks to the recent development in preparing highly isotopically purified samples of \vb-hB${}^{15}$N~\cite{Lee2025, Liu2025}.
We show that a purely in-plane magnetic field, oriented asymmetrically relative to the \highlight{defect} dangling bonds, allows one to individually address each nuclear spin by engineering radio frequency (RF) drives. 
By applying an approperiate combination of RF frequencies, \highlight{and engineering their amplitude and phases}, rotation gates can be prepared in the interaction dressed frame, and with the help of \highlight{a Hahn-echo pulse sequence}, the gates are then transferred to the computational frame.
}

%%%%%%%%%%%%%%%%%%%%%%%%%%%%%%%%%%%%%%%%%%%%%%%%%%%%%%
%%%% FORMULATION
%%%%%%%%%%%%%%%%%%%%%%%%%%%%%%%%%%%%%%%%%%%%%%%%%%%%%%
\begin{figure}[t]
    \centering
    \includegraphics[width=1\columnwidth]{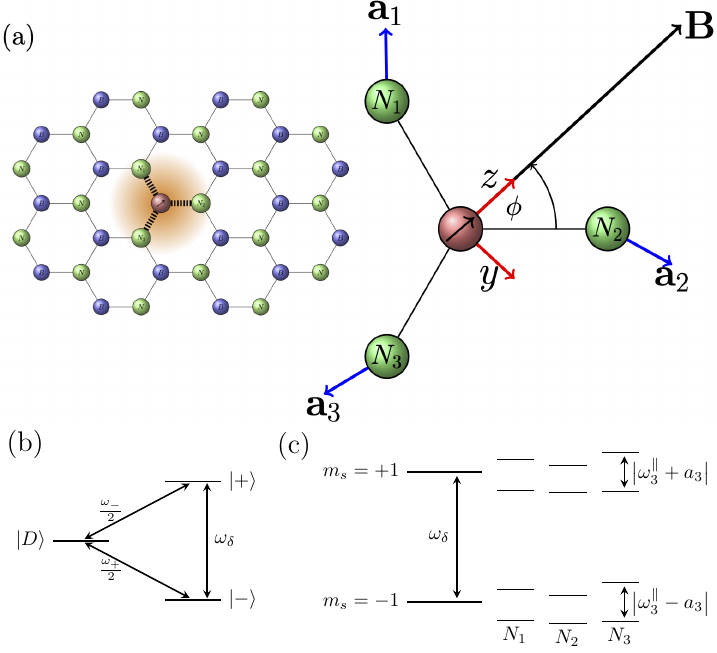}\\
    \caption{
 (a) The boron vacancy center in a hBN lattice is depicted on the left, with a close-up on the right showing the VB3N cluster with the electron spin (red ball) at the vacancy center and three neighboring spins (green balls). An in-plane magnetic field $\mathbf{B}$ is applied at an angle $\phi$ from a dangling bond near the vacancy. The blue arrows indicate the effective hyperfine interaction vectors $\boldsymbol{a}_k$.
 (b) The magnetic field induces a non-degenerate three-level system for the electron Hamiltonian $\hat{H}_e$, allowing for the identification of a two-level subsystem.
 (c) Each electron level's conditional Hamiltonian for the nuclear spins can be derived, with eigenstates from $\hat{H}_{\rm reduced}$ serving as computational bases for nuclear spin qubits. Up to seven control frequencies are utilized to manipulate cluster dynamics and achieve the target gate.
}\label{fig:hBN}%
\end{figure}

%basic model
\section{The Model}
%%%%%
%\emph{\textbf{The Model}.}---
In this work, we consider a VB3N spin cluster in hB${}^{15}$N, consisting of one electron spin associated with the VB-center and \highlight{the three nearest nitrogen $^{15}$N nuclei}.
The Hamiltonian describing the system is $\hat{H}_{\rm sys} = \hat{H}_e + \hat{H}_n + \hat{H}_i$ with the components ($\hbar = 1$): $\hat{H}_e = D_{gs}\hat{S}_x^2 + \gamma_e \mathbf{B}\cdot\mathbf{S},$ where the $x$-axis denotes the axis perpendicular to the plane of hBN layers, while $(z,y)$ are the in-plane coordinates, $D_{gs}$ is the zero-field splitting of VB-center groundstate triplet and $\gamma_e$ is the electron gyromagnetic ratio. Here, $\mathbf{B}$ is the vector of background magnetic field, while $\mathbf{S} {=} (\hat{S}_x,\hat{S}_y,\hat{S}_z)^\mathrm{T}$ is the vector of spin-$1$ operators. 
The Hamiltonian of the nuclear spins reads $\hat{H}_n = \sum_k\gamma_k \mathbf{B}\cdot\mathbf{I}_k$ where $\gamma_k$ and $\mathbf{I}_k$ are the nuclear gyromagnetic ratio and the spin operators vector, respectively. For the ${}^{15}$N nuclear spin, which is spin-$\frac{1}{2},$ there is no electric quadrupole interaction, and the magnetic dipole-dipole interaction can be omitted in our consideration due to its negligibility. 

The electron and the three nearest neighbor nuclei interact dominantly through Fermi contact~\cite{Liu2025}, which is strong compared to the dipolar interaction with other parties. This allows us to consider this cluster isolated from the rest of the spin lattice within the relevant timescale.
The interaction Hamiltonian reads $\hat{H}_i =  \sum_k\mathbf{S} \cdot\mathbb{A}_k\cdot\mathbf{I}_k$ where $\mathbb{A}_k$ is the hyperfine tensor of the $k$th spin, given by
 \begin{equation}
 \mathbb{A}_k = \cv{ \begin{array}{ccc}
 A_k^{zz} & A_k^{zy} & 0\\
 A_k^{zy} & A_k^{yy} & 0\\
   0 & 0 & A_k^{xx}
  \end{array}}\label{eq:A_k}.
 \end{equation}
To explain the system dynamics, we adopt the following conventions:
We express the electron spin operators with respect to the magnetic field direction, i.e., a $z-$axis is given by the magnetic field direction.
Note that due to the geometrical nature of the hyperfine interaction, the geometric $D_{3h}$ symmetry of the point defect is inherited by the hyperfine interaction matrix $\mathbb{A}_k$, see Refs.~\cite{Ivady2020, Gracheva2023}.

%diagonalise and reduce in one paragraph
By diagonalizing the electron Hamiltonian one has $\hat{H}_e \Longrightarrow {\rm diag}\cv{\omega_+/2,D_{gs},-\omega_-/2},$ where $\omega_\pm = \omega_\delta \pm D_{gs},$ $\omega_\delta=\sqrt{D_{gs}^2 + 4\omega_e^2},$ and $\omega_e = \gamma_eB,$ with the eigenstates denoted by $\ket{+},\ket{D}$ and $\ket{-}$, respectively. 
See the level diagram in Fig.~\ref{fig:hBN}(b). 
One can reduce the electron Hilbert space into two by driving the $\ket{+}{\leftrightarrow}\ket{-}$ transition by an appropriate MW field, resulting in the reduced Hamiltonian $\hat{H}_{\rm reduced} = \frac{\omega_\delta}{2}\hat{\sigma}_z + \sum_k\omega_n\cv{\hat{\mathbf{z}}\cdot{\mathbf{I}}_k} + \hat{\sigma}_z\sum_{k}{\boldsymbol{a}}_k\cdot\mathbf{I}_k - \hat{\sigma}_x\sum_{k}{\boldsymbol{c}}_k\cdot\mathbf{I}_k,$
where ${\boldsymbol{a}}_k = \cv{{2\omega_e}/{\omega_\delta}}\hat{\mathbf{z}}\cdot\mathbb{A}_k$ and ${\boldsymbol{c}}_k = \cv{{D_{gs}}/{\omega_\delta}}\hat{\mathbf{z}}\cdot\mathbb{A}_k$ are effective diagonal and off-diagonal hyperfine vectors. \highlight{See Appendices~\ref{appen:system} and \ref{appen:electron_subspace}.}
Finally, by moving to the rotating frame with respect to the electron free Hamiltonian $\frac{\omega_\delta}{2}\hat{\sigma}_z$ and assuming $\omega_\delta{\gg} c_k$ for all $k$, %which holds for large enough magnetic fields, 
we obtain the effective Hamiltonian
 \begin{equation}
 \hat{H}_{\rm eff} =  \sum_k\omega_n\cv{\hat{\mathbf{z}}\cdot{\mathbf{I}}_k} + \hat{\sigma}_z\sum_{k}{\boldsymbol{a}}_k\cdot\mathbf{I}_k \label{eq:H_eff}.
 \end{equation}
Here, we consider a magnetic field strength high enough to guarantee $c_k {\ll} \omega_\delta$, and thus, the validity of Hamiltonian \eqref{eq:H_eff}. 
 
The reduced Hamiltonian $\hat{H}_{\rm reduced}$ allows for the selective control over the nuclei from several perspectives. One may consider the trivial treatment where the system operates at a weak coupling regime and the nuclei' Larmor frequencies are resolved. Nonetheless, neither of these are satisfied in our system, as magnetic fields as high as $B{\sim} 10$ T need to be employed.
\highlight{In this article}, we consider a moderate magnetic field $B{\sim} 0.1{-}1$ T and devise a protocol for gate implementation from the effective Hamiltonian \eqref{eq:H_eff} in spite of a strong interaction condition $\omega_n{\ll} a_k$. 
In the following derivation, the frame rotating with respect to the electron Hamiltonian ($\frac{\omega_\delta}{2}\hat{\sigma}_z$) is referred to as the `original' reference frame. 

%reduced to effective Hamiltonian
\section{Rotation Gates}
%%%%%
%\emph{\textbf{Rotation Gates}.}--- 
We first note that by applying the unitary transformation $\exp\{-it\hat{H}_{\rm prime}\}:=\exp\{-it\hat{\sigma}_z\sum_{k}{\boldsymbol{a}}_k\cdot\mathbf{I}_k\}$ to Eq.~\eqref{eq:H_eff} one arrives at 
\begin{equation}
 \widetilde{H}_{\rm rot}(t) = \sum_k\omega_k^\parallel{\hat{I}}_k^z +\omega_k^\perp\cvb{{\hat{I}}_k^y\cos(a_k t) + \hat{\sigma}_z{\hat{I}}_k^x\sin(a_k t)}
\label{eq:H_rot-in-int},
\end{equation}
where $\omega_k^\parallel = \omega_n\hat{\mathbf{z}}\cdot\hat{\mathbf{a}}_k$, $\omega_k^\perp = \vert\omega_n\hat{\mathbf{z}}-\omega_k^\parallel\hat{\mathbf{a}}_k\vert$, ${\hat{I}}_k^z = \hat{\mathbf{a}}_k\cdot\mathbf{I}_k$, and ${\hat{I}}_k^y = (\hat{\mathbf{z}}\times\hat{\mathbf{a}}_k)\cdot\mathbf{I}_k$ with $\hat{\mathbf{a}}_k=\boldsymbol{a}_k/a_k$.
One, thus, exploits the eigenstates of the operator ${\hat{I}}_k^z$ to form a computational basis for the spin $k$.

% %putting control
We apply an RF drive to manipulate the nuclear spins at each electron energy level, enabled by their distinct transition frequencies due to different diagonal coupling [see Fig.~\ref{fig:hBN}(c)].
In particular, we apply the drive Hamiltonian of the form $\hat{H}_{\rm RF}(t) = 2\sum_k\Omega_k(t)\cos(a_kt)\hat{I}_k^x,$ with the Rabi frequencies $\Omega_k(t)$.
In the rotating frame with respect to $\hat{H}_{\rm prime}$ this reads $\widetilde{H}_{\rm RF}(t){\approx}\sum_k\Omega_k(t)\hat{I}_k^x$, by assuming $\cvv{\Omega_k(t)}{\ll} a_k$.
By further moving to the interaction picture with respect to $\widetilde{H}_{\rm RF}(t)$, one has $\mathcal{T}e^{i\int_0^tdt\widetilde{H}_{\rm RF}(t)}\widetilde{H}_{\rm rot}(t)\mathcal{T}e^{-i\int_0^tdt\widetilde{H}_{\rm RF}(t)} = \sum_k \widetilde{H}_k^{\rm rot}(t),$ where $\mathcal{T}$ is the time-order operator and the spin Hamiltonians in the `RF frame' read 
\begin{align}
 \widetilde{H}_k^{\rm rot}(t)
  &\approx \omega_k^\parallel\cvb{\hat{I}_k^z\cos{g_k(t)} + \hat{I}_k^y\sin{g_k(t)}}\nn\\
  & + \omega_k^\perp\cvb{\hat{I}_k^y\cos{g_k(t)}\cos(a_k t) - \hat{I}_k^z\sin{g_k(t)}\cos(a_k t)}\nn\\
  & + \omega_k^\perp\hat{\sigma}_z\hat{I}_k^x\sin(a_k t) \label{eq:H_RF-rot},
\end{align}
with $g_k(t) = \int_0^tds\ \Omega_k(s)$.
Therefore, the time evolution in the `original' frame is given by
 \begin{equation}
 \hat{U}(t) \approx \prod_k e^{-ita_k\hat{\sigma}_z\hat{I}_k^z}e^{-ig_k(t)\hat{I}_k^x}\cvb{\mathcal{T}e^{-i\int_0^t\widetilde{H}_k^{\rm rot}(t)dt}} \label{eq:eq:U-rot-along-int_mod-g}.
 \end{equation}
Having this decomposition, we accomplish our protocol through three main steps: \textit{i}) removal of the 
contribution from $\widetilde{H}_k^{\rm rot}(t)$, \textit{ii}) adding a shift in the RF Hamiltonian to realize the target gates, and \textit{iii}) transferring the gate to the original frame.

%remove the remnants
\paragraph{Removing the unwanted oscillations:}
From the conditions $\cvv{\Omega_k(t)}{\ll} a_k,$ it can be seen that $g_k(t)$ are slowly changing compared to the oscillatory terms.
Hence, we simply set the duration time $\tau$ much larger than the interaction time, i.e. $\tau {\gg} \min\{a_k^{-1}\}.$ 
Moreover, for moderate magnetic fields we have $|\omega_k^\perp|,|\omega_k^\parallel| \ll a_k,$ allowing us to set $\tau{=}1/\omega_{\min}$, with $\omega_{\min}{=}\min\big\{|\omega_k^\parallel|,|\omega_k^\perp|\big\}$, as the operation timescale. 
For constant Rabi frequencies, as the most trivial choice, one has $g_k(t) = 2\pi t/\tau$.
Therefore, for the evolution time $\tau$, the first line in Eq.~\eqref{eq:H_RF-rot} contributes to a trivial global phase, whereas the dynamical contributions from the second and third lines can be neglected.
Hence, the whole evolution Eq.~\eqref{eq:eq:U-rot-along-int_mod-g} reads
\begin{equation}
\hat{U}(\tau) \approx \hat{U}_\iden(\tau) = \prod_k e^{-ia_k\tau\hat{\sigma}_z\hat{I}_k^z} \label{eq:U-rot-along-int_mod-g_const-g}.
\end{equation}
This sets a blank canvas on which we shall apply an extra manipulation to obtain the target operations.   

%add shifting
\paragraph{Realization of gates in the RF frame:}
We now introduce the extra RF Hamiltonian
 \begin{equation}
 \hat{H}_{\rm shift}(t) = -2\sum_k\cv{{\varphi_k}/\tau}\cos(a_kt)\hat{I}_k^x\label{eq:H_shift}
 \end{equation}
which in the rotating frame with respect to $\hat{H}_{\rm prime}$ is $\widetilde{H}_{\rm shift}(\boldsymbol{\varphi}) {=} \sum_k \widetilde{H}_k^{\rm shift}(\varphi_k) {=} {-}{\sum_k}\cv{{\varphi_k}/{\tau}}\hat{I}_k^x$, approximately. 
We deliberately distinguish this RF drive from that of $\hat{H}_{\rm RF}$ and keep these terms as part of the transformed Hamiltonian for performing the gates.
One finally arrives at $\widetilde{H}_k^{\rm rot}(t) + \widetilde{H}_{\rm shift}(\boldsymbol{\varphi})$ for the gate Hamiltonian in the RF frame. 
Hence, at the time $t{=}\tau$ this gives the rotation gate ${\rm R}_{xxx}(\boldsymbol{\varphi}) {=} \prod_ke^{i\varphi_k\hat{I}_k^x}$ in the RF interaction picture. 
Generically, one can prepare the rotation gate of the form ${\rm R}_{\boldsymbol{\alpha}}(\boldsymbol{\varphi})=\prod_k e^{i\varphi_k\hat{I}_k^{\alpha_{\!k}}}$, where $\boldsymbol{\alpha} = \cv{\alpha_1,\alpha_2,\alpha_3}$ is a vector of axes of rotations with $\alpha_k = x, y$ and $\varphi_k$ denotes the angle of rotation for the spin $k.$

\paragraph{Transferring back to the original frame:} 
To inherit the target gate to the original frame, one can concatenate the evolution $\hat{U}_G(\tau)$ corresponding to the target gate implementation and the reversal of the evolution $\hat{U}_\iden(\tau)$ in Eq.~\eqref{eq:U-rot-along-int_mod-g_const-g}. 
Thanks to the form of $\hat{H}_{\rm prime} = \sum_ka_k\hat{\sigma}_z\hat{I}_k^z,$ the latter component can be prepared by a conjugation by the fast $\pi-$pulses along the $x$-axis on the $\hat{U}_\iden(\tau),$ i.e. $\hat{\sigma}_x\hat{U}_\iden(\tau)\hat{\sigma}_x = \hat{U}_\iden(-\tau).$ 
From the previous discussion, we observe that $\hat{U}_G(\tau)=\hat{U}_\iden(\tau)\hat{G},$ where $\hat{G}$ is the gate prepared in the final rotation frame, and hence
 \begin{equation}
 \hat{\sigma}_x\hat{U}_\iden(\tau)\hat{\sigma}_x\hat{U}_G(\tau) = \hat{U}_\iden(-\tau)\hat{U}_\iden(\tau)\hat{G} = \hat{G} \label{eq:transfer_G}.
 \end{equation}
Note that this is a standard Hahn echo mechanism known in NMR~\cite{Hahn1950}.

\add{
We summarize the steps in our manipulation protocols in terms of signal processing as follows: First, the main RF control $\hat{H}_{\rm RF}$ is employed to destructively interfere with the system dynamics at $t=\tau,$ resulting in an interaction-free dressed frame. Later, an extra shift in frequency, encrypted in $\hat{H}_{\rm shift},$ is added to the RF control, enabling a nontrivial dynamical phase on a target nuclear spin. Lastly, a Hahn-echo is executed to remove the dynamical envelope given by the interaction dressing frame. 
}

%%%%%%%%%%%%%%%%%%%%%%%%%%%%%%%%%%%%%%%%%%%%%%%%%%%%%%%
%%%%% NUMERICAL EXAMPLES
%%%%%%%%%%%%%%%%%%%%%%%%%%%%%%%%%%%%%%%%%%%%%%%%%%%%%%%
\section{Numerical Method}
%%%%%
%\emph{\textbf{Numerical Method}.}---
Now, we illustrate the feasibility of our technique by employing parameters adopted from the experimental data and considering the decoherence \highlight{(see Appendix~\ref{appen:elec_dec})}.
We compute the average gate fidelities defined as \cite{Nielson20022, Nielsen_Chuang_2010, Gilchrist2005}
\begin{equation}
\mathcal{F}(\mathcal{E}, \mathcal{U}) = \frac{\dfrac{1}{d}\tra\cv{{P}^\dagger_{\mathcal{E}}{P}_{\mathcal{U}}}+1}{d+1}\label{eq:fidelity_def},
\end{equation}
where ${P}_\mathcal{C}$ is the Pauli transfer matrix representation of a channel $\mathcal{C}$, given by $\cv{{P}_\mathcal{C}}_{ij} = \frac{1}{d}\tra\cvb{f_i\mathcal{C}(f_j)}$, for a set of $d-$dimensional Pauli operators $f_i$.
Here, $\mathcal{E}$ and $\mathcal{U}$ are the superoperator representations of the (possibly non-unitary) resulting gate and the target gate, respectively.

We account for the environmental influence in our formulation by considering electron spin dephasing, the main decoherence effect, using the Lindblad equation:
\begin{equation}
 \dfrac{d\rho}{dt} = -i\cvb{\hat{H}_{\rm eff}(t),\rho} + \frac{\Gamma}{2}\cv{2\hat{L}\rho\hat{L}^\dagger -\hat{L}^\dagger\hat{L}\rho - \rho\hat{L}^\dagger\hat{L}} \label{eq:dephasing_term},
\end{equation}
where $\hat{L} = \hat{\sigma}_z$ is the electron dephasing generator in the concerned two-level electron spin subspace, and $\Gamma>0$ is the pure dephasing rate of the electron spin. 
At room temperature, the experimentally reported decoherence rate is around $\Gamma = 0.5~{\rm MHz}$ and assumes smaller values for lower temperatures, see e.g. Refs.~\cite{Gottscholl2021, Ramsay2023, Liu2025}. 
All the gates discussed in this section are expressed in the `original' frame with the correction mechanism in Eq.~\eqref{eq:transfer_G}.
The numerical analyses are performed in the same frame using the QuTip package~\cite{QuTip1, QuTip2}.
Here, we assume perfect $\pi-$pulses, while a magnetic field with magnitude $B=250$ mT is applied pointing at $\phi = 2\pi/9.$ 
\add{
These values, which guarantee both the validity of the rotating wave approximations and the distinguishability of the effective hyperfine vectors, are examples for the sake of demonstration. These parameters are not specific and can be optimized for specific experimental setups \highlight{(see Appendix~\ref{appen:freq_dist})}.
}

%unconditional 
\begin{figure}[tb]
    \centering
    \includegraphics[width=0.75\columnwidth]{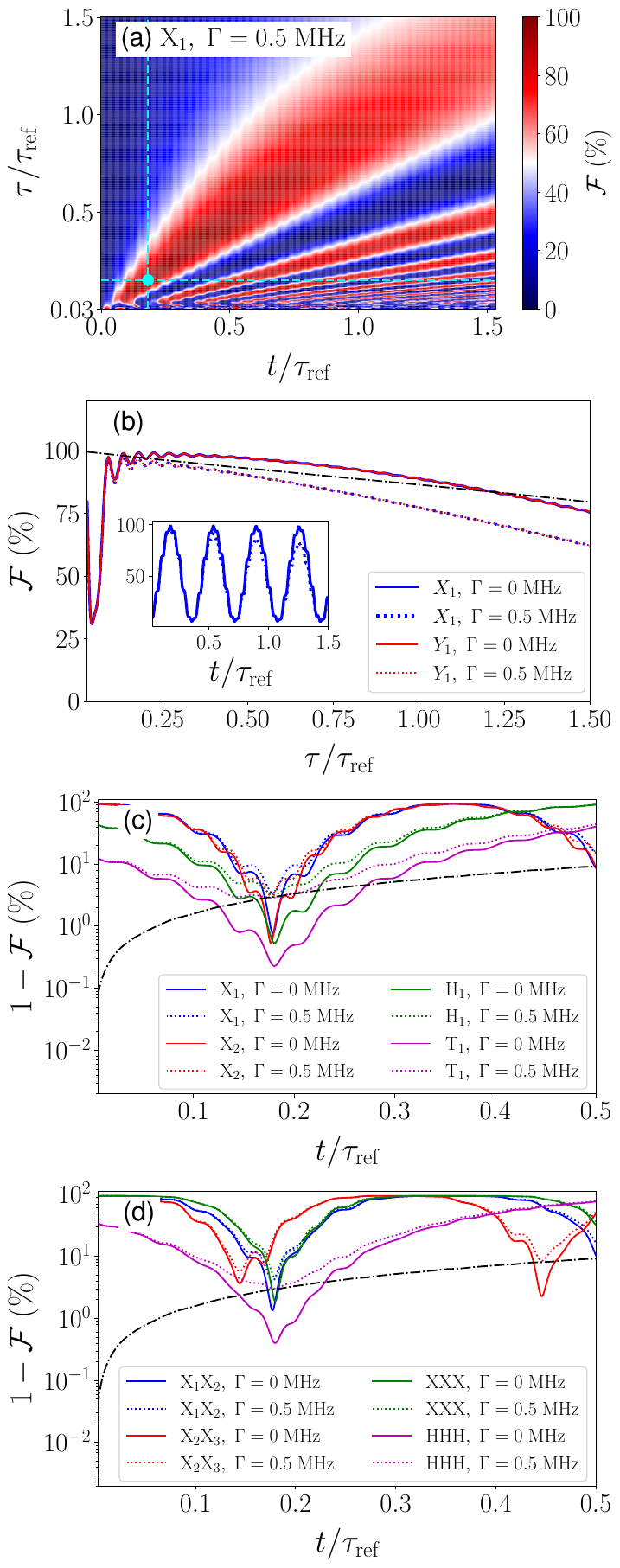}
    \caption{
 (a) Fidelity of gate ${\rm X}_1$ under dephasing as a function of control period $\tau$ and evolution time $t$, with the cyan dot indicating the global maximum.  
 (b) Fidelity of ${\rm X}_1$ and ${\rm Y}_1$ with (dashed) and without (solid) dephasing at the optimal point. The inset shows ${\rm X}_1$ fidelity dynamics for optimal $\tau$.
 In (c) and (d), the infidelity profiles $1-\mathcal{F}$ for various single- and multi-qubit gates are plotted versus evolution time for optimized $\tau$.
 The black dash-dotted lines in (b) to (d) indicate the infidelity of the identity gate as a reference.
 }\label{fig:sample_plot_uncon}%
\end{figure}

%\paragraph{Single-qubit gates:} 
\subsection{Single-qubit gates}
We define the unconditional single-qubit gates generated by $\hat{I}_k^x$ using the following symbols: ${\rm X}_k = e^{i\pi\hat{I}_k^x},$ ${\rm H}_k = e^{i\pi\hat{I}_k^x/2}$ and ${\rm T}_k = e^{i\pi\hat{I}_k^x/4}.$ Similarly, we write ${\rm Y}_k = e^{i\pi\hat{I}_k^y}.$ \add{
These are the elementary Pauli X gate, Hadamard gate, T gate, and Pauli Y gate, which are relevant in quantum computational tasks \cite{Shor1996}.} 
By symmetry, one observes that the dynamics for the X and Y gates are not qualitatively different. 
Therefore, we mainly illustrate the control through $\hat{I}_k^x.$ We set $\tau_{\rm ref} = 1/\omega_{\min} = 509$ ns for the reference of the duration. 
In Fig.~\ref{fig:sample_plot_uncon}, the performance of the gates is summarized. 
First, we investigate the optimal timing for the gate implementation.
In particular, Fig.~\ref{fig:sample_plot_uncon}(a) shows $\mathcal{F}$ for ${\rm X}_1$, from which one observes the first peak in the fidelity at the evolution time $t$ around the control period $\tau$. 
The fidelity limitations are twofold; at the short control and duration times, the rotating wave approximations for the control and the shift Hamiltonians are not yet valid, whereas at longer times, the dephasing effect, as well as the error accumulated from the terms that we had approximately omitted in the analytical part, spoil the fidelity [see Fig.~\ref{fig:sample_plot_uncon}(b)]. 
In the absence of imperfections, the gate fidelities would oscillate towards unity with the control time $\tau$, and the maximal fidelity being attained along the diagonal ridge ($t=\tau$). 
In the realistic case, the decoherence deforms the fidelity landscape, making the maximum peaks deviate from the exact diagonal line. 
Specifically, for $\mathrm{X}_1$ the cyan point in Fig.~\ref{fig:sample_plot_uncon}(a) indicates the global maximum. See also the inset in Fig.~\ref{fig:sample_plot_uncon}(b) for observing the effect of decoherence for a fixed $\tau$ value.
In general, for a given control period $\tau$, the gate time ($t_{\rm gate}$) is found from the dynamics of the average gate fidelity as the first local maximum, which is the global maximum.
To better observe the degradation of the gate performance over time, we plot the infidelities ($1-\mathcal{F}$) in Fig.~\ref{fig:sample_plot_uncon}(c)-(d).
Similarly, the dips in the infidelity profiles indicate the optimal gate time, which is found roughly around $t ~\!{\sim}~\! 0.2\tau_{\rm ref} ~\!{\simeq}~\! 200$ ns, for our parameters.

By comparing the fidelities for ${\rm X}_1$, $\mathrm{H}_1$, and ${\rm T}_1$, one clearly observes that gates with larger angles ($\mathrm{X}_1$) have lower fidelity even in the absence of decoherence.
This is attributed to the extra error accumulation that larger $\varphi_k$ values impose in the dynamics.
The numerical results confirm our expectation that \highlight{this behavior is qualitatively the same for all three nuclear spins with small differences, see Appendix~\ref{appen:numerical_table}.}

%\paragraph{Two- and three-qubit gates:}
\subsection{Two- and three-qubit gates}
Our protocol, in principle, allows for the implementation of various multi-qubit gates. 
For example, one can construct a gate such that it rotates the nuclear spin $N_1$ unconditionally for an angle $\varphi_1,$ while rotating the nuclear spin $N_2$ conditioned on the electron spin state for an angle $\varphi_2$ and keeps the nuclear spin $N_3$ unchanged.
These can be chosen by engineering the shift Hamiltonian \eqref{eq:H_shift} and the control axes in the formalism. 
\add{
The infidelity profiles of ${\rm X}_1{\rm X}_2$ and ${\rm X}_2{\rm X}_3$ for the two-qubit gates, and, ${\rm XXX}$ and ${\rm HHH}$ for three-qubit gates for optimal $\tau$ values are illustrated in Fig.~\ref{fig:sample_plot_uncon}(d).
The general behaviors are similar to the single-qubit cases. 
\highlight{See Appendix~\ref{appen:numerical_table}.}
}
Not surprisingly, both two- and three-qubit gates have lower fidelities than the single-qubit counterparts, and are more sensitive to the dephasing effects.

\section{Conditional Manipulation}
%%%%%
%\emph{\textbf{Conditional Manipulation}.}---
In addition to the unconditional rotation gates, we observe that by tuning the phase of the shift Hamiltonian Eq.~\eqref{eq:H_shift}, it is possible to implement rotations about an arbitrary axis conditioned on the electron spin state. 
This becomes obvious from Eq.~\eqref{eq:H_rot-in-int}, that the local operators, e.g. $\hat{I}_k^x$, are transformed by $e^{it\hat{H}_{\rm prime}}$ to an \emph{apparent} conditional operator, e.g. $\hat{\sigma}_z\hat{I}_k^y$. 
The same argument is applied to the shift Hamiltonian, i.e., by setting
 \begin{equation}
 \hat{H}_{\rm shift}(t) = -2\sum_k\cv{{\varphi_k}/\tau}\sin(a_kt)\hat{I}_k^y\label{eq:H_shift-for-CR},
 \end{equation}
under the condition $\cvv{\varphi_k/\tau} {\ll} \min\{a_k\},$ the relevant term in the rotating frame reads $\widetilde{H}_{\rm shift}(\boldsymbol{\varphi}) {= -}\sum_k\cv{{\varphi_k}/{\tau}}\hat{\sigma}_z\hat{I}_k^x.$ 
The conditional rotation operator ${\rm CR}_{xxx}(\boldsymbol{\varphi}) {=} \prod_ke^{i\varphi_k\hat{\sigma}_z\hat{I}_x}$ is thus similarly derived.
In general, our technique allows us to prepare any non-trivial composite rotation of the following form 
 ${\rm R}_{\boldsymbol{\alpha};\boldsymbol{\nu}}(\boldsymbol{\varphi}) = \prod_ke^{i\varphi_k\hat{\sigma}_{\nu_k}\hat{I}_k^{\alpha_k}},$ 
where $\boldsymbol{\nu} = \cv{\nu_1,\nu_2,\nu_3}$ denotes a vector of conditionality of the rotation of the spin $k$ given by $\nu_k=0,z$ with $\hat{\sigma}_0=\iden$.

%unconditional 
\begin{figure}[tb]
    \centering
    \includegraphics[width=0.75\columnwidth]{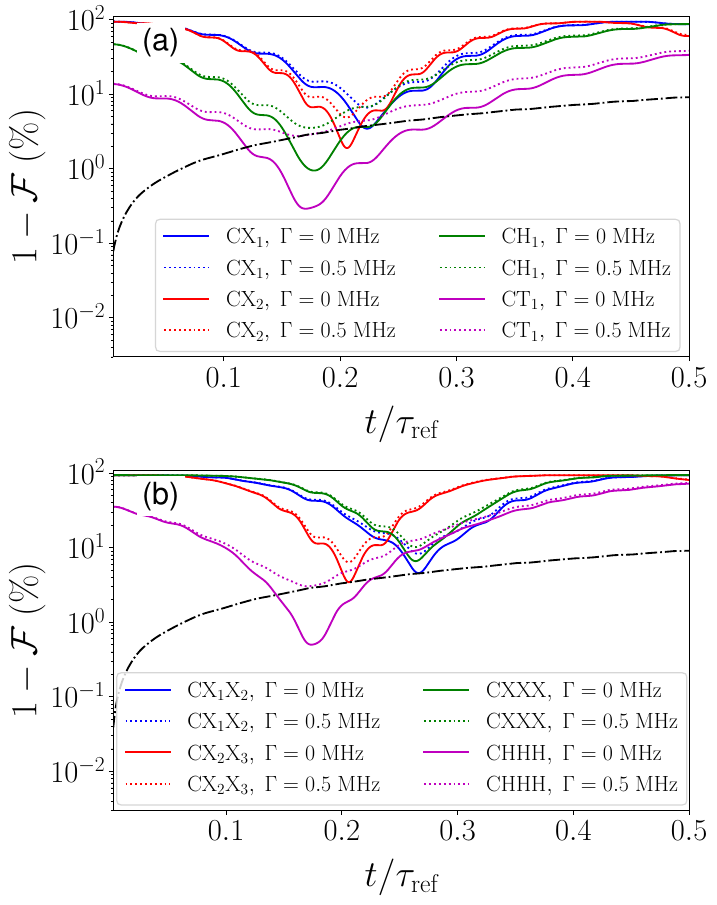}
    \caption{The infidelity $1-\mathcal{F}$ of the single-qubit conditional gates (a) ${\rm CX}_1,$ ${\rm CX}_2,$ ${\rm CH}_1,$ and ${\rm CT}_1,$ (b) the two-qubit gates ${\rm CX}_1{\rm X}_2$ and ${\rm CX}_2{\rm X}_3,$ and the three-qubit gates ${\rm CXXX}$ and \add{
    Hadamard gate} ${\rm CHHH}.$ 
 }\label{fig:sample_plot_con}
\end{figure}

%The numerical results for the unconditional gates are summarized in Table~\ref{tab:table_fidelity}.
The infidelities for \add{the conditional counterparts of} single-qubit gates ${\rm CX}_k = e^{i\pi\hat{\sigma}_z\hat{I}_k^x},$ ${\rm CH}_k = e^{i\pi\hat{\sigma}_z\hat{I}_k^x/2},$ and ${\rm CT}_k = e^{i\pi\hat{\sigma}_z\hat{I}_k^x/4},$ are plotted in Fig.~\ref{fig:sample_plot_con}(a) in the absence and presence of decoherence.
Here, one can observe a substantially lower fidelity since the appearance of $\hat{\sigma}_z$ in the conditional gates, which unlike the unconditional gate, allows the target nuclear spin to interact with the rest of the system via the coupling with the electron spin added in the corresponding $\hat{H}_{\rm shift}(t).$
By a similar reasoning, deviations of the optimal point $(t_{\rm gate}, \tau)$ from the diagonal ridges are expectable in this case.
The similar behavior can also be seen for multi-qubit gates, as illustrated in Fig.~\ref{fig:sample_plot_con}(b). 
 
\add{
The deviation arises from opposite unwanted phase accumulations in different electron spin levels $\ket{\pm},$ which cancel themselves in the unconditional case; introducing $\hat{\sigma}_z$ in conditional gates changes the sign in the $\ket{-}$ branch, leading to greater deviation from the target phase. Similar phenomena occur in Hahn-echo implementations. This effect can be mitigated by using a different pulse profile $\Omega_k(t)$ and applying dynamical decoupling.
}  
 
The gate ${\rm CHHH}$ in particular, is of crucial interest for the preparation of the Greenberger-Horne-Zeilinger (GHZ) states~\cite{Greenberger1990, Sakuldee2025}.
For instance, one easily finds that ${\rm CHHH}\Big[\ket{X_+}\cv{\bigotimes_k\ket{m_I=-\tfrac{1}{2}}_k}\Big]$ is a GHZ state of all the spins in the VB3N cluster, where $\ket{X_\pm}{=}\cv{\ket{+}\pm\ket{-}}/{\sqrt{2}},$ and $\ket{m_I}_k$ is the eigenstate of $\hat{I}_k^z.$ 
A projective measurement on the electron spin state onto either $\ket{m_s=+1}$ or $\ket{m_s=-1}$ will lead to a typical tripartite GHZ state of the three nuclear spins \cite{Sakuldee2025}.
Our protocol provides another tool in preparation for GHZ states in the VB3N cluster with high controllability.

%%%%%%%%%%%%%%%%%%%%%%%%%%%%%%%%%%%%%%%%%%%%%%%%%%%%%%
%%%% CONCLUSIONS    
%%%%%%%%%%%%%%%%%%%%%%%%%%%%%%%%%%%%%%%%%%%%%%%%%%%%%%
\section{Conclusion}
%%%%%
%\textit{\textbf{Conclusion}.}---
\add{
We have developed a protocol for preparing single- and multi-qubit gates on a VB3N spin cluster in \vb-hB${}^{15}$N, utilizing the interaction anisotropy induced by the orientation of the in-plane magnetic field. Such clusters in hBN offer advantages for quantum processing due to their uniform relationship with magnetic fields, isolated interactions within clusters, and the novel use of nuclear spins as computational units while preserving electron spins as auxiliary qubits. 
The method separates strong electron-nuclear spin interactions from the desired evolution by eliminating unwanted residual terms through a continuous control field. An additional RF drive allows for the implementation of target gates, which are eventually transferred to the computational basis via the Hahn echo mechanism. From a numerical calculation of a simple setup, the computed gate fidelities are approximately $96\%$ for unconditional gates, while conditional gates demonstrate lower fidelities. By invoking additional enhancements such as control optimization and decoherence suppression techniques~\cite{Haase2018, Chen2025, Jin2025, Murzakhanov2022}, our protocol sets a promising approach for practical gate implementations in experimental settings.
Finally, it is worth mentioning that under a strong in-plane magnetic field, the optical pumping does not efficiently polarize the electron spin. However, thanks to the symmetry achieved for an in-plane magnetic field, the initialization in our computational basis is plausible by employing a pulsed magnet or an incoherent maser drive \highlight{(see Appendices~\ref{appen:elec_pol_trad}-\ref{appen:gate_assist}).}
}

\begin{acknowledgments}
%\textit{Acknowledgements.}---
This work was partially supported by the Yangyang Development Foundation. 
We are grateful to Benchen Huang, for providing the computationally found hyperfine coupling values for hB$^{15}$N. 
We thank Zhenyu Wang, Sujin Suwanna, and Chanaprom Cholsuk for the fruitful discussions and their comments on our paper.
\end{acknowledgments}

\appendix
%\documentclass[pra, letterpaper, 10pt, onecolumn, longbibliography, superscriptaddress]{revtex4-1} 
%
%\usepackage[T1]{fontenc}
%\usepackage[utf8]{inputenc}
%\usepackage{multirow}
%\usepackage{dcolumn}
%\usepackage{amsmath}
%\usepackage{amsfonts}
%\usepackage{amssymb}
%\usepackage{tabularx}
%\usepackage{bbm, dsfont}
%\usepackage[breaklinks,colorlinks,bookmarks=false,citecolor=blue,linkcolor=blue,urlcolor=blue]{hyperref}
%
%\usepackage{pst-plot}
%\usepackage{enumitem}
%
%\usepackage{orcidlink}
%
%\makeatletter
%\newcommand*{\rom}[1]{\expandafter\@slowromancap\romannumeral#1@}
%\makeatother
%
%\newcommand{\nn}{\nonumber}
%
%\newcommand{\vb}[0]{V$_\text{B}^-$}
%\newcommand{\cv}[1]{\left(#1\right)}
%\newcommand{\cvb}[1]{\left[#1\right]}
%\newcommand{\cvc}[1]{\left\{#1\right\}}
%\newcommand{\cvv}[1]{\left\vert #1\right\vert}
%\newcommand{\ket}[1]{\vert#1\rangle}
%\newcommand{\bra}[1]{\langle#1\vert}
%\newcommand{\ketbra}[2]{\ket{#1}\bra{#2}}
%
%\newcommand{\iden}{\mathbbm{1}}
%\newcommand{\tr}[1]{\text{Tr}\cv{#1}}
%\newcommand{\tra}{\text{Tr}}
%\newcommand{\sinc}{\text{sinc}}
%
%
%
%\begin{document}
%\begin{center}
%{\Large Arbitrary Manipulation of Nuclear Spins in hexagonal Boron Nitride: Supplemental Material}	
%\end{center}

\section{A Generic Setup for the System Hamiltonian}\label{appen:system}
The phenomenological form of the Hamiltonian for the VB3N cluster is given by 
 \begin{equation}
  \hat{H} = \mathbf{S}\cdot\mathbb{D}\cdot\mathbf{S} + \gamma_e\mathbf{B}\cdot\mathbf{S} + \sum_k\gamma_n\mathbf{B}\cdot\mathbf{I}_k + \sum_k\mathbf{S}\cdot\mathbb{A}_k\cdot\mathbf{I}_k \label{eq:H_primitive}.
 \end{equation}
\begin{figure}[b]
	\centering
	\includegraphics[width=0.5\columnwidth]{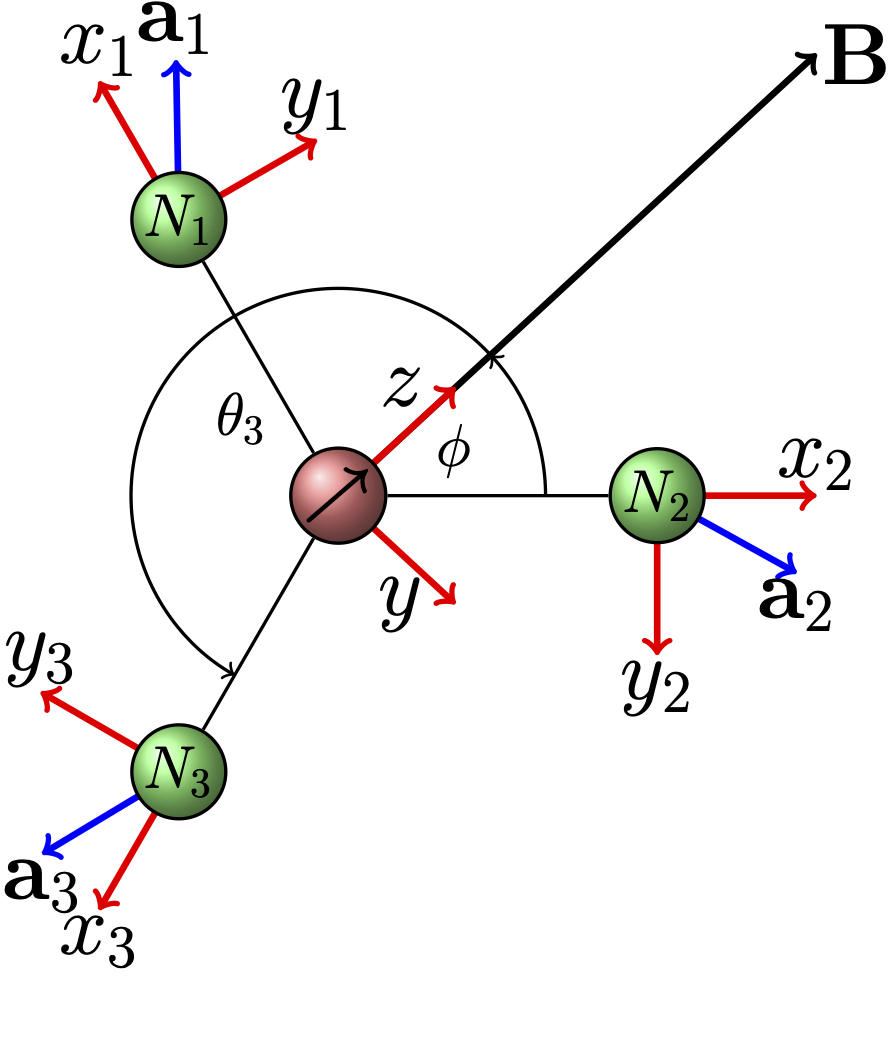}
	\caption{A consideration on the cluster of an electron spin (red ball) residing at the vacancy center and three nearest spins (green balls $N_1$ to $N_3$) embedded in VB-hB${}^{15}$N, is enabled by their strong interactions relative to other terms. An adequately strong magnetic field $\mathbf{B}$ is applied in-plane at the angle $\phi$ from the dangling bond$-2$ connecting the central spin and the nuclear spin $N_2,$ where the effective fields $\mathbf{a}_k$ are given by the blue arrows. Likewise, $\theta_k$ denotes the angles from the magnetic field direction and the dangling bond$-k.$ For the coordinate of the electron spin operator, we define an axis $\hat{z}$ in the direction of the magnetic field, whereas we define spin operators in terms of their local in-plane coordinates $(x_k,y_k)$ (red arrows.)}\label{fig:hBN-appen}
\end{figure}
Conventionally, one chooses the $z$-axis perpendicular to the lattice plane and along the VB-center axis of symmetry, that is the $c$-axis. 
However, as we consider the in plane magnetic field ($\mathbf{B}\perp c$-axis, we define the coordinates such that the direction of the magnetic field defines the $z$-axis and the $x$-axis matches that of $c$-axis.
Hence, $\gamma_e\mathbf{B}\cdot\mathbf{S} = \omega_e\hat{S}_z,$ while $\mathbf{S}\cdot\mathbb{D}\cdot\mathbf{S}\equiv D_{gs}\hat{S}_x^2.$ 
For the nuclear spins, let us begin with the \emph{local coordinates} dictated by the principal axes aligned with the dangling bond between {the vacancy and the nuclear spins.} In particular, let $\hat{J}_k^x$ be the spin operators in such directions and $\hat{J}_k^y$ be those for the in-plane complementary parts. One can also write the electron spin in these three coordinate systems, where in such a case the hyperfine interaction for each spin takes a diagonal form since the $x_k-$axis is also the principal axis for the interaction between electron spin and the nuclear spin $k$ \cite{Gracheva2023}, i.e.
 \begin{equation}
  \mathbf{S}\cdot\mathbb{A}_k\cdot\mathbf{I}_k = \sum_{\alpha=x,y,z}A_k^{\alpha\alpha}\hat{S}_k^\alpha\hat{J}_k^\alpha = \sum_{\alpha=x,y,z}A_{\alpha\alpha}\hat{S}_k^\alpha\hat{J}_k^\alpha \label{eq:HFI-principal-k}.
 \end{equation}
Note that $A_k^{\alpha\alpha}$ are uniform in $k$ by symmetry, where all nuclei are identical in their own local coordinates, and hence we have put $A_k^{\alpha\alpha} = A_{\alpha\alpha}$ in the last step. 
 
%interaction terms
In Eq.~\eqref{eq:HFI-principal-k}, there are three separate representations for the electron spin operators, which is not a convenient choice for the analysis of the system {as a whole}. Therefore, we use the $z$-axis given by the magnetic field direction for the electron's coordinate as the universal quantization axis. {The relations between electron spin's coordinates and nuclear spins' coordinates can be characterized by angles $\theta_1 = 2\pi/3-\phi,$ $\theta_2 = -\phi,$ and $\theta_3 = 4\pi/3-\phi,$ i.e., the angles from the magnetic field direction to the bond$-k.$} 
The transform between these two sets of coordinates is given by
 \begin{align*}
  \hat{S}_k^z &= \hat{S}_z\cos\theta_k + \hat{S}_y\sin\theta_k\\
  \hat{S}_k^y &= \hat{S}_y\cos\theta_k - \hat{S}_z\sin\theta_k\\
  \hat{S}_k^x &= \hat{S}_x.
 \end{align*}
It then follows that
 \begin{align}
  \mathbf{S}&\cdot\mathbb{A}_k\cdot\mathbf{I}_k \nn\\
   &= \cv{ \begin{array}{c}
     \hat{S}_z\\
     \hat{S}_y\\
     \hat{S}_x
    \end{array}}^\mathrm{T}
    \cv{ \begin{array}{ccc}
     A_{xx}\cos\theta_k & -A_{yy}\sin\theta_k & 0\\
     A_{xx}\sin\theta_k & A_{yy}\cos\theta_k & 0\\
     0 & 0 & A_{zz}
    \end{array}}
    \cv{ \begin{array}{c}
     \hat{J}_k^x\\
     \hat{J}_k^y\\
     \hat{J}_k^z
    \end{array}}
      \label{eq:HFI-in-magnetic},
 \end{align}
where the hyperfine tensors $\mathbb{A}_k$ take the generic form in Eq.~(1). Here, one can effectively see the difference in the effective hyperfine interaction tensor, which allows us to selectively address each nuclear spin in our formulation. 

\bigskip
%diagonalization
\section{Electron Subspace for In-plane Magnetic Field}\label{appen:electron_subspace}
Here, we simplify the problem by employing only a two-level subspace of the electron.
To do so, let us begin with the diagonalization of $\hat{H}_e,$ which takes the form 
 \[\hat{H}_e \Longrightarrow \hat{T}^\dagger\hat{H}_e\hat{T} = {\rm diag}\cv{\omega_+/2,D_{gs},-\omega_-/2}\] 
where $\omega_\pm = \omega_\delta \pm D_{gs},$ and $\omega_\delta=\sqrt{D_{gs}^2 + 4\omega_e^2}.$ 
We denote the eigenstates by $\ket{+},\ket{D}$ and $\ket{-}$ with respect to the three consecutive eigenvalues, respectively. 
The diagonalization matrix above is given by
 \begin{equation}
  \hat{T} = \displaystyle \left(\begin{matrix}
    \sqrt{\tfrac{\omega_+}{4\omega_\delta}} & -\tfrac{\sqrt{2}}{2} & \sqrt{\tfrac{\omega_-}{4\omega_\delta}}\\
    \sqrt{\tfrac{\omega_-}{2\omega_\delta}} & 0 & -\sqrt{\tfrac{\omega_+}{2\omega_\delta}}\\
     \sqrt{\tfrac{\omega_+}{4\omega_\delta}} & \tfrac{\sqrt{2}}{2} & \sqrt{\tfrac{\omega_-}{4\omega_\delta}}
   \end{matrix}\right)\label{eq:T}.
 \end{equation}
At high magnetic field $\gamma_e B\gg D_{gs}$, we have $\omega_\delta \approx 2\omega_e$ and $\omega_\pm \approx 2\omega_e,$ leading to 
 \begin{equation*}
  \hat{T} \approx 
   \displaystyle \left(\begin{matrix}
   \tfrac{1}{2} & -\tfrac{\sqrt{2}}{2} & \tfrac{1}{2}\\
   \tfrac{\sqrt{2}}{2} & 0 & -\tfrac{\sqrt{2}}{2}\\
   \tfrac{1}{2} & \tfrac{\sqrt{2}}{2} & \tfrac{1}{2}
   \end{matrix}\right),
 \end{equation*}
which simply transforms the electron spin operators as $\cv{\hat{S}_z,\hat{S}_y,\hat{S}_x}$ to $\cv{\hat{S}_x,\hat{S}_y,-\hat{S}_z},$ i.e. it acts as a relabelling matrix.

%spin operators
By applying the transformation matrix $\hat{T}$ to the spin operator we obtain a new set of operators:
 \begin{subequations}  
 \begin{align}
  \Lambda_z &= \hat{T}^\dagger\hat{S}_z\hat{T} = \cv{\tfrac{2\omega_e}{\omega_\delta}}\hat{\sigma}_z^{+-} -\cv{\tfrac{D_{gs}}{\omega_\delta}}\hat{\sigma}_x^{+-} \label{eq:S_x}\\
  \Lambda_y &= \hat{T}^\dagger\hat{S}_y\hat{T} = \hat{\sigma}_y^{D+}{\sqrt{\tfrac{\omega_-}{2\omega_\delta}}} - \hat{\sigma}_y^{D-}{\sqrt{\tfrac{\omega_+}{2\omega_\delta}}} \label{eq:S_y}\\
  \Lambda_x &= \hat{T}^\dagger\hat{S}_x\hat{T} = -\hat{\sigma}_x^{D-}{\sqrt{\tfrac{\omega_-}{2\omega_\delta}}} - \hat{\sigma}_x^{D+}{\sqrt{\tfrac{\omega_+}{2\omega_\delta}}} \label{eq:S_z},
 \end{align}
 \end{subequations} 
where we have introduced the Pauli operators $\hat{\sigma}_z^{\alpha\beta} = \ketbra{\alpha}{\alpha} - \ketbra{\beta}{\beta},$ $\hat{\sigma}_x^{\alpha\beta} = \ketbra{\alpha}{\beta} + \ketbra{\beta}{\alpha},$ $\hat{\sigma}_y^{\alpha\beta} = -2i\hat{\sigma}_z^{\alpha\beta}\hat{\sigma}_x^{\alpha\beta}$ for $\ket\alpha,\ket\beta = \ket{D},\ket{\pm}.$ 
Remark here that the resulting operators after the transformation $\hat{T}$ are not spin operators, i.e., the symmetric or anti-symmetric combinations of the Gell-Mann matrices, but rather slant combinations of the basis matrices. 

In particular, for the subspace composed of $\{\ket{+},\ket{-}\}$ it follows that
 \begin{subequations}  
 \begin{align}
  \Lambda_z &\rightarrow \cv{\tfrac{2\omega_e}{\omega_\delta}}\hat{\sigma}_z -\cv{\tfrac{D_{gs}}{\omega_\delta}}\hat{\sigma}_x \label{eq:tau_z}\\
  \Lambda_y &\rightarrow 0,~~\Lambda_x \rightarrow 0 \label{eq:tau_xy},
 \end{align}
 \end{subequations} 
where we drop the superscript for simplicity.  
Since 
\[\hat{H}_e - D_{gs}\iden = {\rm diag}\cv{\tfrac{-D_{gs} + \omega_\delta}{2}, 0, \tfrac{-D_{gs} - \omega_\delta}{2}} \equiv \frac{\omega_\delta}{2}\hat{\sigma}_z\] 
on the $+-$ subspace, the total Hamiltonian on such a subspace will then be 
 \begin{align}
  \hat{H}_{\rm reduced} &= \frac{\omega_\delta}{2}\hat{\sigma}_z + \sum_k\omega_n\cv{\hat{\mathbf{z}}\cdot{\mathbf{I}}_k}\nonumber\\
  	&\phantom{=.} + \hat{\sigma}_z\sum_{k}{\boldsymbol{a}}_k\cdot\mathbf{I}_k - \hat{\sigma}_x\sum_{k}{\boldsymbol{c}}_k\cdot\mathbf{I}_k \label{eq:H_reduce-appen},
 \end{align}
where the effective hyperfine interaction vectors are introduced:
 \begin{align}
  {\boldsymbol{a}}_k &= \cv{\tfrac{2\omega_e}{\omega_\delta}}\hat{\mathbf{z}}\cdot\mathbb{A}_k = \cv{\tfrac{2\omega_e}{\omega_\delta}}\cv{\hat{\mathbf{x}}_kA_{xx}\cos\theta_k + \hat{\mathbf{y}}_kA_{yy}\sin\theta_k} \label{eq:def-a_k-appen}\\
  {\boldsymbol{c}}_k &= \cv{\tfrac{D_{gs}}{\omega_\delta}}\hat{\mathbf{z}}\cdot\mathbb{A}_k = \cv{\tfrac{D_{gs}}{\omega_\delta}}\cv{\hat{\mathbf{x}}_kA_{xx}\cos\theta_k + \hat{\mathbf{y}}_kA_{yy}\sin\theta_k}. \label{eq:def-c_k-appen}
 \end{align}

\bigskip
\section{Frequency Discrepancy}\label{appen:freq_dist}
\begin{figure*}[tb]
	\centering
	\includegraphics[width=0.8\textwidth]{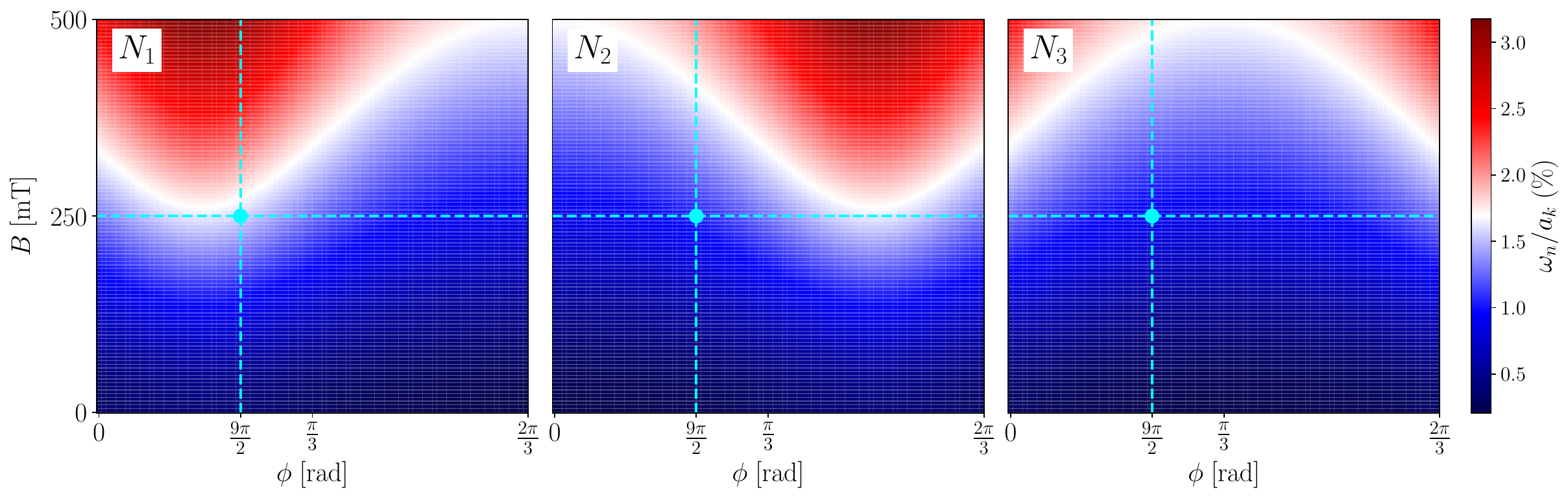}
	\caption{The distributions of the ratio $\omega_n/a_k$ for various magnetic field strengths $B$ and its orientation $\phi$ for nuclear spins $N_1$ to $N_3.$ Here one can see the equivalence among the three spins, e.g., the translation of the angle $\phi$ by $\pi/6$ will result in a permutation of the spin labels $k.$ The cyan dashed lines and dots indicate the points corresponding to the $B = 250$ mT and $\phi = 9\pi/2$ used in the example in our calculation.}\label{fig:w_over_a-profile}
\end{figure*}
Here, we discuss an overview of the parameter resolutions for experimental implementation in future. We note that the effective hyperfine vectors ${\boldsymbol{a}}_k$ are the crucial objects as they set the referent frames for the nuclear spin dynamics. Note that for high enough magnetic field, e.g., $\omega_\delta \gg D_{gs}\cvv{\mathbb{A}_k}$ or numerically $B \gg 100$ mT, the contribution concerning the coefficients ${\boldsymbol{c}}_k$ is negligible and can be omitted. 
The concerns in our protocol mainly comprise the relative strength of the nuclear spin $\omega_n$ to the effective interaction vectors and the distinguishability of the nuclear spins. 
In Fig.~\ref{fig:w_over_a-profile}, we plot the ratio $\omega_n/a_k$ for various magnetic field strengths $B$ and its orientation $\phi.$ One can see that, in the intermediate regime $B {<} 1$~T, although the ratio $\omega_n/a_k$ is small, i.e., $\omega_n/a_k\sim 1-3 \%,$ allowing us to remove the contributions from the terms with $\omega^{\parallel (\perp)}_k,$ {defined by $\omega_k^\parallel = \omega_n\hat{\mathbf{z}}\cdot\hat{\mathbf{a}}_k$, $\omega_k^\perp = \vert\omega_n\hat{\mathbf{z}}-\omega_k^\parallel\hat{\mathbf{a}}_k\vert$,} in the analysis for an appropriate magnetic field parameters and timescales. 
The cyan dashed lines and dots in Fig.~\ref{fig:w_over_a-profile} indicate the points corresponding the $B = 250$ mT and $\phi = 9\pi/2$ used in the example in our numerical calculation, at which the maximum ratio $\max\cvc{\vert\omega^{\parallel}_k\vert, \vert\omega^{\perp}_k\vert}/ a_k \sim 1.5~\%.$

\begin{figure*}[tb]
	\centering
	\includegraphics[width=0.8\textwidth]{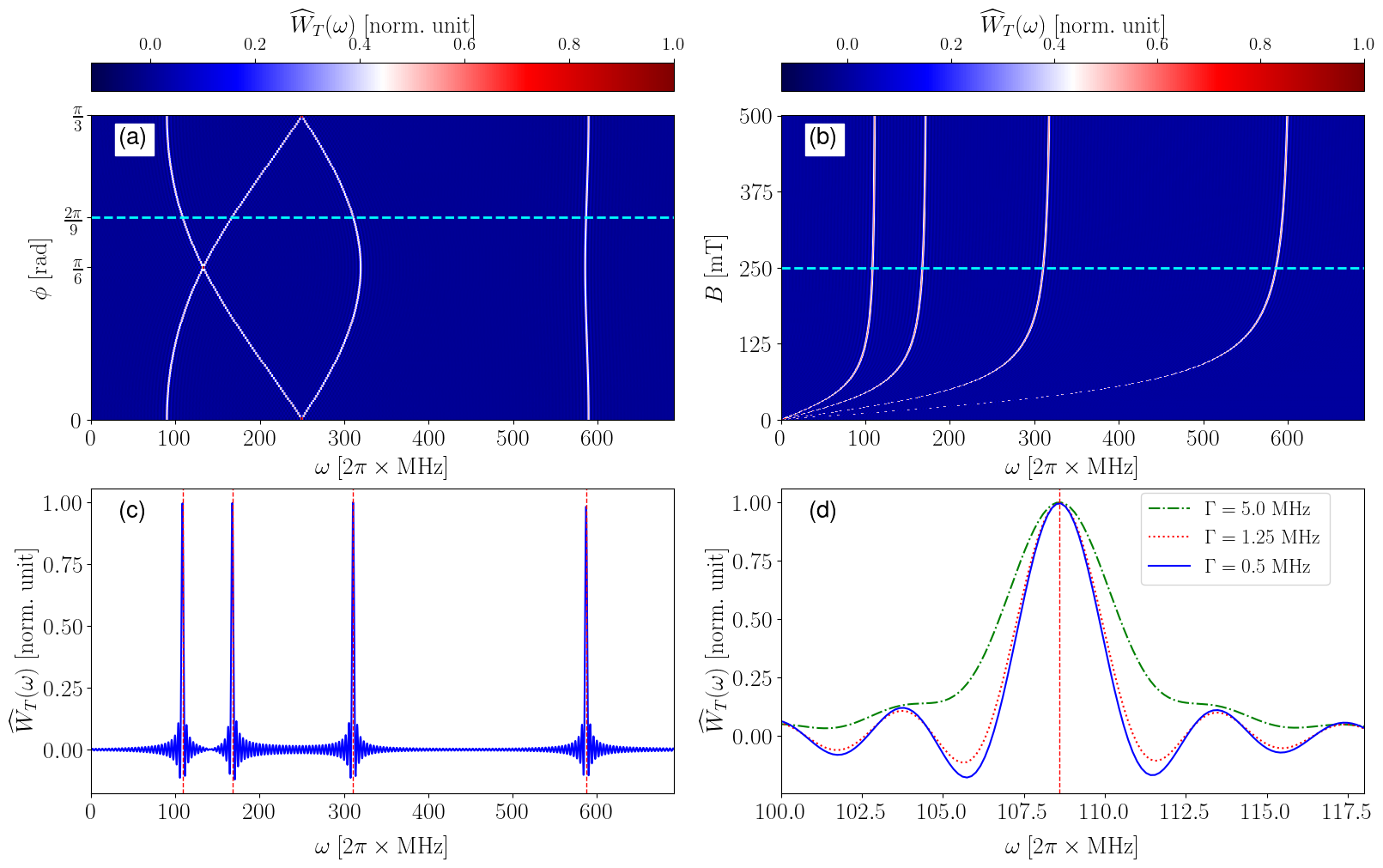}
	\caption{The characteristics of the coherence function in the frequency domain $\widehat{W}_T(\omega)$ with $T\sim 1.6~{\rm \mu s}$ in its normalized unit. For a fixed magnetic field strength $B=250$ mT and the coherence rate $\Gamma = 0.5$ MHz, we can see the distinct peaks (a) for various orientations $\phi$ except at the points where $\phi$ is a multiple of $\pi/6,$ i.e., the magnetic field is parallel or perpendicular to one of the dangling bonds. The distance between two adjacent peaks is larger for the larger magnetic field strength for a fixed angle $\phi = 9\pi/2$ (b), given the coherence rate $\Gamma = 0.5$ MHz, reflecting that the strong magnetic field provides more robust separation between two nuclear spins. The peaks can be estimated by the last expression in Eq.~\eqref{eq:W_f}, where one can see good agreements between the quantities $\cvv{2\sum_{k}m_ka_k}$ and the peaks (c), for our set of parameters $B=250$ mT, $\phi = 9\pi/2,$ and $\Gamma = 0.5$ MHz. The peak separations can be made relatively large in our construction, and the disturbances to the environment become relatively small. For instance, for $B=250$ mT and $\phi = 9\pi/2,$ the decoherence effect on the first peek will alter its width in the order of a few MHz, whereas the peak separations are in the order of several $10$ MHz (d), providing a robustness for determining individual nuclear spins.}\label{fig:Wf_at-selected_params}
\end{figure*}

The parameters $B = 250$ mT and $\phi = 9\pi/2$ also offer an appropriate degree of discrepancies among the three spins. To see that, let us consider the coherent function defined by
	\begin{equation}
		W(t) = \tra\cv{\hat{\sigma}_x e^{-it\hat{H}_{\rm eff}} \rho_0 e^{it\hat{H}_{\rm eff}}} = \frac{1}{16}\tra\cv{\hat{U}^\dagger_{0}\hat{U}_{1} + \hat{U}^\dagger_{1}\hat{U}_{0}}	\label{eq:W_t},
	\end{equation}
where $\hat{H}_{\rm eff} = \ketbra{0}{0}\otimes\hat{H}_0 + \ketbra{1}{1}\otimes\hat{H}_1,$ $\hat{U}_{0 (1)} = e^{-it\hat{H}_{0 (1)}}$ with $\hat{H}_0 = \sum_k\omega_n\cv{\hat{\mathbf{z}}\cdot{\mathbf{I}}_k} + \sum_{k}{\boldsymbol{a}}_k\cdot\mathbf{I}_k$ and $\hat{H}_1 = \sum_k\omega_n\cv{\hat{\mathbf{z}}\cdot{\mathbf{I}}_k} - \sum_{k}{\boldsymbol{a}}_k\cdot\mathbf{I}_k,$ and the initial state is set such that the electron is in a superposition state, while the nuclei are in a fully thermal state. Mathematically, it reads $\rho_0 = (\iden_e+\hat{\sigma}_x)\otimes\iden_N/16,$ with the electron spin and nuclear spin identity operators $\iden_e$ and $\iden_N,$ respectively.
For $a_k\gg\omega_n,$ up to some normalization, the function can be estimated via the relations
	\begin{align*}
		\hat{U}_{0} &= e^{-it\hat{H}_{0}} \approx e^{-it\sum_k\omega_k^\parallel\hat{I}_k^z}e^{-it\sum_ka_k\hat{I}_k^z},\\
		\hat{U}_{1} &= e^{-it\hat{H}_{1}} \approx e^{-it\sum_k\omega_k^\parallel\hat{I}_k^z}e^{it\sum_ka_k\hat{I}_k^z},
	\end{align*}
leading to an estimation \[W(t) \propto  \sum_{m_k=\pm 1}\cos\cv{2t\sum_{k}m_ka_k}.\] 
The Fourier transform of the function reveals the response of the dynamics at different frequencies. Numerically, we can evaluate it by
	\begin{align}
		\widehat{W}_T(\omega) &= \frac{1}{T} \int_0^Tdt \cos(\omega t)W(t)\nonumber\\
			 &\approx \mathcal{N}_T\sum_{m_k=\pm 1}\sinc\cvb{\cv{\omega - 2\sum_{k}m_ka_k}T}	\label{eq:W_f},
	\end{align}
for large enough duration $T$ and upto some normalization factor $\mathcal{N}_T,$ where $ \sinc(x) = \sin(x)/x$ for $x\neq 0$ and $\sinc(0) = 1$ is the sinc function. Ideally, we have $\mathcal{N}_T^{-1}\widehat{W}_T(\omega) \rightarrow \sum_{m_k=\pm 1}\delta\cvb{\cv{\omega - 2\sum_{k}m_ka_k}}$ as ${T\rightarrow\infty},$ where $\delta$ is a Dirac distribution. 
From this, one can expect the peaks of the function around the four combinations $\cvv{2\sum_{k}m_ka_k}$ for the positive frequency $\omega,$ and hence by determining the discrepancy of the peaks, we can infer the distinction in the dynamical response between two nuclear spins in the cluster. 

In Fig.~\ref{fig:Wf_at-selected_params}(a), we can see that at $B=250$ mT with the decoherence rate $\Gamma = 0.5$ MHz, we can see four distinct peak locations at any angle $\phi$ except $\phi = m\pi/6$ for any integer $m,$ at which a partial symmetry is preserved as $a_k=a_l$ for some $k\neq l.$ For a fixed orientation $\phi = 9\pi/2,$ the peak separations can be made larger by increasing the magnetic field strength $B$ as seen in Fig.~\ref{fig:Wf_at-selected_params}(b). In Fig.~\ref{fig:Wf_at-selected_params}(c), for our set of parameters $B=250$ mT, $\phi = 9\pi/2,$ and $\Gamma = 0.5$ MHz, one can observe a clear separation between two adjacent peaks with several $10$ MHz discrepancy, where we can also see that the increasing of the decoherent rate $\Gamma$ will smear the resolution only in the order of few MHz (see Fig.~\ref{fig:Wf_at-selected_params}(d).) This suggests the distinction among the three nuclear spins can be accessed in our protocol, and the individual spin manipulation can be engineered efficiently. 
Furthermore, since the distinguishability here technically involves one of the experiment limitations, i.e., the resolution of the RF generator, one can observe that the high selectivity reflected in the coherence function provides flexibility for the resolution of the driving field. For example, suggested by Fig.~\ref{fig:Wf_at-selected_params}(d), for our set of parameters, the driving frequency resolution in the order of a few MHz can still discriminate two nuclear spins effectively.

\bigskip
\section{Effect of Electron Spin Decoherence}\label{appen:elec_dec}
\begin{figure}[b]
\includegraphics[width=0.47\textwidth]{"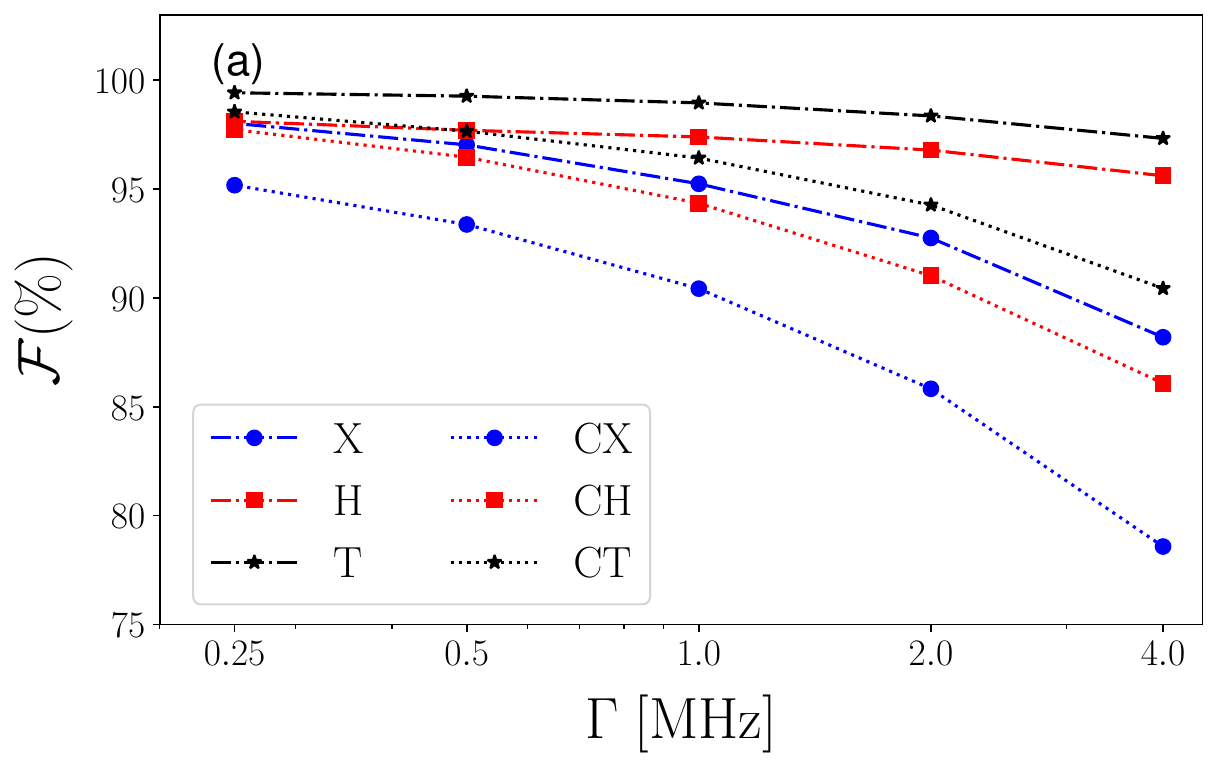"}
\includegraphics[width=0.47\textwidth]{"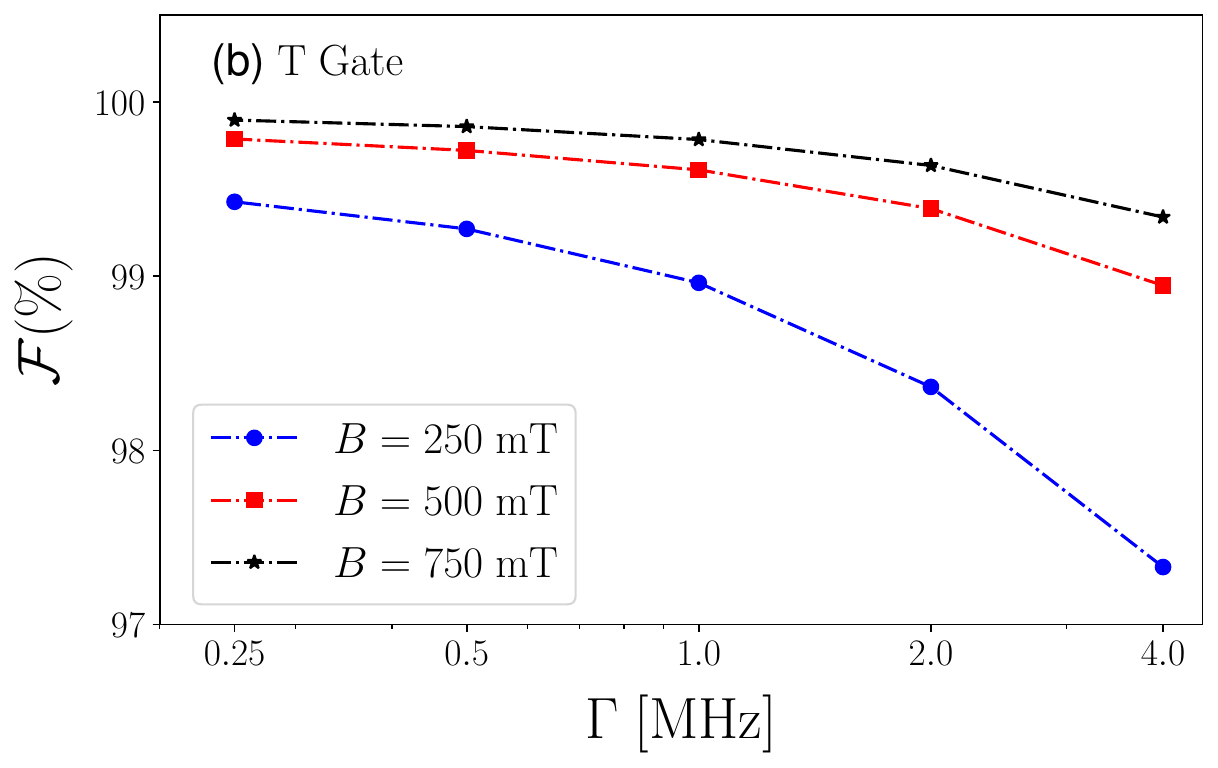"}
\caption{The maximum fidelity of the gates $\rm X,$ $\rm H$ and $\rm T$ and their conditional counterparts $\rm CX,$ $\rm CH$ and $\rm CT$ for nuclear spin $1$ with different coherent rates $\Gamma$, for our choice of magnetic field strength $B=250$ mT (a). For a higher magnetic field, one can expect a higher fidelity of the same gate, e.g., for $\rm T$ gate in (b).}\label{fig:Fidelity_vary_T2}
\end{figure}

%vary Gamma_2 at B = 250 mT
Here, we consider the effect of electron spin decoherence $\Gamma$ on the maximum average gate fidelity. For simplicity, we employ the same calculation and the same set of parameters for the nuclear spins and vary $\Gamma.$ Since to sweep $\Gamma$ it is computationally demanding, here we only choose a few values of decoherence rates, i.e. $\Gamma = 0.25, 0.5, 1.0, 1.25,$ and $4$ MHz, to demonstrate the trend. First, we set $B=250$ mT and calculate the fidelity landscapes for the gates $\rm X,$ $\rm H$ and $\rm T$ and their conditional counterparts $\rm CX,$ $\rm CH$ and $\rm CT,$ and find the maximum points. In Fig.~\ref{fig:Fidelity_vary_T2}(a), one can observe that the relations between unconditional and conditional gates of the same phase $\varphi,$ and the fidelity increases as the phase $\varphi$ decreases for all $\Gamma.$ 
For any gates, it is clear that the fidelity drops substantially when the decoherence rate increases. This particularly is true for conditional gates. For instance, the fidelity of the $\rm X$ gate drops from around $98 \%$ for $\Gamma = 0.25$ MHz to around  $97 \%$ for $\Gamma = 0.5$ MHz in the main text, and to around $88\%$ for $\Gamma = 4$ MHz. We observe that for all gates the optimal periods $\tau$ range between $100$ to $300$ ns and the gate time $t_{\rm gate}$ is in between $70$ to $300$ ns. The latter suggests that the total gate times ($2t_{\rm gate}$) become larger than the decoherence times and hence the decoherence effect can spoil the gate qualities. 
Naively, given that the system is subjected to the same environmental decoherence, one can avoid this issue by tuning the parameters in such a way that the gate duration is short, e.g., within the coherent time window. 

%vary Gamma_2 at various B
The most direct method is to increase the magnetic field strength $B.$ In Fig.~\ref{fig:Fidelity_vary_T2}(b), the fidelity for the same gate, e.g. $\rm T,$ increases as the magnetic field increases. This could be expected as the magnetic field increases, the concerned parameters $\omega_k^\perp$ and $a_k$ increase, leading to the shorter control duration $\tau=1/\min\cvc{\vert\omega_k^\parallel\vert,\cvv{\omega_k^\perp}}$ in our setup; and hence the gate time $t_{\rm gate}\sim \tau,$ in principle, will then be able to be made shorter than the decoherent time. 
For instance concerning Fig.~\ref{fig:Fidelity_vary_T2}(b), at the rate $\Gamma = 4$ MHz, the fidelity of the $\rm T$ gate increases from around $85\%$ for $B=250$ mT with $\tau \approx 180 $ ns at $t_{\rm gate} \approx 145$ ns, to around $90\%$ for $B=750$ mT with $\tau \approx 91 $ ns at $t_{\rm gate} \approx 53$ ns. 
We remark that in our calculation, we assume the Lindblad dynamics for the dephasing mechanism, which only leads to free induction decay. This model cannot be decoupled by our mechanism, i.e., the Hanh-echo is assumed to have no effect on this decoherence. In reality, this can only be treated as a lower bound for the gate fidelity, since one can expect the echo mechanism to remove parts, if not all, of the couplings concerning the decoherence, and then the coherent time can be effectively extended. 
Moreover, apart from the control pulse shaping technique, i.e. the optimization of the function $\Omega_k,$ one can further incorporate the dynamical decoupling pulse in our scheme to remove the unwanted contribution especially the non control-modulated term $\omega_k^\perp\hat{\sigma}_z\hat{I}_k^x$ in Eq.~(4) in the main text, as well as to extend the electron spin coherence time.  
The development in this direction is open for further study.

\bigskip
\section{Numerical Parameters}\label{appen:numerical_table}
The parameters used in our examples, which we have taken from the experimental and \textit{ab initio} data, are given as follows: $D_{gs} = 3.472~2\pi\times{\rm GHz},$ $\gamma_{e} = 27.9925~2\pi\times{\rm GHz},$ $\gamma_{{}^{15}\text{N}} = 4.32~2\pi\times{\rm GHz},$ the magnetic field strength is $B=250$ mT, with only inplane components, pointing with $\phi = 2\pi/9$ from the secondary symmetry axis or the dangling bond$-2,$ and the hyperfine matrices are 
 \begin{align}
  \mathbb{A}_1 &= \left(\begin{array}{ccc}
   113.740773 & 26.063791 & 0\\
   26.063791 & 83.6449 & 0\\
   0 & 0 & 70.20792
  \end{array}\right) \label{eq:A_1}\\
  \mathbb{A}_2 &= \left(\begin{array}{ccc}
   68.59693 & 0 & 0\\
   0 & 128.788676 & 0\\
   0 & 0 & 70.207916
  \end{array}\right) \label{eq:A_2}\\
  \mathbb{A}_3 &= \left(\begin{array}{ccc}
   113.740778 & -26.063791 & 0\\
   -26.063791 & 83.644905 & 0\\
   0 & 0 & 70.20792
  \end{array}\right) \label{eq:A_3},
 \end{align}
where the elements are expressed in the unit of $2\pi\times{\rm MHz}$. 
This angle $\phi = 2\pi/9$ is merely a pedagogical decision for the demonstration; in theory, any angle that is not parallel or perpendicular to any dangling bonds can be utilized in a similar manner, with a different degree of discrepancies of the effective hyperfine vectors Eqs.~\eqref{eq:def-a_k-appen}-\eqref{eq:def-c_k-appen}. 
The convention of the $x-$axis and $y-$axis for the representations of the matrix above is expressed in the system coordinate. Here the $x-$axis is pointing along the secondary symmetry axis $\sigma_v,$ i.e. along the dangling bond$-2$ in Fig.~
\ref{fig:hBN-appen}, the $z$-axis is pointing in the direction of the plane axis, and the $y-$axis is the complement of the two. We are grateful to Benchen Huang from the Department of Chemistry, University of Chicago, for providing the computationally generated values of $\mathbb{A}_k.$ With these parameters, the summary results of the calculation are given in Tab.~\ref{tab:table_fidelity_full}. 
\begin{table}
\caption{\label{tab:table_fidelity_full} {Maximum fidelity $\mathcal{F}$ of various gates and nuclear spins without and with the effect of dephasing. The control period $\tau$ and the optimal gate time $t_{\rm gate}$ are extracted from the maximum on the fidelity landscape, as exemplified for the gate X$_1$ in Fig.~2(a) in the main text, given the decoherence effect $\Gamma = 0.5~{\rm MHz}.$}}
\begin{ruledtabular}
\begin{tabular}{cccccc}
\multirow{2}{*}{Gate}  &   \multirow{2}{*}{N}   &   \multicolumn{2}{c}{Fidelity $(\%)$}  &  \multirow{2}{*}{$\tau~({\rm ns})$}    &   \multirow{2}{*}{$t_{\rm gate}~({\rm ns})$} \\    \cline{3-4}
   &   &  $\Gamma = 0~{\rm MHz}$  &  $\Gamma = 0.5~{\rm MHz}$  &   &  \\
\hline
\multirow{3}{*}{X}   &1    &{$99.23~$} &{$96.59~$} &{$179.78$} &{$182.78$} \\
    &2    &{$99.46~$} &{$96.86~$} &{$179.78$} &{$179.78$} \\
    &3    &{$99.49~$} &{$96.84~$} &{$179.78$} &{$182.78$} \\
\hline
\multirow{3}{*}{H}   &1    &{$99.45~$} &{$96.81~$} &{$182.78$} &{$182.78$} \\
    &2    &{$99.70~$} &{$97.09~$} &{$179.78$} &{$179.78$} \\
    &3    &{$99.74~$} &{$97.09~$} &{$179.78$} &{$182.78$} \\
\hline
\multirow{3}{*}{T}   &1    &{$99.74~$} &{$97.38~$} &{$179.78$} &{$161.80$} \\
    &2    &{$99.60~$} &{$97.33~$} &{$179.78$} &{$155.81$} \\
    &3    &{$99.74~$} &{$97.38~$} &{$179.78$} &{$161.80$} \\
\hline
\multirow{3}{*}{CX}   &1    &{$96.55~$} &{$93.38~$} &{$215.74$} &{$227.72$} \\
    &2    &{$98.10~$} &{$95.13~$} &{$203.75$} &{$209.75$} \\
    &3    &{$98.20~$} &{$95.59~$} &{$179.78$} &{$182.78$} \\
\hline
\multirow{3}{*}{CH}   &1    &{$99.02~$} &{$96.47~$} &{$179.78$} &{$176.79$} \\
    &2    &{$99.58~$} &{$97.01~$} &{$179.78$} &{$176.79$} \\
    &3    &{$99.65~$} &{$97.04~$} &{$179.78$} &{$179.78$} \\
\hline
\multirow{3}{*}{CT}   &1    &{$99.68~$} &{$97.24~$} &{$179.78$} &{$167.80$} \\
    &2    &{$99.80~$} &{$97.35~$} &{$179.78$} &{$167.80$} \\
    &3    &{$99.57~$} &{$97.35~$} &{$179.78$} &{$152.82$} \\
\hline
\multirow{3}{*}{Y}   &1    &{$99.20~$} &{$96.56~$} &{$179.78$} &{$182.78$} \\
    &2    &{$99.49~$} &{$96.88~$} &{$179.78$} &{$179.78$} \\
    &3    &{$99.49~$} &{$96.84~$} &{$179.78$} &{$182.78$} \\
\hline
\multirow{3}{*}{CY}   &1    &{$96.54~$} &{$93.37~$} &{$215.74$} &{$227.72$} \\
    &2    &{$98.11~$} &{$95.13~$} &{$203.75$} &{$209.75$} \\
    &3    &{$98.20~$} &{$95.59~$} &{$179.78$} &{$182.78$} \\
\hline
\multirow{3}{*}{XX}   &12    &{$98.64~$} &{$96.06~$} &{$176.79$} &{$179.78$} \\
    &23    &{$98.45~$} &{$96.26~$} &{$149.82$} &{$152.82$} \\
    &31    &{$98.75~$} &{$96.12~$} &{$179.78$} &{$182.78$} \\
\hline
\multirow{3}{*}{CXX}   &12    &{$95.37~$} &{$91.69~$} &{$260.69$} &{$269.67$} \\
    &23    &{$96.55~$} &{$93.62~$} &{$203.75$} &{$209.75$} \\
    &31    &{$95.76~$} &{$91.43~$} &{$305.63$} &{$317.62$} \\
\hline
\multirow{3}{*}{CYY}   &12    &{$95.40~$} &{$91.71~$} &{$260.69$} &{$269.67$} \\
    &23    &{$96.56~$} &{$93.63~$} &{$203.75$} &{$209.75$} \\
    &31    &{$95.76~$} &{$91.43~$} &{$308.63$} &{$317.62$} \\
\hline
\multirow{3}{*}{CXY}   &12    &{$95.38~$} &{$91.70~$} &{$260.69$} &{$269.67$} \\
    &23    &{$96.55~$} &{$93.62~$} &{$203.75$} &{$209.75$} \\
    &31    &{$95.75~$} &{$91.43~$} &{$305.63$} &{$317.62$} \\
\hline
\multicolumn{2}{c}{HHH}   &{$98.94~$} &{$96.31~$} &{$182.78$} &{$182.78$} \\
\multicolumn{2}{c}{XXX}   &{$98.10~$} &{$95.49~$} &{$182.78$} &{$182.78$} \\
\multicolumn{2}{c}{CHHH}   &{$98.36~$} &{$95.79~$} &{$179.78$} &{$179.78$} \\
\multicolumn{2}{c}{CXXX}   &{$93.36~$} &{$89.76~$} &{$260.69$} &{$269.67$} \\
\end{tabular}
\end{ruledtabular}
\end{table}

\bigskip
\section{Polarization of the Electron Spin}\label{appen:elec_pol_trad}
The in-plane magnetic field that we consider in our protocol is essential for discriminately addressing the individual nuclear spins.
However, a large magnetic field which is not aligned with the defect symmetry axis can mix its electronic states.
In this section, we investigate the possibility of initializing the electron state through the optical pumping in when an in-plane magnetic field is applied.

In what follows, we adopt the model for the optical pump-induced electron spin polarization from Ref.~\cite{Lee2025}. Recall the free electron Hamiltonian for the considered ground state:
 \begin{equation}
  	\hat{H}_e^{\rm gs} = D\hat{S}_x^2 + \omega_e\hat{S}_z \label{eq:He_GS},
 \end{equation}
with a shorthand notation $D = D_{gs},$ $x$- and $z$-axes denote the principal axis and the direction of the magnetic field, respectively. 
In principle, when one considers the polarization process, the pumping to the excited state and the intermediate metastable state will be incorporated, and the Hamiltonian above will be embedded in a seven-level Hamiltonian. The excited-state Hamiltonian also takes the form 
	\begin{align}
		\hat{H}_e^{\rm es} &= D'(\hat{S}'_x)^2 + \omega'_e\hat{S}'_z \label{eq:He_tot-He_ES},
	\end{align}
where $D' = D_{es}$ is the zero-field splitting of the excited state manifold, $\hat{S}'_\alpha$ are spin operators acting on the spin states thereon, and $\omega'_e = \gamma'_eB$ is the excited state Zeeman splitting. 
The metastable state subspace is one-dimensional, and its Hamiltonian is trivial. 
Note that there is also a metastable state sitting in between the excited state and ground state manifolds. 
Define a functional
	\begin{equation}
		\mathcal{D}_{\hat{L}}\cvb{\rho} = \hat{L}\rho\hat{L}^\dagger -\dfrac{1}{2}\cvc{\hat{L}^\dagger\hat{L},\rho} \label{eq:define-D_L},
	\end{equation}
for some Lindblad operators $\hat{L}.$ 
The spin-conserve radiative process is described by the Lindblad operator
	\begin{align}
		\mathcal{L}_{\rm rad}\cvb{\rho} &= \Gamma_{\rm rad}\mathcal{D}_{\hat{L}_{\rm rad}}\cvb{\rho},~~\hat{L}_{\rm rad} = \sum_{m_s = 0,\pm 1} \ketbra{m_s}{m_s'}\label{eq:L_rad},
	\end{align}
where $\ket{m_s = 0,\pm 1}$ and $\ket{m'_s = 0,\pm 1}$ are eigenstates of the operators $\hat{S}_x$ and $\hat{S}'_x,$ respectively. 
Similarly, the pumping process also takes the global operator form, namely
	\begin{align}
		\mathcal{L}_{\rm pump}\cvb{\rho} &= \Gamma_{p}\mathcal{D}_{\hat{L}_{\rm pump}}\cvb{\rho},~~\hat{L}_{\rm pump} = \sum_{m_s = 0,\pm 1} \ketbra{m'_s}{m_s}\label{eq:L_p}.
	\end{align}
The non-radiative decay, however, is described by the spin-dependent local jump operators from the excited state manifold to the metastable state (intersystem crossing or ISC) and the $s-$decay process from the metastable state to the ground state manifolds:
	\begin{align}
		\mathcal{L}_{\rm non-rad}\cvb{\rho} &= \mathcal{L}_{\rm ISC}\cvb{\rho} + \mathcal{L}_{s}\cvb{\rho} \label{eq:L_non-rad},\\
		\mathcal{L}_{\rm ISC}\cvb{\rho} &= \sum_{m_s = 0,\pm 1} \gamma'_{m_s}\mathcal{D}_{\hat{L}'_{m_s}}\cvb{\rho}, ~~\hat{L}'_{m_s} = \ketbra{p}{m_s'}\label{eq:L_ISC},\\
		\mathcal{L}_s\cvb{\rho} &= \sum_{m_s = 0,\pm 1} \gamma_{m_s}\mathcal{D}_{\hat{L}_{m_s}}\cvb{\rho}, ~~\hat{L}_{m_s} = \ketbra{m_s}{p}\label{eq:L_s},
	\end{align}
where $\ket{p}$ denotes the matastable state. Note that $\gamma'_{1}=\gamma'_{-1}$ and $\gamma_{1}=\gamma_{-1},$ and hence the local-decay process above can be characterized by the ratio $r=\gamma'_{0}/\gamma'_{1}$ and $k=\gamma_{1}/\gamma_{0}.$ When there is no magnetic field, the small ISC ratio, with considerable number $k,$ suggests that the population in the spin state $\ket{m'_s=\pm 1}$ are quickly erased by the ISC decay, whereas that in $\ket{m'_s=\pm 0}$ will decay much later; while the populations in the ground state manifold are pumped to the excited state. This combined effect enables the accumulation of the polarization in the ground state spin state in the asymptotic limit \cite{Lee2025}.

To determine the asymptotic behavior, let us consider the steady state $\mathcal{L}_{\rm par}\cvb{\rho^{\rm par}_{ss}} = 0.$ Let $\rho_{m_s},$ $\rho'_{m_s},$ and $\rho_p$ denote the populations of the spin states in the ground state, excited state, and metastable state manifolds, respectively. 
We find that $\rho'_{m_s} = \rho_{m_s}\Gamma_{\rm pump}/(\Gamma_{\rm rad} + \gamma'_{m_s})$ and $\rho_p = \rho'_{m_s}\gamma'_{m_s}/\gamma_{m_s}$ for all $m_s= 0, \pm 1.$ 
The populations in the spin states on the ground state manifold satisfy
	\begin{equation}
		\rho_{\pm 1} = rk\cv{\dfrac{1 + \Gamma_{\rm rad}/\gamma'_1}{r + \Gamma_{\rm rad}/\gamma'_1}}\rho_0 \label{eq:rho_1-over-rho_0_par}.
	\end{equation}	
In the ideal case $r \ll 1,$ and given that $\Gamma_{\rm rad}\sim\gamma'_1$ and $k\leq 0.5,$ it will follow that $\rho_{\pm 1}/\rho_0\ll 1$ or the spin polarization substantially accumulates in the spin ground state $\ket{m_s=0}$ as claimed. 

When the magnetic field is perpendicular to the axis plane, the analysis above does not hold since the Hamiltonian is no longer preserving the diagonal versus off-diagonal separation in the spin basis structure. In general, the perpendicular magnetic field component will mix the spin state during the dynamics and spoil the high selectivity in the spin-dependent decay rate. 
However, for the case with no parallel magnetic field component, we can observe that the mixing effect concerns only a two-dimensional subspace of the spin submanifolds. 
First we recall the following conventions (from the matrix Eq.~\eqref{eq:T}):
	\highlight{	
	\begin{align*}
		\ket{\pm} &= c_\pm\cv{\frac{\ket{+1} + \ket{-1}}{\sqrt{2}}} \mp c_\mp\ket{0} ,\\
		\ket{D} &= -\cv{\frac{\ket{+1} - \ket{-1}}{\sqrt{2}}},
	\end{align*}
or conversely,
	\begin{align*}
		\ket{m_s = \pm 1} &= \frac{\ket{R} \mp \ket{D}}{\sqrt{2}}, \\
		\ket{m_s = 0} &= \ket{L} = c_-\ket{+} - c_+\ket{-},
	\end{align*}
where 
	\begin{align*}
		c_{\pm} &= \sqrt{\frac{\omega_{\pm}}{2\omega_{\delta}}}, ~~\ket{R} = c_+\ket{+} + c_-\ket{-},\\
		 ~\text{and}&~~\ket{L} = c_-\ket{+} - c_+\ket{-}.
	\end{align*}	}
While the states $\ket{0},\ket{\pm 1}$ is independent of the magnetic field strength $B, $ the basis generated by $\cvc{\ket{+}, \ket{D}, \ket{-}}$ is clearly magnetic field strength $B$ dependent. 
Note that this change of basis is only a mathematical transformation for the convenient frame of reference, without any changes in the physical phenomena involved. 
The coherent part described by the Hamiltonians above will exhibit a mixing in the superposition basis generated by $\cvc{\ket{R}, \ket{D}, \ket{L}}:$ 
	\highlight{	
	\begin{equation}
		\hat{H}_e^{\rm gs} = D\cv{\ketbra{R}{R} + \ketbra{D}{D}} + \omega_{e}\cv{\ketbra{R}{L} + \ketbra{L}{R}}. \label{eq:H_gs_e}
	\end{equation}}
Note that the state mixing only occur in the $+-$ subspace (or $RL$ subspace), while the subspace $D$ is invariant. 
Interestingly, within this superposition basis, the non-radiative decays preserve the same form:
	\begin{align}
		\mathcal{L}_{\rm ISC}\cvb{\rho} &= \sum_{s = R, D, L} \gamma'_{s}\mathcal{D}_{\hat{L}'_{s}}\cvb{\rho}, ~~\hat{L}'_{s} = \ketbra{p}{s'},\label{eq:L_ISC_DRL}\\
		\mathcal{L}_s\cvb{\rho} &= \sum_{s = R, D, L} \gamma_{s}\mathcal{D}_{\hat{L}_{s}}\cvb{\rho}, ~~\hat{L}_{s} = \ketbra{s}{p}\label{eq:L_s_DRL},
	\end{align}
where $\gamma'_{L}=\gamma'_{0},$ $\gamma'_{R}=\gamma'_{D}=\gamma'_{1},$ $\gamma_{L}=\gamma_{0},$ and, $\gamma_{R}=\gamma_{D}=\gamma_{1}.$ 
The local form in this basis of the Lindblad operations above can be treated as state-dependent decays in a similar way as spin-dependent decays in the conventional setup discussed above. 
On the other hand, the radiative decay operations and the pump Lindbladians in general do not preserve the structure in this basis. 
Likewise, the Lindblad operators above can be approximated as
	\begin{align}
		\hat{L}_{\rm rad} = \sum_{s = R,D,L} \ketbra{s}{s'},~~ \hat{L}_{\rm pump} = \sum_{s = R,D,L} \ketbra{s'}{s}\label{eq:L_rad-p_DRL},
	\end{align}
i.e., they also take the same form as in the spin basis. 

We follows the same analysis for the steady state $\mathcal{L}\cvb{\rho_{ss}} = 0,$ and assume that the magnetic field is strong such that $\omega_e\gg D.$ We have
	\begin{equation}
		\frac{\rho'_{\pm}}{\rho_{\pm}} = \frac{\Gamma_{\rm pump}}{\Gamma_{\rm rad} + (\gamma'_0 + \gamma'_1)/2}, ~\text{and}~~\frac{\rho'_{D}}{\rho_{D}} = \frac{\Gamma_{\rm pump}}{\Gamma_{\rm rad} + \gamma'_1}  \label{eq:pop_ES-to-GS},
	\end{equation}
in the original nomenclature, while the population in the metastable state follows 
	$\rho_p = [(\gamma_0+\gamma_1)/(\gamma'_0+\gamma'_1)]\rho'_{\pm} =  (\gamma_1/\gamma'_1)\rho'_{D}.$
Here, one can see that the relations among the matrix elements for the strong in-plane magnetic field in our operational basis are similar to those for the conventional magnetic field direction in the spin basis, with different factors given by the combinations of decay rates. Hence, the suppression of the population of in the excited state and meatstable state manifolds can also be achieved when the pumping rate is relative small compared to the effective decay rate $\Gamma_{\rm pump} \ll \Gamma_{\rm rad} + \gamma'_1,$ although it will result in different values according to the different prefactors. 
However, from the same expressions, the population ratio within the ground state subspace will no longer be trivially controlled by the branching ratio as in the conventional case Eq.~\eqref{eq:rho_1-over-rho_0_par}. 
To be precise, we would rather have
	\begin{equation}
		\rho_{\pm} = \cv{1 + \frac{1}{k}}\cv{\frac{1}{1+r}}\cv{\frac{1+r+2\Gamma_r/\gamma'_1}{2 + 2\Gamma_r/\gamma'_1}}\rho_D	\label{eq:rho_pm-over-rho_0}.
	\end{equation}
Unlike the condition Eq.~\eqref{eq:rho_1-over-rho_0_par}, the ideal branching ratio $r \ll 1,$ together with $\Gamma_{\rm rad}\sim\gamma'_1$ and $k\leq 0.5,$ does not lead to the aggregation of the polarization in any certain state, when the in-plane magnetic field is strong. 
The equality of $\rho_{+}$ and $\rho_{-}$ also signifies the mixed polarity under the extremely strong in-plane magnetic field.

\section{Efficient electron spin polarization}\label{appen:elec_pol_eff}
From the analysis and the numerical observations, we can conclude that for the case with the intermediate and strong in-plane magnetic field in the direction perpendicular to the system axis, one cannot efficiently polarize the electron spin using only conventional optical pumping and relying on the intersystem crossing decay. We would mention further that without electron high spin polarization, one may also not be able to polarize nuclear spins in the cluster using a unitary spin transfer mechanism, e.g., by electron-nuclear spin flip-flop \cite{Gao2022, Tabesh2023}. 
The main reason for such failure is due to the fact that the process is trace preserving, or at least non-increasing, and hence does not increase the purity of the overall state. When the electron initial state is fully polarized, this is not a problem since the purity of the three nuclear spins is transferred to the electron state together with their polarization, and the purity of the overall state remains the same before the electron spin reset. If the electron state is in a mixed state, the purity of the overall state cannot increase, i.e., the nuclear spin states cannot be fully polarized. In other words, the electron high spin polarization is essential for the quantum information processing using \vb-hBN. 
To our knowledge, although the orientation of the magnetic field perpendicular to the system principal axis has been used in the characterization of \vb-hBN, e.g., to study the anisotropy of the hyperfine interaction \cite{Gottscholl2020,Gracheva2023}, there is no study on the polarization under a magnetic field in the direction perpendicular to the symmetry axis. 

Here, we propose two possible protocols for the initialization of the electron spin in a pure state, which can be applied for the polarization of the nuclei. 
First, and the most straightforward choice is the application of a pulsed magnetic field. In particular, one can consider different magnetic field strengths between the electron spin initialization and the nuclear spins operation periods. One of the possibilities is the application of a pulsed magnetic field, see e.g. Refs.~\cite{Haase2003,Kuhne2024}, in which it is set close to zero during the electron spin initialization period, and it is active during the nuclear spins manipulation after the initialization. During the initialization, since there is no perturbation from the background field, the optical pump together with the ISC principle will populate the electron spin into $\ket{0}$ state. The rapid change in magnetic field will maintain the electron spin state $\ket{0}$ in the new magnetic field environment, where in the magnetic field dependent basis $+D-$ we know that
	\highlight{	
	\begin{align*}
		\ket{m_s = 0} &= \ket{L} = c_-\ket{+} - c_+\ket{-}\\
		  &= \cv{\sqrt{\frac{\omega_{-}}{2\omega_{\delta}}}}\ket{+} - \cv{\sqrt{\frac{\omega_{+}}{2\omega_{\delta}}}}\ket{-}
	\end{align*}}
for any value of the magnetic field strength $B.$ In this aspect, the electron initial state is prepared fully in our considered electron qubit subspace $+-,$ and the electron spin can be used as an operational qubit alongside the nuclear spins. 
Another advantage of using the pulsed magnets is that the field strength can be made larger than the usual static magnets; hence, the requirement for the moderate field strength in our protocol can be trivially achievable \cite{Haase2003,Kuhne2024}.
	
Another possibility is the modification of the electron spin initialization process. In this aspect, one may consider a long wavelength pump or maser \cite{Chacko2024, Ng2024}, e.g., using a nitrogen vacancy (NV) center in diamond, converting the conventional $532$ nm laser pump to a GHz-regime maser emission \cite{Zollitsch2025,Breeze2018}. 
Suggested in Ref.~\cite{,Zollitsch2025} it is also interesting to consider additional VB in hBN, whose principal axis points in the magnetic direction, as a maser emitter to initialize our concerned center. 
In particular, by replacing the laser pumping term Eq.~\eqref{eq:L_p} by an incoherent maser drive with the wavelength corresponding to the energy difference between the $\ket{D}$ and $\ket{-}$ levels, i.e., $\omega_+/2 =  [D_{gs}+\sqrt{D_{gs}^2 + 4(\gamma_e B)^2}]/2.$ This can be modeled by
	\begin{equation}
		\mathcal{L}_{\rm maser}\cvb{\rho} = \Omega_{\rm maser}\mathcal{D}_{\hat{L}_{\rm maser}}\cvb{\rho} , ~~ \hat{L}_{\rm maser} = \ketbra{D}{-}, \label{eq:L_maser}
	\end{equation}
where $\Omega_{\rm maser}$ is the pumping rate; the decay processes from the excited state and the metastable state subspaces can be omitted due to the lack of laser pump, and one can find that the state $\ketbra{D}{D}$ in the ground state subspace is one of the steady states. 
This observation can also hold in the approximation for intermediate and strong magnetic fields, with the presence of laser pump and the excited state decay processes. 
For instance, with $\Omega_{\rm maser} = 100$ MHz, the steady state in the ground state subspace (normalized in the seven-level space) of the system 
	\begin{equation}
		\rho_{ss, g} = \left(\begin{array}{ccc}
						0.009 & 1.815\times 10^{ -8 } & 1.511\times 10^{ -9 }\\
						1.815\times 10^{ -8 } & 0.937 & 2.746\times 10^{ -9 }\\
						1.511\times 10^{ -9 } & 2.746\times 10^{ -9 } & 0.054
						\end{array}\right) \label{eq:rho_e_g_maser}
	\end{equation}
in the $+D-$ representation.  
In this sense, the electron initial state is completely outside our operational subspace $+-,$ and one can further bring the state $\ket{D}$ into $\ket{+}$ or $\ket{-}$ using a conventional MW drive. 
The principle behind this relies on the separation between the $D$ and $RL$ (or $+-$) subspaces under the intermediate and high in-plane magnetic field, since the coherent part of the dynamics in $\hat{H}^e_{gs},$ Eq.~\eqref{eq:H_gs_e}, mixes only the states in $RL$ subspace and leave the component in $D$ subspace intact. 
This is not true for arbitrary tilted angles from the $c-$axis apart from the perpendicular one, in which the mixing will involve the whole ground state Hilbert space, and the fixed point of either laser- or maser-induced dynamics will never be a rank one projection. 

These two possibilities have not yet been explored either theoretically or experimentally, and are still open for further studies in the future. 
Since the three nuclear spins cannot be naturally distinguished when the magnetic field is parallel to the $c-$axis, the development of the technique in these directions becomes crucial. 
In principle, it is a question whether to preserve the simplicity of conventional initialization and lose the distinguishability of the individual nuclear spin in the cluster, or employ the perpendicular magnetic field and modify the initialization process to gain the selective controllability of the nuclear spins. 
With the use of the pulsed magnets or the application of maser drive, we believe that our protocol is promising for nuclear spin manipulation in practice with the available state-of-the-art technologies. 

\bigskip
\section{Gate Assisted Polarization of the Nuclear Spins}\label{appen:gate_assist}
To understand the dynamical structure of our considered system, let us revisit the nuclear spin polarization in the operational language. The common procedure of nuclear spin polarization typically relies on the combination of perfect electron spin polarization and the electron-nuclear spin transfer. As a pedagogical example, assume that the initial state is given by $\rho_{0} = \ketbra{\psi}{\psi}\otimes\rho_{N,0},$ where $\ket{\psi}$ is an electron polarized state, $\rho_{N,0}$ is a nuclear spin state, and the dynamics is given by a unitary evolution $\hat{U}$ encrypting the interaction between electron and nuclear spins. The nuclear spin reduced state after the first cycle can be written as $\rho_{N,1} = \frac{\mathcal{K}_{\psi}[\rho_{N,0}]}{\tra\cv{\mathcal{K}_{\psi}[\rho_{N,0}]}},$ where $\mathcal{K}_{\psi}[\rho] = \tra_{e}\big[\hat{U}(\ketbra{\psi}{\psi}\otimes\rho)\hat{U}^\dagger\big],$ and $\tra_{e}$ is the partial trace over electron degree of freedom. By repeating this for $n$ cycles, the nuclear spins reduced state will be $\rho_{N,n} = \frac{\mathcal{K}^n_{\psi}[\rho_{N,0}]}{\tra\cv{\mathcal{K}^n_{\psi}[\rho_{N,0}]}},$ where the denominator itself is also the success probability of the procedure. The key element of the polarization protocol is the evolution $\hat{U},$ that made the reduced operation $\mathcal{K}^n_{\psi}$ approaches a projection with some appropriate number of repetition $n.$

For the electron spin defect in solid states, the typical setup for this mechanism would be $\ket{\psi} = \ket{-}$ in the computational basis, while the interaction is given by the flip-flop interaction, $\sum_kJ_k\cv{\hat{\sigma}_+\hat{I}_k^- + \hat{\sigma}_-\hat{I}_k^+},$ with the coupling strengths $J_k.$ 
Note that, in our convention, we write $\hat{\sigma}_\pm = \ketbra{\pm}{\mp}.$ 
This interaction form can be prepared in hBN samples by utilizing the (dipolar) interaction $\sum_kJ_k\hat{\sigma}_x\hat{I}_x$ under a strong magnetic field, and some resonant condition, such that the Larmor frequencies are relatively large compared to the coupling parameters $J_k$ \cite{Tabesh2023}. 
In the large enough time scale the dipolar interaction is dominant and the Fermi contact interaction can be omitted as fast rotating term, and the interactions of the considered cluster of nuclear spins among themselves and with the rest of hBN, i.e., $\sum_{k\neq l}b_{kl}\cvb{\hat{I}_k^z\hat{I}_l^z - (\hat{I}_k^+\hat{I}_l^-+\hat{I}_k^-\hat{I}_l^+)/4},$ will also be incorporated. This latter exchange term will average out the spins in the cluster, while the flip-flip term will transfer the electron spin polarization to the total nuclear spin $\hat{I}_z = \sum_k\hat{I}_k^z,$ and by repeating this procedure, the total nuclear spin polarization will increase \cite{Gao2022, Tabesh2023}.

In our setup, however, since we consider the situation in the intermediate field regime, the Fermi contact will become dominant while the electron-nuclear and nuclear-nuclear dipolar interactions are negligible. The electron spin Larmor frequency will remain relatively larger than the interaction terms, whereas those of nuclear spins are small but yet visible in this regime, i.e., it is no longer a weak coupling regime for the nuclear spin dynamics  (see the main text). The flip-flop type interaction then cannot be conveniently prepared as in the conventional setup, in which the magnetic field direction is along the $c-$axis. However, such interaction can be engineered using the gates prepared by our protocols. 
Let $\ket{\uparrow}_k$ and $\ket{\downarrow}_k$ are eigenstates of $\hat{I}_k^z$ corresponds to the eigenvalues $+1/2$ and $-1/2,$ respectively. Here for the nuclear raising and lowering operators we use the conventions $\hat{I}_k^+ = \hat{I}_k^x + i\hat{I}_k^y = \ket{{\uparrow}}_k\bra{{\downarrow}}$ and $\hat{I}_k^- = \hat{I}_k^x - i\hat{I}_k^y = \ket{{\downarrow}}_k\bra{{\uparrow}}.$ 
First, we observe that $\cvb{\hat{\sigma}_x\hat{I}_k^x,\hat{\sigma}_y\hat{I}_k^y} = 0$ for any nuclear spin $k,$ and hence one can have a decomposition
	\begin{align}
		&\vspace{-0.2cm} e^{2i\varphi_k\cv{\hat{\sigma}_+\hat{I}_k^- + \hat{\sigma}_-\hat{I}_k^+}}\nn\\ 
			&= e^{i\varphi_k\cv{\hat{\sigma}_x\hat{I}_k^x + \hat{\sigma}_y\hat{I}_k^y}} = e^{i\varphi_k\hat{\sigma}_x\hat{I}_k^x}e^{i\varphi_k\hat{\sigma}_y\hat{I}_k^y}\nn\\
			&= e^{-i(\pi/4)\hat{\sigma}_y}{\rm CR}_{k}^x(\varphi_k)e^{i(\pi/4)\hat{\sigma}_y}e^{i(\pi/4)\hat{\sigma}_x}{\rm CR}_{k}^y(\varphi_k)e^{-i(\pi/4)\hat{\sigma}_x} \label{eq:synt_flip-flop}.
	\end{align}
In this sense, for a single nuclear spin, the flip-flop-typed interaction can be \emph{synthesized} from the conditional gates and the electron local rotations. 
Note that for any nuclear spin $k,$ one can further have
	\begin{align}
		e^{i\varphi_k \cv{\hat{\sigma}_+\hat{I}_k^- + \hat{\sigma}_-\hat{I}_k^+}} &= \ketbra{+}{+}\otimes(\hat{P}_k^{\uparrow} + \cos\varphi_k\hat{P}_k^{\downarrow})\nn\\
			&\phantom{.=} + \ketbra{-}{-}\otimes(\hat{P}_k^{\downarrow} + \cos\varphi_k\hat{P}_k^{\uparrow})\nn\\
			&\phantom{.=} + i\sin\varphi_k(\hat{\sigma}_+\hat{I}_k^- + \hat{\sigma}_-\hat{I}_k^+) \label{eq:U_flip-flop_k},
	\end{align}
where $\hat{P}_k^{\uparrow} = \ket{{\uparrow}}_k\bra{{\uparrow}}$ ($\hat{P}_k^{\downarrow} = \ket{{\downarrow}}_k\bra{{\downarrow}}.$) 
With these observations, by setting $\varphi_k = -\pi/2,$ for the polarized electron spin initial state $\ket{\psi} = \ket{-},$ one can achieve at a selective spin transfer:
	\begin{align}
		&\vspace{-0.2cm} e^{-i\pi\cv{\hat{\sigma}_+\hat{I}_k^- + \hat{\sigma}_-\hat{I}_k^+}/2}\ket{-}\otimes(\beta_k^\uparrow\ket{\uparrow} + \beta_k^\downarrow\ket{\downarrow})\nn\\ 
		&= \cvb{\ketbra{+}{+}\otimes\hat{P}_k^{\uparrow} + \ketbra{-}{-}\otimes\hat{P}_k^{\downarrow}  - i(\hat{\sigma}_+\hat{I}_k^- + \hat{\sigma}_-\hat{I}_k^+)}\nn\\
		&\phantom{.=~~}\times\ket{-}\otimes(\beta_k^\uparrow\ket{\uparrow} + \beta_k^\downarrow\ket{\downarrow}) \nn\\
		&=(-i\beta_k^\uparrow\ket{+} + \beta_k^\downarrow\ket{-})\otimes\ket{\downarrow} \label{eq:e-N_k-spin-transfer}.
	\end{align}
By repeating this procedure for all three nuclear spins in the VB3N cluster, our whole system can then be essentially polarized. The identities above are not satisfied for simultaneously polarization of multiple spins since the operators $\hat{\sigma}_+\hat{I}_k^- + \hat{\sigma}_-\hat{I}_k^+$ do not commute for different $k,$ and hence, without an averaging mechanism by intra-cluster interaction in this short time regime, the nuclear spin polarization should be done in a site-by-site manner. 

Note that the swapping between an electron spin state and the nuclear spin state on a particular site can be exploited as a reading mechanism for the nuclear spins. 
For the swap process between the electron spin state $\rho_e$ and the nuclear spin state $\rho_1 = \cv{\begin{array}{cc} r & s\\ s^* & 1-r\end{array}}$ at site $1$ can be written as 
\begin{widetext}
	\begin{equation}
		{\rm CZ}^\dagger_1\cvb{e^{-i\pi\cv{\hat{\sigma}_+\hat{I}_1^- + \hat{\sigma}_-\hat{I}_1^+}/2}\cv{\rho_e\otimes\rho_1\otimes\rho_{23}}e^{i\pi\cv{\hat{\sigma}_+\hat{I}_1^- + \hat{\sigma}_-\hat{I}_1^+}/2}}{\rm CZ}_1 = \cv{\begin{array}{cc} r & s\\ s^* & 1-r\end{array}}\otimes\rho_e\otimes\rho_{23}, \label{eq:local_swap}
	\end{equation}
\end{widetext}
where $\rho_{23}$ is an arbitrary nuclear spin state for the remaining sites, ${\rm CZ}_1 = e^{i\pi\hat{\sigma}_z\hat{I}_1^z/2}= e^{-i\pi\hat{I}_1^y/2}{\rm CX}_1e^{i\pi\hat{I}_1^y/2} = e^{-i\pi\hat{I}_1^y/2}e^{i\pi\hat{\sigma}_z\hat{I}_1^x/2}e^{i\pi\hat{I}_1^y/2}$ is a controlled $Z$ gate in the compultational basis on the site $1,$ and, the flip-flop operation can be synthesized via Eq.~\eqref{eq:synt_flip-flop} with $\boldsymbol{\varphi}_k = (\pi/2, 0, 0).$ 
This selective swapping, together with optically or photoelectrically \cite{Ru2025} detected magnetic resonance reading technique (ODMR and PDMR respectively), the reading protocol for nuclear spin states can then be achieved, while the writing protocol on nuclear spins can be done by using unconditional gates after nuclear spins initialization.

%\newpage
%\bibliography{reference.bib}
%
%
%\end{document}

\bibliography{main.bbl}

\begin{thebibliography}{53}%
\makeatletter
\providecommand \@ifxundefined [1]{%
 \@ifx{#1\undefined}
}%
\providecommand \@ifnum [1]{%
 \ifnum #1\expandafter \@firstoftwo
 \else \expandafter \@secondoftwo
 \fi
}%
\providecommand \@ifx [1]{%
 \ifx #1\expandafter \@firstoftwo
 \else \expandafter \@secondoftwo
 \fi
}%
\providecommand \natexlab [1]{#1}%
\providecommand \enquote  [1]{``#1''}%
\providecommand \bibnamefont  [1]{#1}%
\providecommand \bibfnamefont [1]{#1}%
\providecommand \citenamefont [1]{#1}%
\providecommand \href@noop [0]{\@secondoftwo}%
\providecommand \href [0]{\begingroup \@sanitize@url \@href}%
\providecommand \@href[1]{\@@startlink{#1}\@@href}%
\providecommand \@@href[1]{\endgroup#1\@@endlink}%
\providecommand \@sanitize@url [0]{\catcode `\\12\catcode `\$12\catcode
  `\&12\catcode `\#12\catcode `\^12\catcode `\_12\catcode `\%12\relax}%
\providecommand \@@startlink[1]{}%
\providecommand \@@endlink[0]{}%
\providecommand \url  [0]{\begingroup\@sanitize@url \@url }%
\providecommand \@url [1]{\endgroup\@href {#1}{\urlprefix }}%
\providecommand \urlprefix  [0]{URL }%
\providecommand \Eprint [0]{\href }%
\providecommand \doibase [0]{https://doi.org/}%
\providecommand \selectlanguage [0]{\@gobble}%
\providecommand \bibinfo  [0]{\@secondoftwo}%
\providecommand \bibfield  [0]{\@secondoftwo}%
\providecommand \translation [1]{[#1]}%
\providecommand \BibitemOpen [0]{}%
\providecommand \bibitemStop [0]{}%
\providecommand \bibitemNoStop [0]{.\EOS\space}%
\providecommand \EOS [0]{\spacefactor3000\relax}%
\providecommand \BibitemShut  [1]{\csname bibitem#1\endcsname}%
\let\auto@bib@innerbib\@empty
%</preamble>
\bibitem [{\citenamefont {Tran}\ \emph {et~al.}(2015)\citenamefont {Tran},
  \citenamefont {Bray}, \citenamefont {Ford}, \citenamefont {Toth},\ and\
  \citenamefont {Aharonovich}}]{Tran2015}%
  \BibitemOpen
  \bibfield  {author} {\bibinfo {author} {\bibfnamefont {T.~T.}\ \bibnamefont
  {Tran}}, \bibinfo {author} {\bibfnamefont {K.}~\bibnamefont {Bray}}, \bibinfo
  {author} {\bibfnamefont {M.~J.}\ \bibnamefont {Ford}}, \bibinfo {author}
  {\bibfnamefont {M.}~\bibnamefont {Toth}},\ and\ \bibinfo {author}
  {\bibfnamefont {I.}~\bibnamefont {Aharonovich}},\ }\bibfield  {title}
  {\bibinfo {title} {Quantum emission from hexagonal boron nitride
  monolayers},\ }\href {https://doi.org/10.1038/nnano.2015.242} {\bibfield
  {journal} {\bibinfo  {journal} {Nat. Nanotechnol.}\ }\textbf {\bibinfo
  {volume} {11}},\ \bibinfo {pages} {37} (\bibinfo {year} {2015})}\BibitemShut
  {NoStop}%
\bibitem [{\citenamefont {Gottscholl}\ \emph {et~al.}(2021)\citenamefont
  {Gottscholl}, \citenamefont {Diez}, \citenamefont {Soltamov}, \citenamefont
  {Kasper}, \citenamefont {Sperlich}, \citenamefont {Kianinia}, \citenamefont
  {Bradac}, \citenamefont {Aharonovich},\ and\ \citenamefont
  {Dyakonov}}]{Gottscholl2021}%
  \BibitemOpen
  \bibfield  {author} {\bibinfo {author} {\bibfnamefont {A.}~\bibnamefont
  {Gottscholl}}, \bibinfo {author} {\bibfnamefont {M.}~\bibnamefont {Diez}},
  \bibinfo {author} {\bibfnamefont {V.}~\bibnamefont {Soltamov}}, \bibinfo
  {author} {\bibfnamefont {C.}~\bibnamefont {Kasper}}, \bibinfo {author}
  {\bibfnamefont {A.}~\bibnamefont {Sperlich}}, \bibinfo {author}
  {\bibfnamefont {M.}~\bibnamefont {Kianinia}}, \bibinfo {author}
  {\bibfnamefont {C.}~\bibnamefont {Bradac}}, \bibinfo {author} {\bibfnamefont
  {I.}~\bibnamefont {Aharonovich}},\ and\ \bibinfo {author} {\bibfnamefont
  {V.}~\bibnamefont {Dyakonov}},\ }\bibfield  {title} {\bibinfo {title} {Room
  temperature coherent control of spin defects in hexagonal boron nitride},\
  }\href {https://doi.org/10.1126/sciadv.abf3630} {\bibfield  {journal}
  {\bibinfo  {journal} {Sci. Adv.}\ }\textbf {\bibinfo {volume} {7}},\ \bibinfo
  {pages} {eabf3630} (\bibinfo {year} {2021})}\BibitemShut {NoStop}%
\bibitem [{\citenamefont {Stern}\ \emph {et~al.}(2022)\citenamefont {Stern},
  \citenamefont {Gu}, \citenamefont {Jarman}, \citenamefont {Eizagirre~Barker},
  \citenamefont {Mendelson}, \citenamefont {Chugh}, \citenamefont {Schott},
  \citenamefont {Tan}, \citenamefont {Sirringhaus}, \citenamefont
  {Aharonovich},\ and\ \citenamefont {Atat\"{u}re}}]{Stern2022}%
  \BibitemOpen
  \bibfield  {author} {\bibinfo {author} {\bibfnamefont {H.~L.}\ \bibnamefont
  {Stern}}, \bibinfo {author} {\bibfnamefont {Q.}~\bibnamefont {Gu}}, \bibinfo
  {author} {\bibfnamefont {J.}~\bibnamefont {Jarman}}, \bibinfo {author}
  {\bibfnamefont {S.}~\bibnamefont {Eizagirre~Barker}}, \bibinfo {author}
  {\bibfnamefont {N.}~\bibnamefont {Mendelson}}, \bibinfo {author}
  {\bibfnamefont {D.}~\bibnamefont {Chugh}}, \bibinfo {author} {\bibfnamefont
  {S.}~\bibnamefont {Schott}}, \bibinfo {author} {\bibfnamefont {H.~H.}\
  \bibnamefont {Tan}}, \bibinfo {author} {\bibfnamefont {H.}~\bibnamefont
  {Sirringhaus}}, \bibinfo {author} {\bibfnamefont {I.}~\bibnamefont
  {Aharonovich}},\ and\ \bibinfo {author} {\bibfnamefont {M.}~\bibnamefont
  {Atat\"{u}re}},\ }\bibfield  {title} {\bibinfo {title} {Room-temperature
  optically detected magnetic resonance of single defects in hexagonal boron
  nitride},\ }\href {https://doi.org/10.1038/s41467-022-28169-z} {\bibfield
  {journal} {\bibinfo  {journal} {Nat. Commun.}\ }\textbf {\bibinfo {volume}
  {13}},\ \bibinfo {pages} {618} (\bibinfo {year} {2022})}\BibitemShut
  {NoStop}%
\bibitem [{\citenamefont {Mu}\ \emph {et~al.}(2022)\citenamefont {Mu},
  \citenamefont {Cai}, \citenamefont {Chen}, \citenamefont {Kenny},
  \citenamefont {Jiang}, \citenamefont {Ru}, \citenamefont {Lyu}, \citenamefont
  {Koh}, \citenamefont {Liu}, \citenamefont {Aharonovich},\ and\ \citenamefont
  {Gao}}]{Mu2022}%
  \BibitemOpen
  \bibfield  {author} {\bibinfo {author} {\bibfnamefont {Z.}~\bibnamefont
  {Mu}}, \bibinfo {author} {\bibfnamefont {H.}~\bibnamefont {Cai}}, \bibinfo
  {author} {\bibfnamefont {D.}~\bibnamefont {Chen}}, \bibinfo {author}
  {\bibfnamefont {J.}~\bibnamefont {Kenny}}, \bibinfo {author} {\bibfnamefont
  {Z.}~\bibnamefont {Jiang}}, \bibinfo {author} {\bibfnamefont
  {S.}~\bibnamefont {Ru}}, \bibinfo {author} {\bibfnamefont {X.}~\bibnamefont
  {Lyu}}, \bibinfo {author} {\bibfnamefont {T.~S.}\ \bibnamefont {Koh}},
  \bibinfo {author} {\bibfnamefont {X.}~\bibnamefont {Liu}}, \bibinfo {author}
  {\bibfnamefont {I.}~\bibnamefont {Aharonovich}},\ and\ \bibinfo {author}
  {\bibfnamefont {W.}~\bibnamefont {Gao}},\ }\bibfield  {title} {\bibinfo
  {title} {Excited-state optically detected magnetic resonance of spin defects
  in hexagonal boron nitride},\ }\href
  {https://doi.org/10.1103/PhysRevLett.128.216402} {\bibfield  {journal}
  {\bibinfo  {journal} {Phys. Rev. Lett.}\ }\textbf {\bibinfo {volume} {128}},\
  \bibinfo {pages} {216402} (\bibinfo {year} {2022})}\BibitemShut {NoStop}%
\bibitem [{\citenamefont {Abdi}\ \emph {et~al.}(2018)\citenamefont {Abdi},
  \citenamefont {Chou}, \citenamefont {Gali},\ and\ \citenamefont
  {Plenio}}]{Abdi2018}%
  \BibitemOpen
  \bibfield  {author} {\bibinfo {author} {\bibfnamefont {M.}~\bibnamefont
  {Abdi}}, \bibinfo {author} {\bibfnamefont {J.-P.}\ \bibnamefont {Chou}},
  \bibinfo {author} {\bibfnamefont {A.}~\bibnamefont {Gali}},\ and\ \bibinfo
  {author} {\bibfnamefont {M.~B.}\ \bibnamefont {Plenio}},\ }\bibfield  {title}
  {\bibinfo {title} {Color centers in hexagonal boron nitride monolayers: A
  group theory and ab initio analysis},\ }\href
  {https://doi.org/10.1021/acsphotonics.7b01442} {\bibfield  {journal}
  {\bibinfo  {journal} {ACS Photonics}\ }\textbf {\bibinfo {volume} {5}},\
  \bibinfo {pages} {1967} (\bibinfo {year} {2018})}\BibitemShut {NoStop}%
\bibitem [{\citenamefont {Sajid}\ \emph {et~al.}(2020)\citenamefont {Sajid},
  \citenamefont {Ford},\ and\ \citenamefont {Reimers}}]{Sajid2020}%
  \BibitemOpen
  \bibfield  {author} {\bibinfo {author} {\bibfnamefont {A.}~\bibnamefont
  {Sajid}}, \bibinfo {author} {\bibfnamefont {M.~J.}\ \bibnamefont {Ford}},\
  and\ \bibinfo {author} {\bibfnamefont {J.~R.}\ \bibnamefont {Reimers}},\
  }\bibfield  {title} {\bibinfo {title} {Single-photon emitters in hexagonal
  boron nitride: a review of progress},\ }\href
  {https://doi.org/10.1088/1361-6633/ab6310} {\bibfield  {journal} {\bibinfo
  {journal} {Rep. Prog. Phys.}\ }\textbf {\bibinfo {volume} {83}},\ \bibinfo
  {pages} {044501} (\bibinfo {year} {2020})}\BibitemShut {NoStop}%
\bibitem [{\citenamefont {Gottscholl}\ \emph {et~al.}(2020)\citenamefont
  {Gottscholl}, \citenamefont {Kianinia}, \citenamefont {Soltamov},
  \citenamefont {Orlinskii}, \citenamefont {Mamin}, \citenamefont {Bradac},
  \citenamefont {Kasper}, \citenamefont {Krambrock}, \citenamefont {Sperlich},
  \citenamefont {Toth}, \citenamefont {Aharonovich},\ and\ \citenamefont
  {Dyakonov}}]{Gottscholl2020}%
  \BibitemOpen
  \bibfield  {author} {\bibinfo {author} {\bibfnamefont {A.}~\bibnamefont
  {Gottscholl}}, \bibinfo {author} {\bibfnamefont {M.}~\bibnamefont
  {Kianinia}}, \bibinfo {author} {\bibfnamefont {V.}~\bibnamefont {Soltamov}},
  \bibinfo {author} {\bibfnamefont {S.}~\bibnamefont {Orlinskii}}, \bibinfo
  {author} {\bibfnamefont {G.}~\bibnamefont {Mamin}}, \bibinfo {author}
  {\bibfnamefont {C.}~\bibnamefont {Bradac}}, \bibinfo {author} {\bibfnamefont
  {C.}~\bibnamefont {Kasper}}, \bibinfo {author} {\bibfnamefont
  {K.}~\bibnamefont {Krambrock}}, \bibinfo {author} {\bibfnamefont
  {A.}~\bibnamefont {Sperlich}}, \bibinfo {author} {\bibfnamefont
  {M.}~\bibnamefont {Toth}}, \bibinfo {author} {\bibfnamefont {I.}~\bibnamefont
  {Aharonovich}},\ and\ \bibinfo {author} {\bibfnamefont {V.}~\bibnamefont
  {Dyakonov}},\ }\bibfield  {title} {\bibinfo {title} {Initialization and
  read-out of intrinsic spin defects in a van der waals crystal at room
  temperature},\ }\href {https://doi.org/10.1038/s41563-020-0619-6} {\bibfield
  {journal} {\bibinfo  {journal} {Nat. Mater.}\ }\textbf {\bibinfo {volume}
  {19}},\ \bibinfo {pages} {540} (\bibinfo {year} {2020})}\BibitemShut
  {NoStop}%
\bibitem [{\citenamefont {{\c C}akan}\ \emph {et~al.}(2025)\citenamefont {{\c
  C}akan}, \citenamefont {Cholsuk}, \citenamefont {Gale}, \citenamefont
  {Kianinia}, \citenamefont {Pa{\c c}al}, \citenamefont {Ate{\c s}},
  \citenamefont {Aharonovich}, \citenamefont {Toth},\ and\ \citenamefont
  {Vogl}}]{Cakan2025}%
  \BibitemOpen
  \bibfield  {author} {\bibinfo {author} {\bibfnamefont {A.}~\bibnamefont {{\c
  C}akan}}, \bibinfo {author} {\bibfnamefont {C.}~\bibnamefont {Cholsuk}},
  \bibinfo {author} {\bibfnamefont {A.}~\bibnamefont {Gale}}, \bibinfo {author}
  {\bibfnamefont {M.}~\bibnamefont {Kianinia}}, \bibinfo {author}
  {\bibfnamefont {S.}~\bibnamefont {Pa{\c c}al}}, \bibinfo {author}
  {\bibfnamefont {S.}~\bibnamefont {Ate{\c s}}}, \bibinfo {author}
  {\bibfnamefont {I.}~\bibnamefont {Aharonovich}}, \bibinfo {author}
  {\bibfnamefont {M.}~\bibnamefont {Toth}},\ and\ \bibinfo {author}
  {\bibfnamefont {T.}~\bibnamefont {Vogl}},\ }\bibfield  {title} {\bibinfo
  {title} {Quantum optics applications of hexagonal boron nitride defects},\
  }\href {https://doi.org/https://doi.org/10.1002/adom.202402508} {\bibfield
  {journal} {\bibinfo  {journal} {Advanced Optical Materials}\ }\textbf
  {\bibinfo {volume} {13}},\ \bibinfo {pages} {2402508} (\bibinfo {year}
  {2025})}\BibitemShut {NoStop}%
\bibitem [{\citenamefont {Ahmadi}\ \emph {et~al.}(2024)\citenamefont {Ahmadi},
  \citenamefont {Schwertfeger}, \citenamefont {Werner}, \citenamefont {Wiese},
  \citenamefont {Lester}, \citenamefont {Da~Ros}, \citenamefont {Krause},
  \citenamefont {Ritter}, \citenamefont {Abasifard}, \citenamefont {Cholsuk},
  \citenamefont {Kr\"{a}mer}, \citenamefont {Atzeni}, \citenamefont
  {G\"{u}ndoğan}, \citenamefont {Sachidananda}, \citenamefont {Pardo},
  \citenamefont {Nolte}, \citenamefont {Lohrmann}, \citenamefont {Ling},
  \citenamefont {Bartholom\"{a}us}, \citenamefont {Corrielli}, \citenamefont
  {Krutzik},\ and\ \citenamefont {Vogl}}]{Ahmadi2024}%
  \BibitemOpen
  \bibfield  {author} {\bibinfo {author} {\bibfnamefont {N.}~\bibnamefont
  {Ahmadi}}, \bibinfo {author} {\bibfnamefont {S.}~\bibnamefont
  {Schwertfeger}}, \bibinfo {author} {\bibfnamefont {P.}~\bibnamefont
  {Werner}}, \bibinfo {author} {\bibfnamefont {L.}~\bibnamefont {Wiese}},
  \bibinfo {author} {\bibfnamefont {J.}~\bibnamefont {Lester}}, \bibinfo
  {author} {\bibfnamefont {E.}~\bibnamefont {Da~Ros}}, \bibinfo {author}
  {\bibfnamefont {J.}~\bibnamefont {Krause}}, \bibinfo {author} {\bibfnamefont
  {S.}~\bibnamefont {Ritter}}, \bibinfo {author} {\bibfnamefont
  {M.}~\bibnamefont {Abasifard}}, \bibinfo {author} {\bibfnamefont
  {C.}~\bibnamefont {Cholsuk}}, \bibinfo {author} {\bibfnamefont {R.~G.}\
  \bibnamefont {Kr\"{a}mer}}, \bibinfo {author} {\bibfnamefont
  {S.}~\bibnamefont {Atzeni}}, \bibinfo {author} {\bibfnamefont
  {M.}~\bibnamefont {G\"{u}ndoğan}}, \bibinfo {author} {\bibfnamefont
  {S.}~\bibnamefont {Sachidananda}}, \bibinfo {author} {\bibfnamefont
  {D.}~\bibnamefont {Pardo}}, \bibinfo {author} {\bibfnamefont
  {S.}~\bibnamefont {Nolte}}, \bibinfo {author} {\bibfnamefont
  {A.}~\bibnamefont {Lohrmann}}, \bibinfo {author} {\bibfnamefont
  {A.}~\bibnamefont {Ling}}, \bibinfo {author} {\bibfnamefont {J.}~\bibnamefont
  {Bartholom\"{a}us}}, \bibinfo {author} {\bibfnamefont {G.}~\bibnamefont
  {Corrielli}}, \bibinfo {author} {\bibfnamefont {M.}~\bibnamefont {Krutzik}},\
  and\ \bibinfo {author} {\bibfnamefont {T.}~\bibnamefont {Vogl}},\ }\bibfield
  {title} {\bibinfo {title} {Quick3$^3$ ‐ design of a satellite‐based
  quantum light source for quantum communication and extended physical theory
  tests in space},\ }\href {https://doi.org/10.1002/qute.202300343} {\bibfield
  {journal} {\bibinfo  {journal} {Adv. Quantum Technol.}\ }\textbf {\bibinfo
  {volume} {7}},\ \bibinfo {pages} {2300343} (\bibinfo {year}
  {2024})}\BibitemShut {NoStop}%
\bibitem [{\citenamefont {Cholsuk}\ \emph {et~al.}(2022)\citenamefont
  {Cholsuk}, \citenamefont {Suwanna},\ and\ \citenamefont
  {Vogl}}]{Cholsuk2022}%
  \BibitemOpen
  \bibfield  {author} {\bibinfo {author} {\bibfnamefont {C.}~\bibnamefont
  {Cholsuk}}, \bibinfo {author} {\bibfnamefont {S.}~\bibnamefont {Suwanna}},\
  and\ \bibinfo {author} {\bibfnamefont {T.}~\bibnamefont {Vogl}},\ }\bibfield
  {title} {\bibinfo {title} {Tailoring the emission wavelength of color centers
  in hexagonal boron nitride for quantum applications},\ }\href
  {https://doi.org/10.3390/nano12142427} {\bibfield  {journal} {\bibinfo
  {journal} {Nanomaterials}\ }\textbf {\bibinfo {volume} {12}},\ \bibinfo
  {pages} {2427} (\bibinfo {year} {2022})}\BibitemShut {NoStop}%
\bibitem [{\citenamefont {Wolfowicz}\ \emph {et~al.}(2021)\citenamefont
  {Wolfowicz}, \citenamefont {Heremans}, \citenamefont {Anderson},
  \citenamefont {Kanai}, \citenamefont {Seo}, \citenamefont {Gali},
  \citenamefont {Galli},\ and\ \citenamefont {Awschalom}}]{Wolfowicz2021}%
  \BibitemOpen
  \bibfield  {author} {\bibinfo {author} {\bibfnamefont {G.}~\bibnamefont
  {Wolfowicz}}, \bibinfo {author} {\bibfnamefont {F.~J.}\ \bibnamefont
  {Heremans}}, \bibinfo {author} {\bibfnamefont {C.~P.}\ \bibnamefont
  {Anderson}}, \bibinfo {author} {\bibfnamefont {S.}~\bibnamefont {Kanai}},
  \bibinfo {author} {\bibfnamefont {H.}~\bibnamefont {Seo}}, \bibinfo {author}
  {\bibfnamefont {A.}~\bibnamefont {Gali}}, \bibinfo {author} {\bibfnamefont
  {G.}~\bibnamefont {Galli}},\ and\ \bibinfo {author} {\bibfnamefont {D.~D.}\
  \bibnamefont {Awschalom}},\ }\bibfield  {title} {\bibinfo {title} {Quantum
  guidelines for solid-state spin defects},\ }\href
  {https://doi.org/10.1038/s41578-021-00306-y} {\bibfield  {journal} {\bibinfo
  {journal} {Nat. Rev. Mater.}\ }\textbf {\bibinfo {volume} {6}},\ \bibinfo
  {pages} {906} (\bibinfo {year} {2021})}\BibitemShut {NoStop}%
\bibitem [{\citenamefont {Chejanovsky}\ \emph {et~al.}(2016)\citenamefont
  {Chejanovsky}, \citenamefont {Rezai}, \citenamefont {Paolucci}, \citenamefont
  {Kim}, \citenamefont {Rendler}, \citenamefont {Rouabeh}, \citenamefont
  {F{\'a}varo~de Oliveira}, \citenamefont {Herlinger}, \citenamefont
  {Denisenko}, \citenamefont {Yang}, \citenamefont {Gerhardt}, \citenamefont
  {Finkler}, \citenamefont {Smet},\ and\ \citenamefont
  {Wrachtrup}}]{Chejanovsky2016}%
  \BibitemOpen
  \bibfield  {author} {\bibinfo {author} {\bibfnamefont {N.}~\bibnamefont
  {Chejanovsky}}, \bibinfo {author} {\bibfnamefont {M.}~\bibnamefont {Rezai}},
  \bibinfo {author} {\bibfnamefont {F.}~\bibnamefont {Paolucci}}, \bibinfo
  {author} {\bibfnamefont {Y.}~\bibnamefont {Kim}}, \bibinfo {author}
  {\bibfnamefont {T.}~\bibnamefont {Rendler}}, \bibinfo {author} {\bibfnamefont
  {W.}~\bibnamefont {Rouabeh}}, \bibinfo {author} {\bibfnamefont
  {F.}~\bibnamefont {F{\'a}varo~de Oliveira}}, \bibinfo {author} {\bibfnamefont
  {P.}~\bibnamefont {Herlinger}}, \bibinfo {author} {\bibfnamefont
  {A.}~\bibnamefont {Denisenko}}, \bibinfo {author} {\bibfnamefont
  {S.}~\bibnamefont {Yang}}, \bibinfo {author} {\bibfnamefont {I.}~\bibnamefont
  {Gerhardt}}, \bibinfo {author} {\bibfnamefont {A.}~\bibnamefont {Finkler}},
  \bibinfo {author} {\bibfnamefont {J.~H.}\ \bibnamefont {Smet}},\ and\
  \bibinfo {author} {\bibfnamefont {J.}~\bibnamefont {Wrachtrup}},\ }\bibfield
  {title} {\bibinfo {title} {Structural attributes and photodynamics of visible
  spectrum quantum emitters in hexagonal boron nitride},\ }\href
  {https://doi.org/10.1021/acs.nanolett.6b03268} {\bibfield  {journal}
  {\bibinfo  {journal} {Nano Lett.}\ }\textbf {\bibinfo {volume} {16}},\
  \bibinfo {pages} {7037} (\bibinfo {year} {2016})}\BibitemShut {NoStop}%
\bibitem [{\citenamefont {Proscia}\ \emph {et~al.}(2018)\citenamefont
  {Proscia}, \citenamefont {Shotan}, \citenamefont {Jayakumar}, \citenamefont
  {Reddy}, \citenamefont {Cohen}, \citenamefont {Dollar}, \citenamefont
  {Alkauskas}, \citenamefont {Doherty}, \citenamefont {Meriles},\ and\
  \citenamefont {Menon}}]{Proscia2018}%
  \BibitemOpen
  \bibfield  {author} {\bibinfo {author} {\bibfnamefont {N.~V.}\ \bibnamefont
  {Proscia}}, \bibinfo {author} {\bibfnamefont {Z.}~\bibnamefont {Shotan}},
  \bibinfo {author} {\bibfnamefont {H.}~\bibnamefont {Jayakumar}}, \bibinfo
  {author} {\bibfnamefont {P.}~\bibnamefont {Reddy}}, \bibinfo {author}
  {\bibfnamefont {C.}~\bibnamefont {Cohen}}, \bibinfo {author} {\bibfnamefont
  {M.}~\bibnamefont {Dollar}}, \bibinfo {author} {\bibfnamefont
  {A.}~\bibnamefont {Alkauskas}}, \bibinfo {author} {\bibfnamefont
  {M.}~\bibnamefont {Doherty}}, \bibinfo {author} {\bibfnamefont {C.~A.}\
  \bibnamefont {Meriles}},\ and\ \bibinfo {author} {\bibfnamefont {V.~M.}\
  \bibnamefont {Menon}},\ }\bibfield  {title} {\bibinfo {title}
  {Near-deterministic activation of room-temperature quantum emitters in
  hexagonal boron nitride},\ }\href {https://doi.org/10.1364/OPTICA.5.001128}
  {\bibfield  {journal} {\bibinfo  {journal} {Optica}\ }\textbf {\bibinfo
  {volume} {5}},\ \bibinfo {pages} {1128} (\bibinfo {year} {2018})}\BibitemShut
  {NoStop}%
\bibitem [{\citenamefont {Exarhos}\ \emph {et~al.}(2019)\citenamefont
  {Exarhos}, \citenamefont {Hopper}, \citenamefont {Patel}, \citenamefont
  {Doherty},\ and\ \citenamefont {Bassett}}]{Exarhos2019}%
  \BibitemOpen
  \bibfield  {author} {\bibinfo {author} {\bibfnamefont {A.~L.}\ \bibnamefont
  {Exarhos}}, \bibinfo {author} {\bibfnamefont {D.~A.}\ \bibnamefont {Hopper}},
  \bibinfo {author} {\bibfnamefont {R.~N.}\ \bibnamefont {Patel}}, \bibinfo
  {author} {\bibfnamefont {M.~W.}\ \bibnamefont {Doherty}},\ and\ \bibinfo
  {author} {\bibfnamefont {L.~C.}\ \bibnamefont {Bassett}},\ }\bibfield
  {title} {\bibinfo {title} {Magnetic-field-dependent quantum emission in
  hexagonal boron nitride at room temperature},\ }\href
  {https://doi.org/10.1038/s41467-018-08185-8} {\bibfield  {journal} {\bibinfo
  {journal} {Nat. Commun.}\ }\textbf {\bibinfo {volume} {10}},\ \bibinfo
  {pages} {222} (\bibinfo {year} {2019})}\BibitemShut {NoStop}%
\bibitem [{\citenamefont {Tabesh}\ \emph {et~al.}(2021)\citenamefont {Tabesh},
  \citenamefont {Hassanzada}, \citenamefont {Hadian}, \citenamefont {Hashemi},
  \citenamefont {Abdolhosseini~Sarsari},\ and\ \citenamefont
  {Abdi}}]{Tabesh2021}%
  \BibitemOpen
  \bibfield  {author} {\bibinfo {author} {\bibfnamefont {F.~T.}\ \bibnamefont
  {Tabesh}}, \bibinfo {author} {\bibfnamefont {Q.}~\bibnamefont {Hassanzada}},
  \bibinfo {author} {\bibfnamefont {M.}~\bibnamefont {Hadian}}, \bibinfo
  {author} {\bibfnamefont {A.}~\bibnamefont {Hashemi}}, \bibinfo {author}
  {\bibfnamefont {I.}~\bibnamefont {Abdolhosseini~Sarsari}},\ and\ \bibinfo
  {author} {\bibfnamefont {M.}~\bibnamefont {Abdi}},\ }\bibfield  {title}
  {\bibinfo {title} {Strain induced coupling and quantum information processing
  with hexagonal boron nitride quantum emitters},\ }\href
  {https://doi.org/10.1088/2058-9565/ac2f4d} {\bibfield  {journal} {\bibinfo
  {journal} {Quantum Sci. Technol.}\ }\textbf {\bibinfo {volume} {7}},\
  \bibinfo {pages} {015002} (\bibinfo {year} {2021})}\BibitemShut {NoStop}%
\bibitem [{\citenamefont {Liu}\ and\ \citenamefont {Hersam}(2019)}]{Liu2019}%
  \BibitemOpen
  \bibfield  {author} {\bibinfo {author} {\bibfnamefont {X.}~\bibnamefont
  {Liu}}\ and\ \bibinfo {author} {\bibfnamefont {M.~C.}\ \bibnamefont
  {Hersam}},\ }\bibfield  {title} {\bibinfo {title} {2d materials for quantum
  information science},\ }\href {https://doi.org/10.1038/s41578-019-0136-x}
  {\bibfield  {journal} {\bibinfo  {journal} {Nat. Rev. Mater.}\ }\textbf
  {\bibinfo {volume} {4}},\ \bibinfo {pages} {669} (\bibinfo {year}
  {2019})}\BibitemShut {NoStop}%
\bibitem [{\citenamefont {Abdi}(2021)}]{Abdi2021}%
  \BibitemOpen
  \bibfield  {author} {\bibinfo {author} {\bibfnamefont {M.}~\bibnamefont
  {Abdi}},\ }\bibfield  {title} {\bibinfo {title} {Continuous-variable
  multipartite vibrational entanglement},\ }\href
  {https://doi.org/10.1103/physreva.103.043520} {\bibfield  {journal} {\bibinfo
   {journal} {Phys. Rev. A}\ }\textbf {\bibinfo {volume} {103}},\ \bibinfo
  {pages} {043520} (\bibinfo {year} {2021})}\BibitemShut {NoStop}%
\bibitem [{\citenamefont {Kianinia}\ \emph {et~al.}(2020)\citenamefont
  {Kianinia}, \citenamefont {White}, \citenamefont {Fr\"{o}ch}, \citenamefont
  {Bradac},\ and\ \citenamefont {Aharonovich}}]{Kianinia2020}%
  \BibitemOpen
  \bibfield  {author} {\bibinfo {author} {\bibfnamefont {M.}~\bibnamefont
  {Kianinia}}, \bibinfo {author} {\bibfnamefont {S.}~\bibnamefont {White}},
  \bibinfo {author} {\bibfnamefont {J.~E.}\ \bibnamefont {Fr\"{o}ch}}, \bibinfo
  {author} {\bibfnamefont {C.}~\bibnamefont {Bradac}},\ and\ \bibinfo {author}
  {\bibfnamefont {I.}~\bibnamefont {Aharonovich}},\ }\bibfield  {title}
  {\bibinfo {title} {Generation of spin defects in hexagonal boron nitride},\
  }\href {https://doi.org/10.1021/acsphotonics.0c00614} {\bibfield  {journal}
  {\bibinfo  {journal} {ACS Photonics}\ }\textbf {\bibinfo {volume} {7}},\
  \bibinfo {pages} {2147} (\bibinfo {year} {2020})}\BibitemShut {NoStop}%
\bibitem [{\citenamefont {Shaik}\ and\ \citenamefont
  {Palla}(2021)}]{Shaik2021}%
  \BibitemOpen
  \bibfield  {author} {\bibinfo {author} {\bibfnamefont {A.~B.}\ \bibnamefont
  {Shaik}}\ and\ \bibinfo {author} {\bibfnamefont {P.}~\bibnamefont {Palla}},\
  }\bibfield  {title} {\bibinfo {title} {Optical quantum technologies with
  hexagonal boron nitride single photon sources},\ }\href
  {https://doi.org/10.1038/s41598-021-90804-4} {\bibfield  {journal} {\bibinfo
  {journal} {Sci. Rep.}\ }\textbf {\bibinfo {volume} {11}},\ \bibinfo {pages}
  {12285} (\bibinfo {year} {2021})}\BibitemShut {NoStop}%
\bibitem [{\citenamefont {Gao}\ \emph {et~al.}(2021)\citenamefont {Gao},
  \citenamefont {Jiang}, \citenamefont {Llacsahuanga~Allcca}, \citenamefont
  {Shen}, \citenamefont {Sadi}, \citenamefont {Solanki}, \citenamefont {Ju},
  \citenamefont {Xu}, \citenamefont {Upadhyaya}, \citenamefont {Chen},
  \citenamefont {Bhave},\ and\ \citenamefont {Li}}]{Gao2021}%
  \BibitemOpen
  \bibfield  {author} {\bibinfo {author} {\bibfnamefont {X.}~\bibnamefont
  {Gao}}, \bibinfo {author} {\bibfnamefont {B.}~\bibnamefont {Jiang}}, \bibinfo
  {author} {\bibfnamefont {A.~E.}\ \bibnamefont {Llacsahuanga~Allcca}},
  \bibinfo {author} {\bibfnamefont {K.}~\bibnamefont {Shen}}, \bibinfo {author}
  {\bibfnamefont {M.~A.}\ \bibnamefont {Sadi}}, \bibinfo {author}
  {\bibfnamefont {A.~B.}\ \bibnamefont {Solanki}}, \bibinfo {author}
  {\bibfnamefont {P.}~\bibnamefont {Ju}}, \bibinfo {author} {\bibfnamefont
  {Z.}~\bibnamefont {Xu}}, \bibinfo {author} {\bibfnamefont {P.}~\bibnamefont
  {Upadhyaya}}, \bibinfo {author} {\bibfnamefont {Y.~P.}\ \bibnamefont {Chen}},
  \bibinfo {author} {\bibfnamefont {S.~A.}\ \bibnamefont {Bhave}},\ and\
  \bibinfo {author} {\bibfnamefont {T.}~\bibnamefont {Li}},\ }\bibfield
  {title} {\bibinfo {title} {High-contrast plasmonic-enhanced shallow spin
  defects in hexagonal boron nitride for quantum sensing},\ }\href
  {https://doi.org/10.1021/acs.nanolett.1c02495} {\bibfield  {journal}
  {\bibinfo  {journal} {Nano Lett.}\ }\textbf {\bibinfo {volume} {21}},\
  \bibinfo {pages} {7708} (\bibinfo {year} {2021})}\BibitemShut {NoStop}%
\bibitem [{\citenamefont {Mendelson}\ \emph {et~al.}(2021)\citenamefont
  {Mendelson}, \citenamefont {Ritika}, \citenamefont {Kianinia}, \citenamefont
  {Scott}, \citenamefont {Kim}, \citenamefont {Fr\"{o}ch}, \citenamefont
  {Gazzana}, \citenamefont {Westerhausen}, \citenamefont {Xiao}, \citenamefont
  {Mohajerani}, \citenamefont {Strauf}, \citenamefont {Toth}, \citenamefont
  {Aharonovich},\ and\ \citenamefont {Xu}}]{Mendelson2021}%
  \BibitemOpen
  \bibfield  {author} {\bibinfo {author} {\bibfnamefont {N.}~\bibnamefont
  {Mendelson}}, \bibinfo {author} {\bibfnamefont {R.}~\bibnamefont {Ritika}},
  \bibinfo {author} {\bibfnamefont {M.}~\bibnamefont {Kianinia}}, \bibinfo
  {author} {\bibfnamefont {J.}~\bibnamefont {Scott}}, \bibinfo {author}
  {\bibfnamefont {S.}~\bibnamefont {Kim}}, \bibinfo {author} {\bibfnamefont
  {J.~E.}\ \bibnamefont {Fr\"{o}ch}}, \bibinfo {author} {\bibfnamefont
  {C.}~\bibnamefont {Gazzana}}, \bibinfo {author} {\bibfnamefont
  {M.}~\bibnamefont {Westerhausen}}, \bibinfo {author} {\bibfnamefont
  {L.}~\bibnamefont {Xiao}}, \bibinfo {author} {\bibfnamefont {S.~S.}\
  \bibnamefont {Mohajerani}}, \bibinfo {author} {\bibfnamefont
  {S.}~\bibnamefont {Strauf}}, \bibinfo {author} {\bibfnamefont
  {M.}~\bibnamefont {Toth}}, \bibinfo {author} {\bibfnamefont {I.}~\bibnamefont
  {Aharonovich}},\ and\ \bibinfo {author} {\bibfnamefont {Z.}~\bibnamefont
  {Xu}},\ }\bibfield  {title} {\bibinfo {title} {Coupling spin defects in a
  layered material to nanoscale plasmonic cavities},\ }\href
  {https://doi.org/10.1002/adma.202106046} {\bibfield  {journal} {\bibinfo
  {journal} {Adv. Mater.}\ }\textbf {\bibinfo {volume} {34}},\ \bibinfo {pages}
  {2106046} (\bibinfo {year} {2021})}\BibitemShut {NoStop}%
\bibitem [{\citenamefont {Guo}\ \emph {et~al.}(2022)\citenamefont {Guo},
  \citenamefont {Liu}, \citenamefont {Li}, \citenamefont {Yang}, \citenamefont
  {Yu}, \citenamefont {Meng}, \citenamefont {Wang}, \citenamefont {Zeng},
  \citenamefont {Yan}, \citenamefont {Li}, \citenamefont {Wang}, \citenamefont
  {Xu}, \citenamefont {Wang}, \citenamefont {Tang}, \citenamefont {Li},\ and\
  \citenamefont {Guo}}]{Guo2022}%
  \BibitemOpen
  \bibfield  {author} {\bibinfo {author} {\bibfnamefont {N.-J.}\ \bibnamefont
  {Guo}}, \bibinfo {author} {\bibfnamefont {W.}~\bibnamefont {Liu}}, \bibinfo
  {author} {\bibfnamefont {Z.-P.}\ \bibnamefont {Li}}, \bibinfo {author}
  {\bibfnamefont {Y.-Z.}\ \bibnamefont {Yang}}, \bibinfo {author}
  {\bibfnamefont {S.}~\bibnamefont {Yu}}, \bibinfo {author} {\bibfnamefont
  {Y.}~\bibnamefont {Meng}}, \bibinfo {author} {\bibfnamefont {Z.-A.}\
  \bibnamefont {Wang}}, \bibinfo {author} {\bibfnamefont {X.-D.}\ \bibnamefont
  {Zeng}}, \bibinfo {author} {\bibfnamefont {F.-F.}\ \bibnamefont {Yan}},
  \bibinfo {author} {\bibfnamefont {Q.}~\bibnamefont {Li}}, \bibinfo {author}
  {\bibfnamefont {J.-F.}\ \bibnamefont {Wang}}, \bibinfo {author}
  {\bibfnamefont {J.-S.}\ \bibnamefont {Xu}}, \bibinfo {author} {\bibfnamefont
  {Y.-T.}\ \bibnamefont {Wang}}, \bibinfo {author} {\bibfnamefont {J.-S.}\
  \bibnamefont {Tang}}, \bibinfo {author} {\bibfnamefont {C.-F.}\ \bibnamefont
  {Li}},\ and\ \bibinfo {author} {\bibfnamefont {G.-C.}\ \bibnamefont {Guo}},\
  }\bibfield  {title} {\bibinfo {title} {Generation of spin defects by ion
  implantation in hexagonal boron nitride},\ }\href
  {https://doi.org/10.1021/acsomega.1c04564} {\bibfield  {journal} {\bibinfo
  {journal} {ACS Omega}\ }\textbf {\bibinfo {volume} {7}},\ \bibinfo {pages}
  {1733} (\bibinfo {year} {2022})}\BibitemShut {NoStop}%
\bibitem [{\citenamefont {Vaidya}\ \emph {et~al.}(2023)\citenamefont {Vaidya},
  \citenamefont {Gao}, \citenamefont {Dikshit}, \citenamefont {Aharonovich},\
  and\ \citenamefont {Li}}]{Vaidya2023}%
  \BibitemOpen
  \bibfield  {author} {\bibinfo {author} {\bibfnamefont {S.}~\bibnamefont
  {Vaidya}}, \bibinfo {author} {\bibfnamefont {X.}~\bibnamefont {Gao}},
  \bibinfo {author} {\bibfnamefont {S.}~\bibnamefont {Dikshit}}, \bibinfo
  {author} {\bibfnamefont {I.}~\bibnamefont {Aharonovich}},\ and\ \bibinfo
  {author} {\bibfnamefont {T.}~\bibnamefont {Li}},\ }\bibfield  {title}
  {\bibinfo {title} {Quantum sensing and imaging with spin defects in hexagonal
  boron nitride},\ }\href {https://doi.org/10.1080/23746149.2023.2206049}
  {\bibfield  {journal} {\bibinfo  {journal} {Adv. Phys.: X}\ }\textbf
  {\bibinfo {volume} {8}},\ \bibinfo {pages} {2206049} (\bibinfo {year}
  {2023})}\BibitemShut {NoStop}%
\bibitem [{\citenamefont {Das}\ \emph {et~al.}(2024)\citenamefont {Das},
  \citenamefont {Melendez}, \citenamefont {Kao}, \citenamefont
  {Garc\'{\i}a-Monge}, \citenamefont {Russell}, \citenamefont {Li},
  \citenamefont {Watanabe}, \citenamefont {Taniguchi}, \citenamefont {Edgar},
  \citenamefont {Katoch}, \citenamefont {Yang}, \citenamefont {Hammel},\ and\
  \citenamefont {Singh}}]{Das2024}%
  \BibitemOpen
  \bibfield  {author} {\bibinfo {author} {\bibfnamefont {S.}~\bibnamefont
  {Das}}, \bibinfo {author} {\bibfnamefont {A.~L.}\ \bibnamefont {Melendez}},
  \bibinfo {author} {\bibfnamefont {I.-H.}\ \bibnamefont {Kao}}, \bibinfo
  {author} {\bibfnamefont {J.~A.}\ \bibnamefont {Garc\'{\i}a-Monge}}, \bibinfo
  {author} {\bibfnamefont {D.}~\bibnamefont {Russell}}, \bibinfo {author}
  {\bibfnamefont {J.}~\bibnamefont {Li}}, \bibinfo {author} {\bibfnamefont
  {K.}~\bibnamefont {Watanabe}}, \bibinfo {author} {\bibfnamefont
  {T.}~\bibnamefont {Taniguchi}}, \bibinfo {author} {\bibfnamefont {J.~H.}\
  \bibnamefont {Edgar}}, \bibinfo {author} {\bibfnamefont {J.}~\bibnamefont
  {Katoch}}, \bibinfo {author} {\bibfnamefont {F.}~\bibnamefont {Yang}},
  \bibinfo {author} {\bibfnamefont {P.~C.}\ \bibnamefont {Hammel}},\ and\
  \bibinfo {author} {\bibfnamefont {S.}~\bibnamefont {Singh}},\ }\bibfield
  {title} {\bibinfo {title} {Quantum sensing of spin dynamics using
  boron-vacancy centers in hexagonal boron nitride},\ }\href
  {https://doi.org/10.1103/PhysRevLett.133.166704} {\bibfield  {journal}
  {\bibinfo  {journal} {Phys. Rev. Lett.}\ }\textbf {\bibinfo {volume} {133}},\
  \bibinfo {pages} {166704} (\bibinfo {year} {2024})}\BibitemShut {NoStop}%
\bibitem [{\citenamefont {Wong}\ \emph {et~al.}(2015)\citenamefont {Wong},
  \citenamefont {Velasco}, \citenamefont {Ju}, \citenamefont {Lee},
  \citenamefont {Kahn}, \citenamefont {Tsai}, \citenamefont {Germany},
  \citenamefont {Taniguchi}, \citenamefont {Watanabe}, \citenamefont {Zettl},
  \citenamefont {Wang},\ and\ \citenamefont {Crommie}}]{Wong2015}%
  \BibitemOpen
  \bibfield  {author} {\bibinfo {author} {\bibfnamefont {D.}~\bibnamefont
  {Wong}}, \bibinfo {author} {\bibfnamefont {J.}~\bibnamefont {Velasco}},
  \bibinfo {author} {\bibfnamefont {L.}~\bibnamefont {Ju}}, \bibinfo {author}
  {\bibfnamefont {J.}~\bibnamefont {Lee}}, \bibinfo {author} {\bibfnamefont
  {S.}~\bibnamefont {Kahn}}, \bibinfo {author} {\bibfnamefont {H.-Z.}\
  \bibnamefont {Tsai}}, \bibinfo {author} {\bibfnamefont {C.}~\bibnamefont
  {Germany}}, \bibinfo {author} {\bibfnamefont {T.}~\bibnamefont {Taniguchi}},
  \bibinfo {author} {\bibfnamefont {K.}~\bibnamefont {Watanabe}}, \bibinfo
  {author} {\bibfnamefont {A.}~\bibnamefont {Zettl}}, \bibinfo {author}
  {\bibfnamefont {F.}~\bibnamefont {Wang}},\ and\ \bibinfo {author}
  {\bibfnamefont {M.~F.}\ \bibnamefont {Crommie}},\ }\bibfield  {title}
  {\bibinfo {title} {Characterization and manipulation of individual defects in
  insulating hexagonal boron nitride using scanning tunnelling microscopy},\
  }\href {https://doi.org/10.1038/nnano.2015.188} {\bibfield  {journal}
  {\bibinfo  {journal} {Nat. Nanotechnol.}\ }\textbf {\bibinfo {volume} {10}},\
  \bibinfo {pages} {949} (\bibinfo {year} {2015})}\BibitemShut {NoStop}%
\bibitem [{\citenamefont {Gao}\ \emph {et~al.}(2022)\citenamefont {Gao},
  \citenamefont {Vaidya}, \citenamefont {Li}, \citenamefont {Ju}, \citenamefont
  {Jiang}, \citenamefont {Xu}, \citenamefont {Allcca}, \citenamefont {Shen},
  \citenamefont {Taniguchi}, \citenamefont {Watanabe}, \citenamefont {Bhave},
  \citenamefont {Chen}, \citenamefont {Ping},\ and\ \citenamefont
  {Li}}]{Gao2022}%
  \BibitemOpen
  \bibfield  {author} {\bibinfo {author} {\bibfnamefont {X.}~\bibnamefont
  {Gao}}, \bibinfo {author} {\bibfnamefont {S.}~\bibnamefont {Vaidya}},
  \bibinfo {author} {\bibfnamefont {K.}~\bibnamefont {Li}}, \bibinfo {author}
  {\bibfnamefont {P.}~\bibnamefont {Ju}}, \bibinfo {author} {\bibfnamefont
  {B.}~\bibnamefont {Jiang}}, \bibinfo {author} {\bibfnamefont
  {Z.}~\bibnamefont {Xu}}, \bibinfo {author} {\bibfnamefont {A.~E.~L.}\
  \bibnamefont {Allcca}}, \bibinfo {author} {\bibfnamefont {K.}~\bibnamefont
  {Shen}}, \bibinfo {author} {\bibfnamefont {T.}~\bibnamefont {Taniguchi}},
  \bibinfo {author} {\bibfnamefont {K.}~\bibnamefont {Watanabe}}, \bibinfo
  {author} {\bibfnamefont {S.~A.}\ \bibnamefont {Bhave}}, \bibinfo {author}
  {\bibfnamefont {Y.~P.}\ \bibnamefont {Chen}}, \bibinfo {author}
  {\bibfnamefont {Y.}~\bibnamefont {Ping}},\ and\ \bibinfo {author}
  {\bibfnamefont {T.}~\bibnamefont {Li}},\ }\bibfield  {title} {\bibinfo
  {title} {Nuclear spin polarization and control in hexagonal boron nitride},\
  }\href {https://doi.org/10.1038/s41563-022-01329-8} {\bibfield  {journal}
  {\bibinfo  {journal} {Nat. Mater.}\ }\textbf {\bibinfo {volume} {21}},\
  \bibinfo {pages} {1024} (\bibinfo {year} {2022})}\BibitemShut {NoStop}%
\bibitem [{\citenamefont {Tabesh}\ \emph {et~al.}(2023)\citenamefont {Tabesh},
  \citenamefont {Fani}, \citenamefont {Pedernales}, \citenamefont {Plenio},\
  and\ \citenamefont {Abdi}}]{Tabesh2023}%
  \BibitemOpen
  \bibfield  {author} {\bibinfo {author} {\bibfnamefont {F.~T.}\ \bibnamefont
  {Tabesh}}, \bibinfo {author} {\bibfnamefont {M.}~\bibnamefont {Fani}},
  \bibinfo {author} {\bibfnamefont {J.~S.}\ \bibnamefont {Pedernales}},
  \bibinfo {author} {\bibfnamefont {M.~B.}\ \bibnamefont {Plenio}},\ and\
  \bibinfo {author} {\bibfnamefont {M.}~\bibnamefont {Abdi}},\ }\bibfield
  {title} {\bibinfo {title} {Active hyperpolarization of the nuclear spin
  lattice: Application to hexagonal boron nitride color centers},\ }\href
  {https://doi.org/10.1103/PhysRevB.107.214307} {\bibfield  {journal} {\bibinfo
   {journal} {Phys. Rev. B}\ }\textbf {\bibinfo {volume} {107}},\ \bibinfo
  {pages} {214307} (\bibinfo {year} {2023})}\BibitemShut {NoStop}%
\bibitem [{\citenamefont {Sakuldee}\ and\ \citenamefont
  {Abdi}(2025)}]{Sakuldee2025}%
  \BibitemOpen
  \bibfield  {author} {\bibinfo {author} {\bibfnamefont {F.}~\bibnamefont
  {Sakuldee}}\ and\ \bibinfo {author} {\bibfnamefont {M.}~\bibnamefont
  {Abdi}},\ }\bibfield  {title} {\bibinfo {title} {Synchronous manipulation of
  nuclear spins via boron-vacancy centers in hexagonal boron nitride},\ }\href
  {https://doi.org/10.1103/vlqs-h75g} {\bibfield  {journal} {\bibinfo
  {journal} {Phys. Rev. Appl.}\ }\textbf {\bibinfo {volume} {24}},\ \bibinfo
  {pages} {024036} (\bibinfo {year} {2025})}\BibitemShut {NoStop}%
\bibitem [{\citenamefont {Lee}\ \emph {et~al.}(2025)\citenamefont {Lee},
  \citenamefont {Liu}, \citenamefont {Zhang}, \citenamefont {Kim},
  \citenamefont {Gong}, \citenamefont {Du}, \citenamefont {Pham}, \citenamefont
  {Poirier}, \citenamefont {Hao}, \citenamefont {Edgar}, \citenamefont {Kim},
  \citenamefont {Zu}, \citenamefont {Davis},\ and\ \citenamefont
  {Yao}}]{Lee2025}%
  \BibitemOpen
  \bibfield  {author} {\bibinfo {author} {\bibfnamefont {W.}~\bibnamefont
  {Lee}}, \bibinfo {author} {\bibfnamefont {V.~S.}\ \bibnamefont {Liu}},
  \bibinfo {author} {\bibfnamefont {Z.}~\bibnamefont {Zhang}}, \bibinfo
  {author} {\bibfnamefont {S.}~\bibnamefont {Kim}}, \bibinfo {author}
  {\bibfnamefont {R.}~\bibnamefont {Gong}}, \bibinfo {author} {\bibfnamefont
  {X.}~\bibnamefont {Du}}, \bibinfo {author} {\bibfnamefont {K.}~\bibnamefont
  {Pham}}, \bibinfo {author} {\bibfnamefont {T.}~\bibnamefont {Poirier}},
  \bibinfo {author} {\bibfnamefont {Z.}~\bibnamefont {Hao}}, \bibinfo {author}
  {\bibfnamefont {J.~H.}\ \bibnamefont {Edgar}}, \bibinfo {author}
  {\bibfnamefont {P.}~\bibnamefont {Kim}}, \bibinfo {author} {\bibfnamefont
  {C.}~\bibnamefont {Zu}}, \bibinfo {author} {\bibfnamefont {E.~J.}\
  \bibnamefont {Davis}},\ and\ \bibinfo {author} {\bibfnamefont {N.~Y.}\
  \bibnamefont {Yao}},\ }\bibfield  {title} {\bibinfo {title} {Intrinsic
  high-fidelity spin polarization of charged vacancies in hexagonal boron
  nitride},\ }\href {https://doi.org/10.1103/PhysRevLett.134.096202} {\bibfield
   {journal} {\bibinfo  {journal} {Phys. Rev. Lett.}\ }\textbf {\bibinfo
  {volume} {134}},\ \bibinfo {pages} {096202} (\bibinfo {year}
  {2025})}\BibitemShut {NoStop}%
\bibitem [{\citenamefont {Liu}\ \emph {et~al.}(2025)\citenamefont {Liu},
  \citenamefont {Gong}, \citenamefont {Huang}, \citenamefont {Jin},
  \citenamefont {Du}, \citenamefont {He}, \citenamefont {Janzen}, \citenamefont
  {Yang}, \citenamefont {Henriksen}, \citenamefont {Edgar}, \citenamefont
  {Galli},\ and\ \citenamefont {Zu}}]{Liu2025}%
  \BibitemOpen
  \bibfield  {author} {\bibinfo {author} {\bibfnamefont {Z.}~\bibnamefont
  {Liu}}, \bibinfo {author} {\bibfnamefont {R.}~\bibnamefont {Gong}}, \bibinfo
  {author} {\bibfnamefont {B.}~\bibnamefont {Huang}}, \bibinfo {author}
  {\bibfnamefont {Y.}~\bibnamefont {Jin}}, \bibinfo {author} {\bibfnamefont
  {X.}~\bibnamefont {Du}}, \bibinfo {author} {\bibfnamefont {G.}~\bibnamefont
  {He}}, \bibinfo {author} {\bibfnamefont {E.}~\bibnamefont {Janzen}}, \bibinfo
  {author} {\bibfnamefont {L.}~\bibnamefont {Yang}}, \bibinfo {author}
  {\bibfnamefont {E.~A.}\ \bibnamefont {Henriksen}}, \bibinfo {author}
  {\bibfnamefont {J.~H.}\ \bibnamefont {Edgar}}, \bibinfo {author}
  {\bibfnamefont {G.}~\bibnamefont {Galli}},\ and\ \bibinfo {author}
  {\bibfnamefont {C.}~\bibnamefont {Zu}},\ }\bibfield  {title} {\bibinfo
  {title} {Temperature-dependent spin-phonon coupling of boron-vacancy centers
  in hexagonal boron nitride},\ }\href
  {https://doi.org/10.1103/PhysRevB.111.024108} {\bibfield  {journal} {\bibinfo
   {journal} {Phys. Rev. B}\ }\textbf {\bibinfo {volume} {111}},\ \bibinfo
  {pages} {024108} (\bibinfo {year} {2025})}\BibitemShut {NoStop}%
\bibitem [{\citenamefont {Iv\'{a}dy}\ \emph {et~al.}(2020)\citenamefont
  {Iv\'{a}dy}, \citenamefont {Barcza}, \citenamefont {Thiering}, \citenamefont
  {Li}, \citenamefont {Hamdi}, \citenamefont {Chou}, \citenamefont {Legeza},\
  and\ \citenamefont {Gali}}]{Ivady2020}%
  \BibitemOpen
  \bibfield  {author} {\bibinfo {author} {\bibfnamefont {V.}~\bibnamefont
  {Iv\'{a}dy}}, \bibinfo {author} {\bibfnamefont {G.}~\bibnamefont {Barcza}},
  \bibinfo {author} {\bibfnamefont {G.}~\bibnamefont {Thiering}}, \bibinfo
  {author} {\bibfnamefont {S.}~\bibnamefont {Li}}, \bibinfo {author}
  {\bibfnamefont {H.}~\bibnamefont {Hamdi}}, \bibinfo {author} {\bibfnamefont
  {J.-P.}\ \bibnamefont {Chou}}, \bibinfo {author} {\bibfnamefont
  {O.}~\bibnamefont {Legeza}},\ and\ \bibinfo {author} {\bibfnamefont
  {A.}~\bibnamefont {Gali}},\ }\bibfield  {title} {\bibinfo {title} {Ab initio
  theory of the negatively charged boron vacancy qubit in hexagonal boron
  nitride},\ }\href {https://doi.org/10.1038/s41524-020-0305-x} {\bibfield
  {journal} {\bibinfo  {journal} {npj Comput. Mater.}\ }\textbf {\bibinfo
  {volume} {6}},\ \bibinfo {pages} {41} (\bibinfo {year} {2020})}\BibitemShut
  {NoStop}%
\bibitem [{\citenamefont {Gracheva}\ \emph {et~al.}(2023)\citenamefont
  {Gracheva}, \citenamefont {Murzakhanov}, \citenamefont {Mamin}, \citenamefont
  {Sadovnikova}, \citenamefont {Gabbasov}, \citenamefont {Mokhov},\ and\
  \citenamefont {Gafurov}}]{Gracheva2023}%
  \BibitemOpen
  \bibfield  {author} {\bibinfo {author} {\bibfnamefont {I.~N.}\ \bibnamefont
  {Gracheva}}, \bibinfo {author} {\bibfnamefont {F.~F.}\ \bibnamefont
  {Murzakhanov}}, \bibinfo {author} {\bibfnamefont {G.~V.}\ \bibnamefont
  {Mamin}}, \bibinfo {author} {\bibfnamefont {M.~A.}\ \bibnamefont
  {Sadovnikova}}, \bibinfo {author} {\bibfnamefont {B.~F.}\ \bibnamefont
  {Gabbasov}}, \bibinfo {author} {\bibfnamefont {E.~N.}\ \bibnamefont
  {Mokhov}},\ and\ \bibinfo {author} {\bibfnamefont {M.~R.}\ \bibnamefont
  {Gafurov}},\ }\bibfield  {title} {\bibinfo {title} {Symmetry of the hyperfine
  and quadrupole interactions of boron vacancies in a hexagonal boron
  nitride},\ }\href {https://doi.org/10.1021/acs.jpcc.2c08716} {\bibfield
  {journal} {\bibinfo  {journal} {J. Phys. Chem. C}\ }\textbf {\bibinfo
  {volume} {127}},\ \bibinfo {pages} {3634} (\bibinfo {year}
  {2023})}\BibitemShut {NoStop}%
\bibitem [{\citenamefont {Sakuldee}\ and\ \citenamefont {Abdi}()}]{supp}%
  \BibitemOpen
  \bibfield  {author} {\bibinfo {author} {\bibfnamefont {F.}~\bibnamefont
  {Sakuldee}}\ and\ \bibinfo {author} {\bibfnamefont {M.}~\bibnamefont
  {Abdi}},\ }\href@noop {} {\bibinfo {title} {Arbitrary manipulation of nuclear
  spins in hexagonal boron nitride: Supplemental material}},\ \bibinfo
  {howpublished} {\url{URL_will_be_inserted_by_publisher}}\BibitemShut
  {NoStop}%
\bibitem [{\citenamefont {Hahn}(1950)}]{Hahn1950}%
  \BibitemOpen
  \bibfield  {author} {\bibinfo {author} {\bibfnamefont {E.~L.}\ \bibnamefont
  {Hahn}},\ }\bibfield  {title} {\bibinfo {title} {Spin echoes},\ }\href
  {https://doi.org/10.1103/PhysRev.80.580} {\bibfield  {journal} {\bibinfo
  {journal} {Phys. Rev.}\ }\textbf {\bibinfo {volume} {80}},\ \bibinfo {pages}
  {580} (\bibinfo {year} {1950})}\BibitemShut {NoStop}%
\bibitem [{\citenamefont {Nielsen}(2002)}]{Nielson20022}%
  \BibitemOpen
  \bibfield  {author} {\bibinfo {author} {\bibfnamefont {M.~A.}\ \bibnamefont
  {Nielsen}},\ }\bibfield  {title} {\bibinfo {title} {A simple formula for the
  average gate fidelity of a quantum dynamical operation},\ }\href
  {https://doi.org/https://doi.org/10.1016/S0375-9601(02)01272-0} {\bibfield
  {journal} {\bibinfo  {journal} {Phys. Lett. A}\ }\textbf {\bibinfo {volume}
  {303}},\ \bibinfo {pages} {249} (\bibinfo {year} {2002})}\BibitemShut
  {NoStop}%
\bibitem [{\citenamefont {Nielsen}\ and\ \citenamefont
  {Chuang}(2010)}]{Nielsen_Chuang_2010}%
  \BibitemOpen
  \bibfield  {author} {\bibinfo {author} {\bibfnamefont {M.~A.}\ \bibnamefont
  {Nielsen}}\ and\ \bibinfo {author} {\bibfnamefont {I.~L.}\ \bibnamefont
  {Chuang}},\ }\href@noop {} {\emph {\bibinfo {title} {Quantum Computation and
  Quantum Information: 10th Anniversary Edition}}}\ (\bibinfo  {publisher}
  {Cambridge University Press},\ \bibinfo {year} {2010})\BibitemShut {NoStop}%
\bibitem [{\citenamefont {Gilchrist}\ \emph {et~al.}(2005)\citenamefont
  {Gilchrist}, \citenamefont {Langford},\ and\ \citenamefont
  {Nielsen}}]{Gilchrist2005}%
  \BibitemOpen
  \bibfield  {author} {\bibinfo {author} {\bibfnamefont {A.}~\bibnamefont
  {Gilchrist}}, \bibinfo {author} {\bibfnamefont {N.~K.}\ \bibnamefont
  {Langford}},\ and\ \bibinfo {author} {\bibfnamefont {M.~A.}\ \bibnamefont
  {Nielsen}},\ }\bibfield  {title} {\bibinfo {title} {Distance measures to
  compare real and ideal quantum processes},\ }\href
  {https://doi.org/10.1103/PhysRevA.71.062310} {\bibfield  {journal} {\bibinfo
  {journal} {Phys. Rev. A}\ }\textbf {\bibinfo {volume} {71}},\ \bibinfo
  {pages} {062310} (\bibinfo {year} {2005})}\BibitemShut {NoStop}%
\bibitem [{\citenamefont {Ramsay}\ \emph {et~al.}(2023)\citenamefont {Ramsay},
  \citenamefont {Hekmati}, \citenamefont {Patrickson}, \citenamefont {Baber},
  \citenamefont {Arvidsson-Shukur}, \citenamefont {Bennett},\ and\
  \citenamefont {Luxmoore}}]{Ramsay2023}%
  \BibitemOpen
  \bibfield  {author} {\bibinfo {author} {\bibfnamefont {A.~J.}\ \bibnamefont
  {Ramsay}}, \bibinfo {author} {\bibfnamefont {R.}~\bibnamefont {Hekmati}},
  \bibinfo {author} {\bibfnamefont {C.~J.}\ \bibnamefont {Patrickson}},
  \bibinfo {author} {\bibfnamefont {S.}~\bibnamefont {Baber}}, \bibinfo
  {author} {\bibfnamefont {D.~R.~M.}\ \bibnamefont {Arvidsson-Shukur}},
  \bibinfo {author} {\bibfnamefont {A.~J.}\ \bibnamefont {Bennett}},\ and\
  \bibinfo {author} {\bibfnamefont {I.~J.}\ \bibnamefont {Luxmoore}},\
  }\bibfield  {title} {\bibinfo {title} {Coherence protection of spin qubits in
  hexagonal boron nitride},\ }\href
  {https://doi.org/10.1038/s41467-023-36196-7} {\bibfield  {journal} {\bibinfo
  {journal} {Nat. Commun.}\ }\textbf {\bibinfo {volume} {14}},\ \bibinfo
  {pages} {461} (\bibinfo {year} {2023})}\BibitemShut {NoStop}%
\bibitem [{\citenamefont {Johansson}\ \emph {et~al.}(2012)\citenamefont
  {Johansson}, \citenamefont {Nation},\ and\ \citenamefont {Nori}}]{QuTip1}%
  \BibitemOpen
  \bibfield  {author} {\bibinfo {author} {\bibfnamefont {J.}~\bibnamefont
  {Johansson}}, \bibinfo {author} {\bibfnamefont {P.}~\bibnamefont {Nation}},\
  and\ \bibinfo {author} {\bibfnamefont {F.}~\bibnamefont {Nori}},\ }\bibfield
  {title} {\bibinfo {title} {Qutip: An open-source python framework for the
  dynamics of open quantum systems},\ }\href
  {https://doi.org/https://doi.org/10.1016/j.cpc.2012.02.021} {\bibfield
  {journal} {\bibinfo  {journal} {Comput. Phys. Commun.}\ }\textbf {\bibinfo
  {volume} {183}},\ \bibinfo {pages} {1760} (\bibinfo {year}
  {2012})}\BibitemShut {NoStop}%
\bibitem [{\citenamefont {Johansson}\ \emph {et~al.}(2013)\citenamefont
  {Johansson}, \citenamefont {Nation},\ and\ \citenamefont {Nori}}]{QuTip2}%
  \BibitemOpen
  \bibfield  {author} {\bibinfo {author} {\bibfnamefont {J.}~\bibnamefont
  {Johansson}}, \bibinfo {author} {\bibfnamefont {P.}~\bibnamefont {Nation}},\
  and\ \bibinfo {author} {\bibfnamefont {F.}~\bibnamefont {Nori}},\ }\bibfield
  {title} {\bibinfo {title} {Qutip 2: A python framework for the dynamics of
  open quantum systems},\ }\href
  {https://doi.org/https://doi.org/10.1016/j.cpc.2012.11.019} {\bibfield
  {journal} {\bibinfo  {journal} {Comput. Phys. Commun.}\ }\textbf {\bibinfo
  {volume} {184}},\ \bibinfo {pages} {1234} (\bibinfo {year}
  {2013})}\BibitemShut {NoStop}%
\bibitem [{\citenamefont {Shor}(1996)}]{Shor1996}%
  \BibitemOpen
  \bibfield  {author} {\bibinfo {author} {\bibfnamefont {P.}~\bibnamefont
  {Shor}},\ }\bibfield  {title} {\bibinfo {title} {Fault-tolerant quantum
  computation},\ }in\ \href {https://doi.org/10.1109/SFCS.1996.548464} {\emph
  {\bibinfo {booktitle} {Proceedings of 37th Conference on Foundations of
  Computer Science}}}\ (\bibinfo {year} {1996})\ pp.\ \bibinfo {pages}
  {56--65}\BibitemShut {NoStop}%
\bibitem [{\citenamefont {Greenberger}\ \emph {et~al.}(1990)\citenamefont
  {Greenberger}, \citenamefont {Horne}, \citenamefont {Shimony},\ and\
  \citenamefont {Zeilinger}}]{Greenberger1990}%
  \BibitemOpen
  \bibfield  {author} {\bibinfo {author} {\bibfnamefont {D.~M.}\ \bibnamefont
  {Greenberger}}, \bibinfo {author} {\bibfnamefont {M.~A.}\ \bibnamefont
  {Horne}}, \bibinfo {author} {\bibfnamefont {A.}~\bibnamefont {Shimony}},\
  and\ \bibinfo {author} {\bibfnamefont {A.}~\bibnamefont {Zeilinger}},\
  }\bibfield  {title} {\bibinfo {title} {Bell's theorem without inequalities},\
  }\href {https://doi.org/10.1119/1.16243} {\bibfield  {journal} {\bibinfo
  {journal} {Am. J. Phys.}\ }\textbf {\bibinfo {volume} {58}},\ \bibinfo
  {pages} {1131} (\bibinfo {year} {1990})}\BibitemShut {NoStop}%
\bibitem [{\citenamefont {Haase}\ \emph {et~al.}(2018)\citenamefont {Haase},
  \citenamefont {Wang}, \citenamefont {Casanova},\ and\ \citenamefont
  {Plenio}}]{Haase2018}%
  \BibitemOpen
  \bibfield  {author} {\bibinfo {author} {\bibfnamefont {J.~F.}\ \bibnamefont
  {Haase}}, \bibinfo {author} {\bibfnamefont {Z.-Y.}\ \bibnamefont {Wang}},
  \bibinfo {author} {\bibfnamefont {J.}~\bibnamefont {Casanova}},\ and\
  \bibinfo {author} {\bibfnamefont {M.~B.}\ \bibnamefont {Plenio}},\ }\bibfield
   {title} {\bibinfo {title} {Soft quantum control for highly selective
  interactions among joint quantum systems},\ }\href
  {https://doi.org/10.1103/PhysRevLett.121.050402} {\bibfield  {journal}
  {\bibinfo  {journal} {Phys. Rev. Lett.}\ }\textbf {\bibinfo {volume} {121}},\
  \bibinfo {pages} {050402} (\bibinfo {year} {2018})}\BibitemShut {NoStop}%
\bibitem [{\citenamefont {Chen}\ and\ \citenamefont {Cory}(2025)}]{Chen2025}%
  \BibitemOpen
  \bibfield  {author} {\bibinfo {author} {\bibfnamefont {J.}~\bibnamefont
  {Chen}}\ and\ \bibinfo {author} {\bibfnamefont {D.}~\bibnamefont {Cory}},\
  }\bibfield  {title} {\bibinfo {title} {Engineering precise and robust
  effective hamiltonians},\ }\Eprint {https://arxiv.org/abs/2506.20730}
  {arXiv:2506.20730 [quant-ph]}  (\bibinfo {year} {2025})\BibitemShut {NoStop}%
\bibitem [{\citenamefont {Jin}\ and\ \citenamefont {Jing}(2025)}]{Jin2025}%
  \BibitemOpen
  \bibfield  {author} {\bibinfo {author} {\bibfnamefont {Z.-Y.}\ \bibnamefont
  {Jin}}\ and\ \bibinfo {author} {\bibfnamefont {J.}~\bibnamefont {Jing}},\
  }\bibfield  {title} {\bibinfo {title} {Universal perspective on nonadiabatic
  quantum control},\ }\href {https://doi.org/10.1103/PhysRevA.111.012406}
  {\bibfield  {journal} {\bibinfo  {journal} {Phys. Rev. A}\ }\textbf {\bibinfo
  {volume} {111}},\ \bibinfo {pages} {012406} (\bibinfo {year}
  {2025})}\BibitemShut {NoStop}%
\bibitem [{\citenamefont {Murzakhanov}\ \emph {et~al.}(2022)\citenamefont
  {Murzakhanov}, \citenamefont {Mamin}, \citenamefont {Orlinskii},
  \citenamefont {Gerstmann}, \citenamefont {Schmidt}, \citenamefont
  {Biktagirov}, \citenamefont {Aharonovich}, \citenamefont {Gottscholl},
  \citenamefont {Sperlich}, \citenamefont {Dyakonov},\ and\ \citenamefont
  {Soltamov}}]{Murzakhanov2022}%
  \BibitemOpen
  \bibfield  {author} {\bibinfo {author} {\bibfnamefont {F.~F.}\ \bibnamefont
  {Murzakhanov}}, \bibinfo {author} {\bibfnamefont {G.~V.}\ \bibnamefont
  {Mamin}}, \bibinfo {author} {\bibfnamefont {S.~B.}\ \bibnamefont
  {Orlinskii}}, \bibinfo {author} {\bibfnamefont {U.}~\bibnamefont
  {Gerstmann}}, \bibinfo {author} {\bibfnamefont {W.~G.}\ \bibnamefont
  {Schmidt}}, \bibinfo {author} {\bibfnamefont {T.}~\bibnamefont {Biktagirov}},
  \bibinfo {author} {\bibfnamefont {I.}~\bibnamefont {Aharonovich}}, \bibinfo
  {author} {\bibfnamefont {A.}~\bibnamefont {Gottscholl}}, \bibinfo {author}
  {\bibfnamefont {A.}~\bibnamefont {Sperlich}}, \bibinfo {author}
  {\bibfnamefont {V.}~\bibnamefont {Dyakonov}},\ and\ \bibinfo {author}
  {\bibfnamefont {V.~A.}\ \bibnamefont {Soltamov}},\ }\bibfield  {title}
  {\bibinfo {title} {Electron--nuclear coherent coupling and nuclear spin
  readout through optically polarized vb--spin states in hbn},\ }\href
  {https://doi.org/10.1021/acs.nanolett.1c04610} {\bibfield  {journal}
  {\bibinfo  {journal} {Nano Lett.}\ }\textbf {\bibinfo {volume} {22}},\
  \bibinfo {pages} {2718} (\bibinfo {year} {2022})}\BibitemShut {NoStop}%
\bibitem [{\citenamefont {Haase}\ \emph {et~al.}(2003)\citenamefont {Haase},
  \citenamefont {Eckert}, \citenamefont {Siegel}, \citenamefont {Eschrig},
  \citenamefont {Müller},\ and\ \citenamefont {Steglich}}]{Haase2003}%
  \BibitemOpen
  \bibfield  {author} {\bibinfo {author} {\bibfnamefont {J.}~\bibnamefont
  {Haase}}, \bibinfo {author} {\bibfnamefont {D.}~\bibnamefont {Eckert}},
  \bibinfo {author} {\bibfnamefont {H.}~\bibnamefont {Siegel}}, \bibinfo
  {author} {\bibfnamefont {H.}~\bibnamefont {Eschrig}}, \bibinfo {author}
  {\bibfnamefont {K.}~\bibnamefont {Müller}},\ and\ \bibinfo {author}
  {\bibfnamefont {F.}~\bibnamefont {Steglich}},\ }\bibfield  {title} {\bibinfo
  {title} {High-field nmr in pulsed magnets},\ }\href
  {https://doi.org/https://doi.org/10.1016/S0926-2040(03)00015-8} {\bibfield
  {journal} {\bibinfo  {journal} {Solid State Nuclear Magnetic Resonance}\
  }\textbf {\bibinfo {volume} {23}},\ \bibinfo {pages} {263} (\bibinfo {year}
  {2003})}\BibitemShut {NoStop}%
\bibitem [{\citenamefont {K{\"u}hne}\ and\ \citenamefont
  {Ihara}(2024)}]{Kuhne2024}%
  \BibitemOpen
  \bibfield  {author} {\bibinfo {author} {\bibfnamefont {H.}~\bibnamefont
  {K{\"u}hne}}\ and\ \bibinfo {author} {\bibfnamefont {Y.}~\bibnamefont
  {Ihara}},\ }\bibfield  {title} {\bibinfo {title} {Nuclear magnetic resonance
  spectroscopy in pulsed magnetic fields},\ }\href
  {https://doi.org/10.1080/00107514.2024.2393009} {\bibfield  {journal}
  {\bibinfo  {journal} {Contemporary Physics}\ }\textbf {\bibinfo {volume}
  {65}},\ \bibinfo {pages} {40} (\bibinfo {year} {2024})}\BibitemShut {NoStop}%
\bibitem [{\citenamefont {Chacko}\ \emph {et~al.}(2024)\citenamefont {Chacko},
  \citenamefont {Louis-Joseph},\ and\ \citenamefont {Abergel}}]{Chacko2024}%
  \BibitemOpen
  \bibfield  {author} {\bibinfo {author} {\bibfnamefont {V.~F. T.~J.}\
  \bibnamefont {Chacko}}, \bibinfo {author} {\bibfnamefont {A.}~\bibnamefont
  {Louis-Joseph}},\ and\ \bibinfo {author} {\bibfnamefont {D.}~\bibnamefont
  {Abergel}},\ }\bibfield  {title} {\bibinfo {title} {Multimode masers of
  thermally polarized nuclear spins in solution nmr},\ }\href
  {https://doi.org/10.1103/PhysRevLett.133.158001} {\bibfield  {journal}
  {\bibinfo  {journal} {Phys. Rev. Lett.}\ }\textbf {\bibinfo {volume} {133}},\
  \bibinfo {pages} {158001} (\bibinfo {year} {2024})}\BibitemShut {NoStop}%
\bibitem [{\citenamefont {Ng}\ \emph {et~al.}(2024)\citenamefont {Ng},
  \citenamefont {Wen}, \citenamefont {Attwood}, \citenamefont {Jones},
  \citenamefont {Oxborrow}, \citenamefont {Alford},\ and\ \citenamefont
  {Arroo}}]{Ng2024}%
  \BibitemOpen
  \bibfield  {author} {\bibinfo {author} {\bibfnamefont {W.}~\bibnamefont
  {Ng}}, \bibinfo {author} {\bibfnamefont {Y.}~\bibnamefont {Wen}}, \bibinfo
  {author} {\bibfnamefont {M.}~\bibnamefont {Attwood}}, \bibinfo {author}
  {\bibfnamefont {D.~C.}\ \bibnamefont {Jones}}, \bibinfo {author}
  {\bibfnamefont {M.}~\bibnamefont {Oxborrow}}, \bibinfo {author}
  {\bibfnamefont {N.~M.}\ \bibnamefont {Alford}},\ and\ \bibinfo {author}
  {\bibfnamefont {D.~M.}\ \bibnamefont {Arroo}},\ }\bibfield  {title} {\bibinfo
  {title} {“maser-in-a-shoebox”: A portable plug-and-play maser device at
  room temperature and zero magnetic field},\ }\href
  {https://doi.org/10.1063/5.0181318} {\bibfield  {journal} {\bibinfo
  {journal} {Applied Physics Letters}\ }\textbf {\bibinfo {volume} {124}},\
  \bibinfo {pages} {044004} (\bibinfo {year} {2024})}\BibitemShut {NoStop}%
\bibitem [{\citenamefont {Zollitsch}\ and\ \citenamefont
  {Breeze}(2025)}]{Zollitsch2025}%
  \BibitemOpen
  \bibfield  {author} {\bibinfo {author} {\bibfnamefont {C.~W.}\ \bibnamefont
  {Zollitsch}}\ and\ \bibinfo {author} {\bibfnamefont {J.~D.}\ \bibnamefont
  {Breeze}},\ }\bibfield  {title} {\bibinfo {title} {Quantum theory of the
  diamond maser: Stimulated and superradiant emission},\ }\href
  {https://doi.org/10.1103/PhysRevA.111.053714} {\bibfield  {journal} {\bibinfo
   {journal} {Phys. Rev. A}\ }\textbf {\bibinfo {volume} {111}},\ \bibinfo
  {pages} {053714} (\bibinfo {year} {2025})}\BibitemShut {NoStop}%
\bibitem [{\citenamefont {Breeze}\ \emph {et~al.}(2018)\citenamefont {Breeze},
  \citenamefont {Salvadori}, \citenamefont {Sathian}, \citenamefont {Alford},\
  and\ \citenamefont {Kay}}]{Breeze2018}%
  \BibitemOpen
  \bibfield  {author} {\bibinfo {author} {\bibfnamefont {J.~D.}\ \bibnamefont
  {Breeze}}, \bibinfo {author} {\bibfnamefont {E.}~\bibnamefont {Salvadori}},
  \bibinfo {author} {\bibfnamefont {J.}~\bibnamefont {Sathian}}, \bibinfo
  {author} {\bibfnamefont {N.~M.}\ \bibnamefont {Alford}},\ and\ \bibinfo
  {author} {\bibfnamefont {C.~W.~M.}\ \bibnamefont {Kay}},\ }\bibfield  {title}
  {\bibinfo {title} {Continuous-wave room-temperature diamond maser},\ }\href
  {https://doi.org/10.1038/nature25970} {\bibfield  {journal} {\bibinfo
  {journal} {Nature}\ }\textbf {\bibinfo {volume} {555}},\ \bibinfo {pages}
  {493} (\bibinfo {year} {2018})}\BibitemShut {NoStop}%
\bibitem [{\citenamefont {Ru}\ \emph {et~al.}(2025)\citenamefont {Ru},
  \citenamefont {An}, \citenamefont {Liang}, \citenamefont {Jiang},
  \citenamefont {Li}, \citenamefont {Lyu}, \citenamefont {Zhou}, \citenamefont
  {Cai}, \citenamefont {Yang}, \citenamefont {He}, \citenamefont {Cernansky},
  \citenamefont {Teo}, \citenamefont {Mukherjee}, \citenamefont {Bettiol},
  \citenamefont {Z\'u\~niga Perez}, \citenamefont {Jelezko},\ and\
  \citenamefont {Gao}}]{Ru2025}%
  \BibitemOpen
  \bibfield  {author} {\bibinfo {author} {\bibfnamefont {S.}~\bibnamefont
  {Ru}}, \bibinfo {author} {\bibfnamefont {L.}~\bibnamefont {An}}, \bibinfo
  {author} {\bibfnamefont {H.}~\bibnamefont {Liang}}, \bibinfo {author}
  {\bibfnamefont {Z.}~\bibnamefont {Jiang}}, \bibinfo {author} {\bibfnamefont
  {Z.}~\bibnamefont {Li}}, \bibinfo {author} {\bibfnamefont {X.}~\bibnamefont
  {Lyu}}, \bibinfo {author} {\bibfnamefont {F.}~\bibnamefont {Zhou}}, \bibinfo
  {author} {\bibfnamefont {H.}~\bibnamefont {Cai}}, \bibinfo {author}
  {\bibfnamefont {Y.}~\bibnamefont {Yang}}, \bibinfo {author} {\bibfnamefont
  {R.}~\bibnamefont {He}}, \bibinfo {author} {\bibfnamefont {R.}~\bibnamefont
  {Cernansky}}, \bibinfo {author} {\bibfnamefont {E.~H.~T.}\ \bibnamefont
  {Teo}}, \bibinfo {author} {\bibfnamefont {M.}~\bibnamefont {Mukherjee}},
  \bibinfo {author} {\bibfnamefont {A.~A.}\ \bibnamefont {Bettiol}}, \bibinfo
  {author} {\bibfnamefont {J.}~\bibnamefont {Z\'u\~niga Perez}}, \bibinfo
  {author} {\bibfnamefont {F.}~\bibnamefont {Jelezko}},\ and\ \bibinfo {author}
  {\bibfnamefont {W.}~\bibnamefont {Gao}},\ }\bibfield  {title} {\bibinfo
  {title} {Room-temperature electrical readout of spin defects in van der waals
  materials},\ }\href {https://doi.org/10.1103/dlzw-dhsr} {\bibfield  {journal}
  {\bibinfo  {journal} {Phys. Rev. Lett.}\ }\textbf {\bibinfo {volume} {135}},\
  \bibinfo {pages} {220802} (\bibinfo {year} {2025})}\BibitemShut {NoStop}%
\end{thebibliography}%

\end{document}